\documentclass[%
 reprint,
superscriptaddress,
 prb,
]{revtex4-2}
\usepackage{graphicx}% Include figure files
\usepackage{dcolumn}% Align table columns on decimal point
\usepackage{bm}% bold math
\usepackage[utf8]{inputenc} % allow utf8 unicode characters
\usepackage{float}
\usepackage{appendix, comment}
\usepackage{chemformula}
\usepackage{amsfonts}

\usepackage[colorlinks=true,linkcolor=blue,anchorcolor=red,citecolor=blue,urlcolor=blue]{hyperref}

\usepackage[mathlines]{lineno}% Enable numbering of text and display math
\begin{document}

%\title{Haldane Valleytronics: \\Logic Gate and Berry Curvature Memory}
%\title{Haldane-Modified Haldane Valley Gapless Semiconductor}
\title{Valley Gapless Semiconductor: Models and Applications}

\author{Kok Wai Lee}
\affiliation{Science, Mathematics and Technology, Singapore University of Technology and Design, Singapore 487372, Singapore}

\author{Pei-Hao Fu}
\email{phy.phfu@gmail.com}
\affiliation{School of Physics and Materials Science, Guangzhou University, Guangzhou 510006, China}
\affiliation{Science, Mathematics and Technology, Singapore University of Technology and Design, Singapore 487372, Singapore}

\author{Jun-Feng Liu}
\affiliation{School of Physics and Materials Science, Guangzhou University, Guangzhou 510006, China}

\author{Ching Hua Lee}
\affiliation{Department of Physics, National University of Singapore, Singapore 117542}

\author{Yee Sin Ang}
\email{yeesin\_ang@sutd.edu.sg}
\affiliation{Science, Mathematics and Technology, Singapore University of Technology and Design, Singapore 487372, Singapore}

%%%%%%%%%%%%%%%%%%%%%%%%%%%%%%%%%%%%%%%%%%%%%%%%%%%%%%%%%%%%%%%%%%%%%%%%%%%%%
\begin{abstract}
The emerging field of valleytronics harnesses the valley degree of freedom of electrons, akin to how electronic and spintronic devices utilize the charge and spin degrees of freedom of electrons respectively.
The engineering of valleytronic devices typically relies on the coupling between valley and other degrees of freedom such as spin, giving rise to valley-spintronics where an external magnetic field manipulates the information stored in valleys.
Here, a valley gapless semiconductor is proposed as a potential electrically controlled valleytronic platform because the valley degree of freedom is coupled to the carrier type, i.e., electrons and holes.
The valley degree of freedom can be electrically controlled by tuning the carrier type via the device gate voltage. 
We demonstrate the proposal for realizing a valley gapless semiconductor in the honeycomb lattice with the Haldane and modified Haldane models.
The system's valley-carrier coupling is further studied for its transport properties in an all-electrically controlled valley filter device setting. 
Our work highlights the significance of the valley gapless semiconductor for valleytronic devices.
\end{abstract}

\maketitle

 %\caption{The realization of VGS and the relevant phase in (top panel) honeycomb lattice. Top panel: (a) a schematic for hopping in the Haldane model ($t_H$) and the modified Haldane model ($t_{MH}$), which determine the system to be (b) an indirect-gap semiconductor with $t_{MH}<t_{H}$, (c) VGS with $t_{MH} = t_{H}$, and (d) metal with $t_{MH}>t_{H}$. The Haldane hopping strength is $t_{H} = 0.1t$ where the  nearest-neighbor hopping strength serves as the energy unit ($t = 1$). The momentum is exhibited in the unit of $k_0 = 2\pi/\sqrt{3}a$. (a) indirect-gap semiconductors with $\left\vert \Delta _{2}\right\vert<\left\vert \Delta _{1}\right\vert $, (b) valley gapless semiconductor with $\left\vert \Delta _{2}\right\vert=\left\vert \Delta _{1}\right\vert $ and metal with $\left\vert \Delta _{2}\right\vert>\left\vert \Delta _{1}\right\vert $.}

\section{Introduction}
The emerging field of valleytronics \cite{10.1063/5.0112893, liu2019valleytronics, schaibley2016valleytronics, yu2015valley, xu2014spin, luo2024valleytronics, https://doi.org/10.1002/adom.202302900,  tyulnev2024valleytronics, doi:10.1021/acsaelm.4c02276, doi:10.1021/acsnano.4c12812, jiang2025chiral, PhysRevB.111.L140404, PhysRevB.111.L121201, PhysRevB.111.075421, herrmann2025nonlinear, PhysRevLett.134.026904, PhysRevB.111.165307, PhysRevB.111.L140416}
harnesses the valley degree of freedom of electrons, similar to how electronic and spintronic \cite{RevModPhys.76.323} devices utilize the charge and spin degrees of freedom of electrons \cite{xu2014spin} respectively.
Valleytronic devices are found in materials such as \ch{AlAs} \cite{PhysRevLett.97.186404,PhysRevB.78.233306,PhysRevLett.121.036802}, ferrovalley materials \cite{tong2016concepts,Shen_2018,lai2019two, hu2020concepts, doi:10.1021/acsanm.4c06703}, transition metal dichalcogenide (TMDC) monolayers \cite{cao2012valley,zeng2012valley,jones2013optical,suzuki2014valley,aivazian2015magnetic,wang2017valley,doi:10.1126/sciadv.1603113,PhysRevLett.119.137401,wu2019intrinsic,li2020room,PhysRevLett.108.196802, https://doi.org/10.1002/smll.202402139} and bilayer systems \cite{gong2013magnetoelectric,wu2013electrical,jones2014spin,sui2015gate,shimazaki2015generation,lee2016electrical,PhysRevLett.124.217403,PhysRevLett.124.037701,zhang2023layer}. 
Such materials exhibit band structures with two distinct local extrema in the first Brillouin zone, representing degenerate and nonequivalent valleys. 
The significant separation between valleys effectively suppresses their coupling %CH: "Interaction" refers to 4-operator processes
\cite{PhysRevLett.97.146805,PhysRevLett.97.016801,PhysRevLett.97.196804}, thus making the valley index an independent degree of freedom similar to the internal degrees of freedom of electrons. 
This unique feature opens up vast possibilities for functional valleytronic devices capable of storing and processing digital information via selective manipulation of electron occupation in specific valleys which have potential applications in quantum computation \cite{gong2013magnetoelectric}, reversible logic circuits \cite{PhysRevB.96.245410}, low-power neuromorphic computing \cite{chen2022room}, thermoelectric materials \cite{xin2018valleytronics}, supercurrent \cite{PhysRevResearch.5.023029,PhysRevB.97.241403,PhysRevB.101.245428}, Cooper pair splitter \cite{PhysRevB.101.125406} and other
novel devices \cite{PhysRevB.92.155427,PhysRevLett.118.096602,PhysRevLett.106.176802,Qiu_2019,LIU2023108550, doi:10.1021/acs.nanolett.2c03947, PhysRevLett.132.156301}. 

The key challenge to designing valleytronic devices involves generating and manipulating valley-polarized carriers. 
The degeneracy of the two valleys, protected by both the time-reversal and inversion symmetries, limits their direct application in information storage. 
To address this limitation, methods have been proposed to break the inversion symmetry and induce a valley Hall effect \cite{PhysRevLett.99.236809, PhysRevLett.132.096302, PhysRevLett.133.196603, PhysRevLett.131.246301} and a valley-polarized quantum anomalous Hall effect \cite{PhysRevLett.119.046403,PhysRevLett.112.106802,PhysRevB.92.041404,PhysRevB.91.045404}, accumulating valley-polarized carriers perpendicular to an external in-plane electric field. 
In zigzag graphene nanoribbons, the existence of robust edge states enables the proposal of a valley filter and valley valve where the gates in the point contact control the longitudinal edge current polarization \cite{rycerz2007valley}. 
Recent studies on TMDC monolayers also have demonstrated valley-spin coupling due to the broken inversion symmetry \cite{suzuki2014valley}. 
This allows an indirect manipulation of valley-polarized carriers via external fields such as circularly polarized optical excitation \cite{cao2012valley,zeng2012valley,jones2013optical}, magnetic fields and proximity effects \cite{PhysRevLett.108.196802,suzuki2014valley,aivazian2015magnetic,wang2017valley,doi:10.1126/sciadv.1603113,PhysRevLett.119.137401}. 
In bilayer systems, the valley-layer coupling allows electrically controlled valley polarization \cite{sui2015gate,PhysRevLett.124.037701,zhang2023layer}.
Out-of-plane electric polarization or external electric fields can disrupt valley degeneracy in different layers which facilitates electrically controlled valley polarization.

Based on the proposed spin gapless semiconductor \cite{PhysRevLett.100.156404}, the recent concept of the valley gapless semiconductor (VGS) offers another viable approach to manipulating valley-polarized carriers \cite{guo2024proposal}.
The VGS is the valley analogue of the spin gapless semiconductor. Both possess gapless band structures at the Fermi level which enable the excitation of charge carriers without threshold energy. The spin gapless semiconductor involves electron-hole pairs fully spin-polarized whereas the VGS involves electron-hole pairs fully valley-polarized.
Generally, the VGS consists of two types. The first type is characterized by a gapless dispersion in one valley and a semiconducting dispersion in the other valley, leading to valley polarization within the semiconducting gap. On the other hand, the second type is characterized by the touching between the conduction band edge of one valley and the valence band edge of the other valley at the Fermi level. 
Here, we focus on the latter where the valley degree of freedom is coupled to the carrier type. As a result, the electron carriers are contributed by one valley whereas the hole carriers are contributed by the other valley. This unique valley-carrier coupling enables the VGS to serve as a potential platform for electrically controlled valleytronics. 
The valley-carrier coupling feature has been experimentally observed in the strain-induced bipolar quantum Hall phase of the \ch{Pb_{1-x}Sn_{x}Se} Dirac system where electron and hole chiral states emerge within different valleys and coexist in a single quantum well without interference \cite{PhysRevLett.132.166601}. Nevertheless, we wish to highlight that our work explicitly proposes the valley-carrier coupling by introducing a theoretical model of the valley gapless semiconductor which is distinct from that reported in \cite{PhysRevLett.132.166601}.

In this work, we introduce the concept of the VGS via a two-band Dirac-type Hamiltonian where a valley-dependent scalar potential term induces the VGS by shifting the energies of the Dirac points. We further demonstrate its realization in the honeycomb lattice where its emergence is governed by the competition between the Haldane \cite{PhysRevLett.61.2015} and modified Haldane terms \cite{PhysRevLett.120.086603}. Appropriate parameter choices lead to the coupling between the valley degree of freedom and carrier type, paving the way for valleytronic device applications. To this end, we characterize its transport properties in an all-electrically controlled valley filter device setting by computing the valley-dependent conductance and valley polarization efficiency.

%In this work, we \textcolor{blue}{demonstrate} that the \textcolor{blue}{graphene honeycomb} lattice with the Haldane and modified Haldane terms leads to a variety of valleytronic phases. 
%a low-energy two-band Dirac-type Hamiltonian and explore their applications in all-electrically controlled valleytronics, including valley filters, valley valves, and valley-selective reversible logic gates. 
%By inducing a valley-dependent scalar potential that shifts the energies of the Dirac points, we realize VGSs. 
%Our low-energy Hamiltonian captures the behavior of a broad range of semiconductors in the honeycomb lattice. 
%
%By extending this idea to a quantum anomalous Hall insulator Hamiltonian \cite{PhysRevLett.113.136403}, we also demonstrate the existence of similar VGS states in the square lattice. 
%
%Within this lattice system, appropriate parameter choices enable the coupling between the valley degree of freedom and carriers.
%paving the way for valleytronic devices.
%To achieve valley polarization, we investigate single- and double-barrier structures with gate potentials which serve as potential valley filters and valley valves respectively. 
%We analyze the gate-controlled valley-dependent conductance and assess the efficiency of valley polarization. 
%
\begin{figure*}
    \centering
    \includegraphics[width = \textwidth]{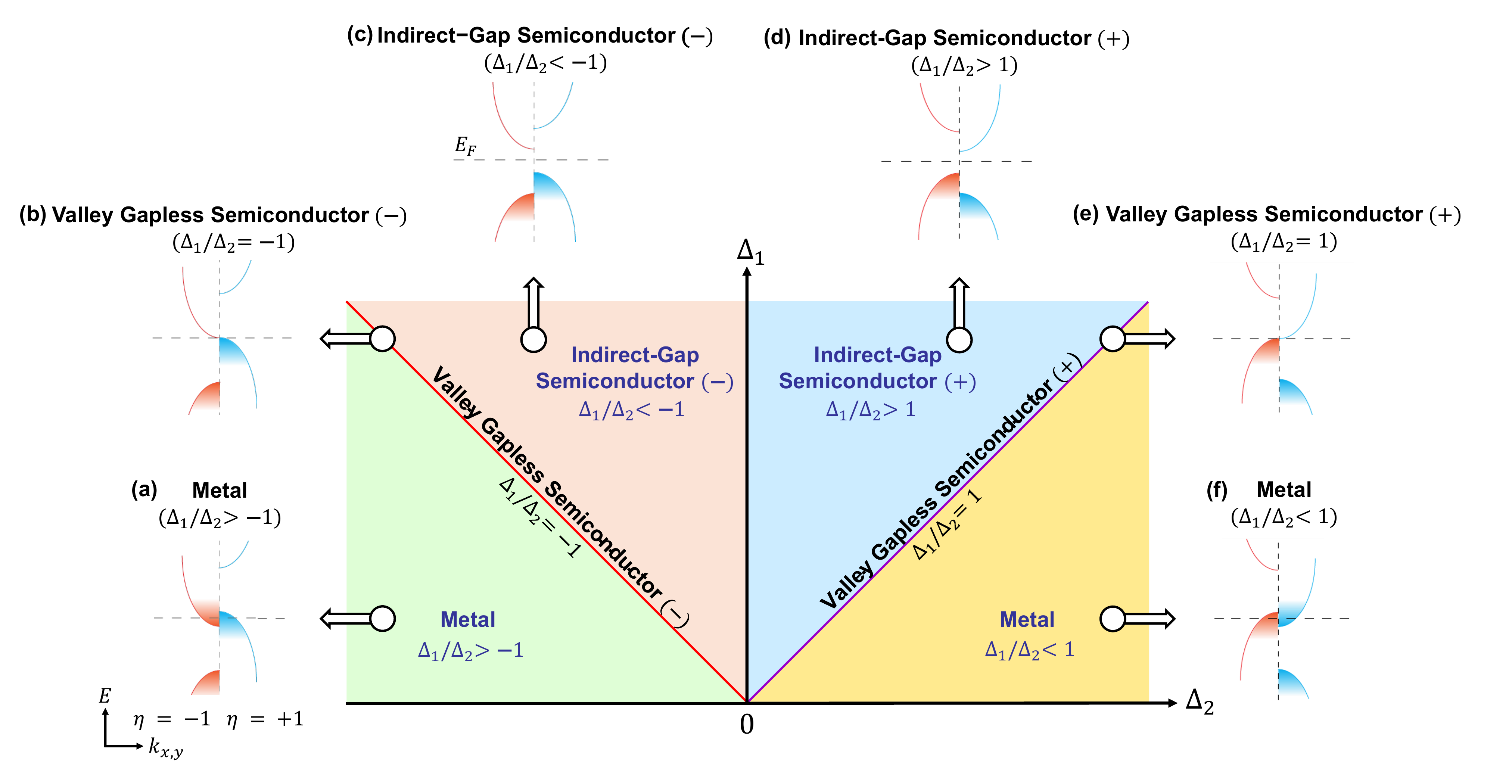}
    \caption{Schematic phase diagram of the system in the $\Delta_{1}$-$\Delta_{2}$ plane with insets (a)-(f) depicting schematic band structures for each regime and $(\pm)$ indicates the sign of $\Delta_{2}$. For $\Delta_{2} < 0$, the ranges of values are $\Delta_{1}/\Delta_{2} < -1$, $\Delta_{1}/\Delta_{2} = -1$ and $\Delta_{1}/\Delta_{2} > -1$ for indirect-gap semiconductor $(-)$, valley gapless semiconductor $(-)$ and metal respectively. For $\Delta_{2} > 0$, the ranges of values are $\Delta_{1}/\Delta_{2} > 1$, $\Delta_{1}/\Delta_{2} = 1$ and $\Delta_{1}/\Delta_{2} < 1$ for indirect-gap semiconductor $(+)$, valley gapless semiconductor $(+)$ and metal respectively.}
    \label{band}
\end{figure*}
The remainder of this paper is organized as follows. 
In Sec. \ref{concept}, we introduce the concept of the VGS and discuss its characteristics. Sec. \ref{model} presents the model
for the VGS in the honeycomb lattice. In Sec. \ref{device}, we delve into all-electrically controlled valleytronics, focusing on the valley filter. 
The results are supported by details which include spinor components and scattering coefficients in Appendices \ref{tunneling}-\ref{transmission}. 
Finally, the results are summarized in Sec. \ref{conclusion}.

\section{Valley Gapless Semiconductor}\label{concept} 
We demonstrate the concept of a two-dimensional VGS via the following two-band valley-dependent Dirac-type Hamiltonian:
\begin{equation}
 \mathcal{H}_{\eta }\left( \bm{k}\right) =\eta \hbar v_{F_{x}}k_{x}\tau
_{x}+\hbar v_{F_{y}}k_{y}\tau _{y} 
+ \Delta_{1} \tau _{z} - \eta \Delta _{2}\tau_{0}
\text{,}  
\label{heta}
\end{equation}
where $\bm{k} = \left(k_{x},k_{y}\right) $,
%\eta = \pm 1$ denotes the valley index 
$v_{F_{x\left(y\right) }}$ is the Fermi velocity along the $x$($y$)-direction,
$\eta = +1$ $(-1)$ represents the $K$ $(K^{\prime})$ valley and $\tau_{0}$ $[\bm{\tau} = (\tau_{x}, \tau_{y}, \tau_{z})]$ is the unit (Pauli) matrix representing a two-state degree of freedom. For instance, the two-state degree of freedom is the sublattice pseudospin for the honeycomb lattice and spin for the Qi-Wu-Zhang model \cite{PhysRevB.74.085308}.
The last two terms are the Dirac mass term, $\Delta _{1}$ and valley-dependent scalar potential, $\Delta _{2}$ which shift the energies of the Dirac points.
Here, we assume that the intervalley interaction is negligible.

The Dirac mass term, $\Delta_{1}$ can be realized by breaking the time-reversal symmetry 
\cite{PhysRevLett.61.2015, PhysRevB.87.155415}. On the other hand, realizing the valley-dependent scalar potential, $\Delta_{2}$ requires breaking the inversion and time-reversal symmetries 
\cite{PhysRevLett.120.086603, PhysRevB.87.155415}. To demonstrate this, Eq. (\ref{heta}) 
can be rewritten explicitly as the following $4 \times 4$ Hamiltonian:
\begin{equation}
    \mathcal{H}(\bm{k}) = \hbar v_{F} (k_{x} \eta_{z} \tau_{x} + k_{y} \eta_{0} \tau_{y} ) + \Delta_{1} \eta_{z} \tau_{z} - \Delta_{2} \eta_{z} \tau_{0}
    \text{,}
    \label{4x4}
\end{equation}
where $\eta_{0}$ $[\bm{\eta} = (\eta_{x}, \eta_{y}, \eta_{z})]$ and $\tau_{0}$  $[\bm{\tau} = (\tau_{x}, \tau_{y}, \tau_{z})]$ are the unit (Pauli) matrices denoting the valley and two-state degrees of freedom respectively. The time-reversal symmetry operator is an antiunitary operator defined as $T = \eta_{x} \tau_{0} \hat{K}$ where $\hat{K}$ is the complex conjugation operator. The time-reversal symmetry is preserved if  $\mathcal{H}(\bm{k})$ satisfies $T\mathcal{H}(\bm{k}) T^{-1} = \mathcal{H}(-\bm{k})$. 
The inversion symmetry operator is a unitary operator defined as $I = \eta_{x} \tau_{x}$. Likewise, the inversion symmetry is preserved if $\mathcal{H}(\bm{k})$ satisfies 
$I\mathcal{H}(\bm{k}) I^{-1} = \mathcal{H}(-\bm{k})$. 
The following $3$ cases are considered: $1$. When $\Delta_1 = \Delta_{2} = 0$, $T\mathcal{H}(\bm{k}) T^{-1} = \mathcal{H}(-\bm{k})$ and $I\mathcal{H}(\bm{k}) I^{-1} = \mathcal{H}(-\bm{k})$. Both the time-reversal and inversion symmetries are preserved. $2$. When $\Delta_{1} \neq 0$ and  $\Delta_{2} = 0$, $T\mathcal{H}(\bm{k}) T^{-1} \neq \mathcal{H}(-\bm{k})$ and $I\mathcal{H}(\bm{k}) I^{-1} = \mathcal{H}(-\bm{k})$. The time-reversal symmetry is broken whereas the inversion symmetry is preserved. $3$. When $\Delta_{1} = 0$ and $\Delta_{2} \neq 0$, $T\mathcal{H}(\bm{k}) T^{-1} \neq \mathcal{H}(-\bm{k})$ and $I\mathcal{H}(\bm{k}) I^{-1} \neq \mathcal{H}(-\bm{k})$. Both the time-reversal and inversion symmetries are broken. 
$\Delta_{1}$ has been experimentally demonstrated by periodically modulating ultracold atoms in optical lattices \cite{jotzu2014experimental}, driving monolayer graphene with a femtosecond pulse of circularly polarized light \cite{mciver2020light} and applying an out-of-plane electric field on AB-stacked \ch{MoTe_{2} /WSe_{2}} heterobilayers \cite{li2021quantum}. $\Delta_{2}$ also has been experimentally accomplished by incorporating on-site magnetization modulation in a microwave-scale gyromagnetic photonic crystal \cite{PhysRevLett.125.263603} and designing braided interconnections in an LC circuit lattice \cite{yang2021observation}.

The competition between $\Delta _{1}$ and $\Delta _{2}$ is crucial for determining the phases of the system as indicated by the following eigenvalues of Eq. (\ref{heta}):
%
%around each valley
%
\begin{equation}
    E_{\eta ,\pm }\left( \bm{k}\right) =-\eta \Delta _{2}\pm \sqrt{\left(
    \hbar v_{F}\left\vert \bm{k}\right\vert \right) ^{2}+\Delta _{1}^{2}}%
    \text{.}  
    \label{ek}
\end{equation}%
For $\Delta _{1} = \Delta _{2} = 0$, the system is gapless which is known as a Dirac semimetal, supporting two Dirac points localized at the same energy. 
When $\Delta _{1} \neq 0$, the Dirac points are destroyed and the band structures illustrated in Fig. \ref{band} reveal an intravalley gap, denoted as $\Delta _{intra}$, which represents the energy difference between the conduction and valence band edges within each individual valley. 
The intravalley gap is related to the Dirac mass term via
\begin{equation}
    \Delta _{intra}=\left\vert E_{\eta
    ,+}-E_{\eta ,-}\right\vert =2\left\vert \Delta _{1}\right\vert \text{.}
    \label{intravalley gap}
\end{equation}
The system manifests a semiconducting phase when the Fermi level lies within this gap.

When both $\Delta _{1}$ and $\Delta _{2} \neq 0$, the system's classification as a semiconductor or a metal depends on the relative strength between $\Delta _{1}$ and $\Delta _{2}$. 
In this regard, an intervalley gap, denoted as $\Delta_{inter}$, is defined to be the minimum energy difference between the edges of the conduction band in one valley and the valence band in the other valley which is expressed as  
\begin{equation}
    \Delta_{inter} = \min \left[ \Delta _{+},\Delta _{-}\right] \text{,}  
    \label{intervalley gap}
\end{equation}
where
\begin{equation}
    \Delta _{\eta} = E_{\eta,+} - E_{\bar{\eta},-} = 2\left( \left\vert \Delta _{1}\right\vert
    -\eta \Delta _{2} \right) \text{,}
\end{equation}
and $\bar{\eta} = - \eta$.
Generally, we can have either $\Delta_{2} > 0$ or $\Delta_{2} < 0$, bringing us a total of $6$ phases [Figs. \ref{band}(a) to (f)].
For $\Delta_{2} < 0$, 
the phases are termed
indirect-gap semiconductor $(-)$, valley gapless semiconductor $(-)$ and metal. 
For $\Delta_{2} > 0$, the 
phases are termed indirect-gap semiconductor $(+)$, valley gapless semiconductor $(+)$ and metal. 
In both cases, $(\pm)$ indicates the sign of $\Delta_{2}$. 
Since the value of $\Delta_{2}$ does not affect the phase transition conditions, the subsequent analysis focuses on $\Delta_{2} > 0$. 
By default, we refer to the $(+)$ phases and drop $(\pm)$ unless otherwise stated.

When $\Delta_{1}/\Delta_{2} > 1$, 
the system is an indirect-gap semiconductor [Fig. \ref{band}(d)] with both non-zero intravalley gap ($\Delta_{intra} \not = 0$) and intervalley gap ($\Delta _{inter} > 
0$). On the other hand, the system becomes metallic when $\Delta_{1}/\Delta_{2} < 1$ even if an indirect gap between the valleys exists [Fig. \ref{band}(f)] because it is now negative ($\Delta_{inter} < 0 $). 
In the regime of $\left\vert \Delta _{2}\right\vert - \left\vert \Delta _{1}\right\vert <  E < \left\vert \Delta _{2}\right\vert +\left\vert \Delta _{1}\right\vert $ and $-\left( \left\vert \Delta _{2}\right\vert + \left\vert \Delta _{1}\right\vert \right) < E <\left\vert \Delta _{1}\right\vert -\left\vert \Delta _{2}\right\vert $, the system is of valley-half-metallic feature, the electrons and holes are contributed by the $\eta = +1$ ($-1$) and  $\eta =-1$ ($+1$)\ valleys for positive (negative) $\Delta _{2}$ respectively. 
This phase resembles a spinful half-metal in ferromagnets where the carrier is contributed by one of the spin subbands.
At $\Delta_{1}/\Delta_{2} = 1$, the intervalley gap vanishes ($ \Delta_{inter} = 0 $).
Consequently, this critical phase between the two aforementioned phases is dubbed as the VGS [Fig. \ref{band}(e)]. 
For $-2\left\vert \Delta _{1}\right\vert < E < 2\left\vert \Delta _{1}\right\vert $, the charge carriers are valley-polarized. 
Similar to the spin gapless semiconductor, the VGS supports electrons with $\eta = +1$ ($-1$) valley polarization and holes with $\eta = -1$ ($+1$) valley polarization for positive (negative) $\Delta _{2}$ respectively.
The VGS, being characterized by a continuous energy distribution of valley-polarized charge carriers as compared to the indirect-gap semiconducting and metallic phases, 
exhibits potential applications in electrically controlled valleytronics.

\begin{figure*}
    \centering
    \includegraphics[width = \textwidth]{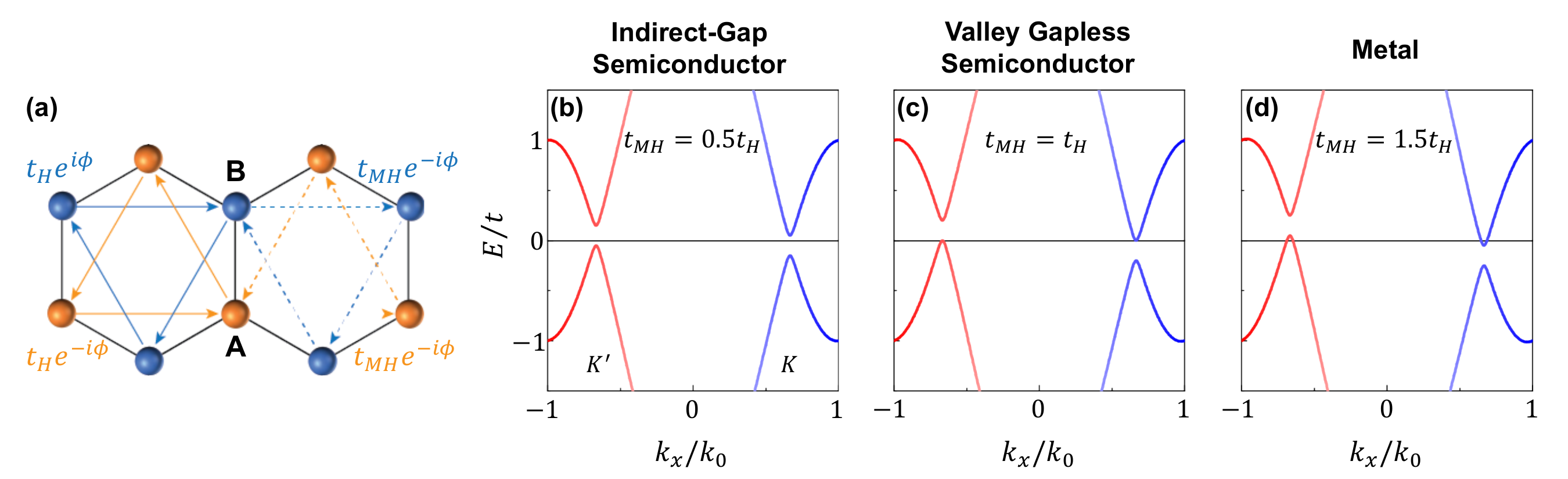}
    \caption{(a) Schematic of the Haldane ($t_{H}$) and modified Haldane ($t_{MH}$) next-nearest-neighbour hoppings where their competition determines the phase of the system, i.e., (b) indirect-gap semiconductor when $t_{MH} < t_{H}$, 
    (c) valley gapless semiconductor when $t_{MH} = t_{H}$ and 
    (d) metal when $t_{MH} > t_{H}$.
    The nearest-neighbor hopping strength $t$ serves as the energy unit $(t = 1)$ of $t_{H}$ and $t_{MH}$.  
    For (b) to (d), the Haldane hopping strength is set at $t_{H} = 0.1t$. 
    The unit of $k_{x}$ is $k_0 = 2\pi/\sqrt{3}a$ where $a$ is the graphene lattice constant taken to be 1.}
    \label{lattice}
\end{figure*}

\section{VGS Realization in the Honeycomb Lattice}
\label{model}
%Before designing valleytronic devices which make use of the VGS, 
We demonstrate the realization of the VGS, semiconducting and metallic phases in the honeycomb lattice subject to the Haldane and modified Haldane terms \cite{PhysRevLett.120.086603,PhysRevB.99.161404} [Fig. \ref{lattice}(a)] which is described by the following Hamiltonian: 
\begin{eqnarray}
    \mathcal{H}_{hc} &=& - t \sum_{\left\langle i,j\right\rangle }c_{i}^{\dagger}c_{j}  \nonumber \\
    && +\frac{1}{3\sqrt{3}}\sum_{\left\langle \left\langle i,j\right\rangle
    \right\rangle }\left( t_{H}e^{-iv_{ij}\phi }+t_{MH}e^{-iv_{ij}^{\prime }\phi
    ^{\prime }}\right) c_{i}^{\dagger }c_{j}  \nonumber \\
    && + H.c.
    \label{hc}
\end{eqnarray}
Here, $c_{i}^{\dag} (c_{i})$ is the spinless fermionic creation (annihilation) operator acting at the $i$th site, the summation of $\langle ij\rangle $ ($\langle \langle ij\rangle \rangle $) runs over all the nearest (next-nearest)-neighbor sites and $H.c.$ denotes the Hermitian conjugate.
The first term describes the nearest-neighbor hopping with strength $t$. 
The second term describes the Haldane (modified Haldane) next-nearest-neighbor hopping with strength $t_{H}$ ($t_{MH}$) \cite{PhysRevLett.61.2015,PhysRevLett.120.086603}. 
The nearest-neighbor hopping strength $t$ serves as the energy unit $(t = 1)$ of $t_{H}$ and $t_{MH}$.  
$v_{ij} = +1$ $(-1)$ denotes the counterclockwise (clockwise) hopping for both sublattices A and B. 
On the other hand, $v_{ij}^{\prime } = \pm 1$ ($v_{ij}^{\prime } = \mp 1$) for sublattice A (B) as illustrated in Fig. \ref{lattice}(a).
The competition between the Haldane and modified Haldane terms determines the emergence of the VGS.

The electronic states in the vicinity of the $\bm{K}_{\eta = \pm 1} = [\eta 4\pi/ (3\sqrt{3}a), 0]$ points are described by the following low-energy effective Hamiltonian:
\begin{equation}
 \mathcal{H}_{hc}^{\eta }\left( \bm{k}\right) = \mathcal{H}_{\eta }\left( \bm{k%
}\right) -\frac{1}{\sqrt{3}}\left( t_{H}\cos \phi + t_{MH}\cos \phi ^{\prime
}\right) \tau _{0}\text{,}  \label{hkhc}
\end{equation}%
where $\hbar v_{F_{x}}=\hbar v_{F_{y}} = 3t/2a$ ($a = 1$ is the honeycomb lattice constant), $\Delta _{1}=\eta t_{H}\sin \phi $, $\Delta _{2} = t_{MH}\sin \phi ^{\prime}$. 
For the honeycomb lattice, the two-state degree of freedom is the sublattice pseudospin denoted by the unit (Pauli) matrix, $\tau_{0}$ $[\bm{\tau} = (\tau_{x}, \tau_{y}, \tau_{z})]$ \cite{PhysRevB.87.155415, PhysRevLett.110.026603}.
%
%\textcolor{blue}{$\tau _{0}$ $[\bm{\tau} = (\tau_{x}, \tau_{y}, \tau_{z})]$ is the unit (Pauli) matrix representing the sublattice pseudospin degree of freedom \cite{PhysRevB.87.155415,PhysRevLett.110.026603}.} 
%
For $\phi = 0$ or $\phi ^{\prime } = 0$, the two Haldane terms behave as conventional next-nearest-neighbor hopping terms which shift the two Dirac cones equally in energy scale. 
We focus on the case where $\phi =\phi ^{\prime } = \pi/2$ and we also assume that both $t_{H}$ and $t_{MH} > 0$ without loss of generality.
The Haldane term breaks the time-reversal symmetry which therefore acts as a Dirac mass term. The modified Haldane term, otherwise known as the staggered Haldane term \cite{PhysRevB.87.155415}, breaks both the inversion and time-reversal symmetries. Thus, the modified Haldane term induces a valley-dependent scalar potential shifting the energies of the Dirac cones 
\cite{PhysRevLett.120.086603}. 
Both terms are necessary to realize the VGS and the other two phases.

Following the discussion in Sec. \ref{concept}, the VGS occurs when $\Delta_{intra}=2\left\vert \Delta _{1}\right\vert =2t_{H}\not=0$ and $\Delta_{inter}=2\left( t_{H}-t_{MH}\right) =0$ or $t_{H}=t_{MH}$. On the other hand, the indirect-gap semiconducting and metallic phases are obtained when $t_{MH} < t_{H}$ and $t_{MH} > t_{H}$ respectively as depicted in Figs. \ref{lattice}(b) to (d).

The Haldane model has been experimentally realized via ultracold atoms in an optical lattice \cite{jotzu2014experimental}, 
AB-stacked \ch{MoTe_{2}/WSe_{2}} moiré bilayers
\cite{zhao2024realization}, 
light-wave-controlled monolayer hexagonal boron nitride \cite{mitra2024light} and 
monolayer graphene optically driven by a circularly polarized light \cite{PhysRevB.79.081406, mciver2020light}.
Meanwhile, the modified Haldane model has been experimentally observed in several artificial systems such as a microwave-scale gyromagnetic photonic crystal \cite{PhysRevLett.125.263603}, topolectrical circuits 
\cite{yang2021observation, PhysRevLett.122.247702}, a 3D layer-stacked photonic metacrystal \cite{liu2023antichiral} and a magnetic Weyl photonic crystal 
\cite{xi2023topological}.

In particular, in TMDC monolayers such as MoS$_{2}$ \cite{PhysRevLett.120.086603,PhysRevLett.108.196802}, the VGS is expected in one of the spin sub-bands because the spin-orbit coupling produces both the Haldane and modified Haldane terms.
The following low-energy effective Hamiltonian is proposed for TMDC monolayers \cite{PhysRevLett.120.086603, PhysRevLett.108.196802}:
\begin{equation} 
    \mathcal{H}(\bm{k}) = \hbar v_{F}(k_{x} \tau_{x} \eta_{z} + k_{y} \tau_{y} \eta_{0}) 
    - \lambda_{SOC} \eta_{z}\frac{\tau_{z} - \tau_{0}}{2}s_{z}  
     + m_{S} \tau_{z} 
    \text{,}
    \label{tmdc hamiltonian}
\end{equation}
where $\lambda_{SOC}$ is the SOC parameter, $m_{S}$ is the Semenoff mass and $s_{z}$ represents the spin degree of freedom. By comparing Eq. (\ref{tmdc hamiltonian}) with Eq. (\ref{4x4}), it can be found that $\Delta_{1}$ and $\Delta_{2}$ are equivalent to $\lambda_{SOC}/2$. For $m_{S} = 0$, Eq. (\ref{tmdc hamiltonian}) exactly describes the VGS phase of our system. It also obtains the metallic and semiconducting phases of our system by tuning $m_{S}$ for $m_{S} \neq 0$. In TMDC monolayers, $\lambda_{SOC}$ can be tuned via an external electric field \cite{yuan2013zeeman}, proximity effects \cite{doi:10.1021/acs.nanolett.8b00691} and strain engineering \cite{PhysRevB.88.155404}.

\begin{figure*}
    \centering
    \includegraphics[width = \textwidth]{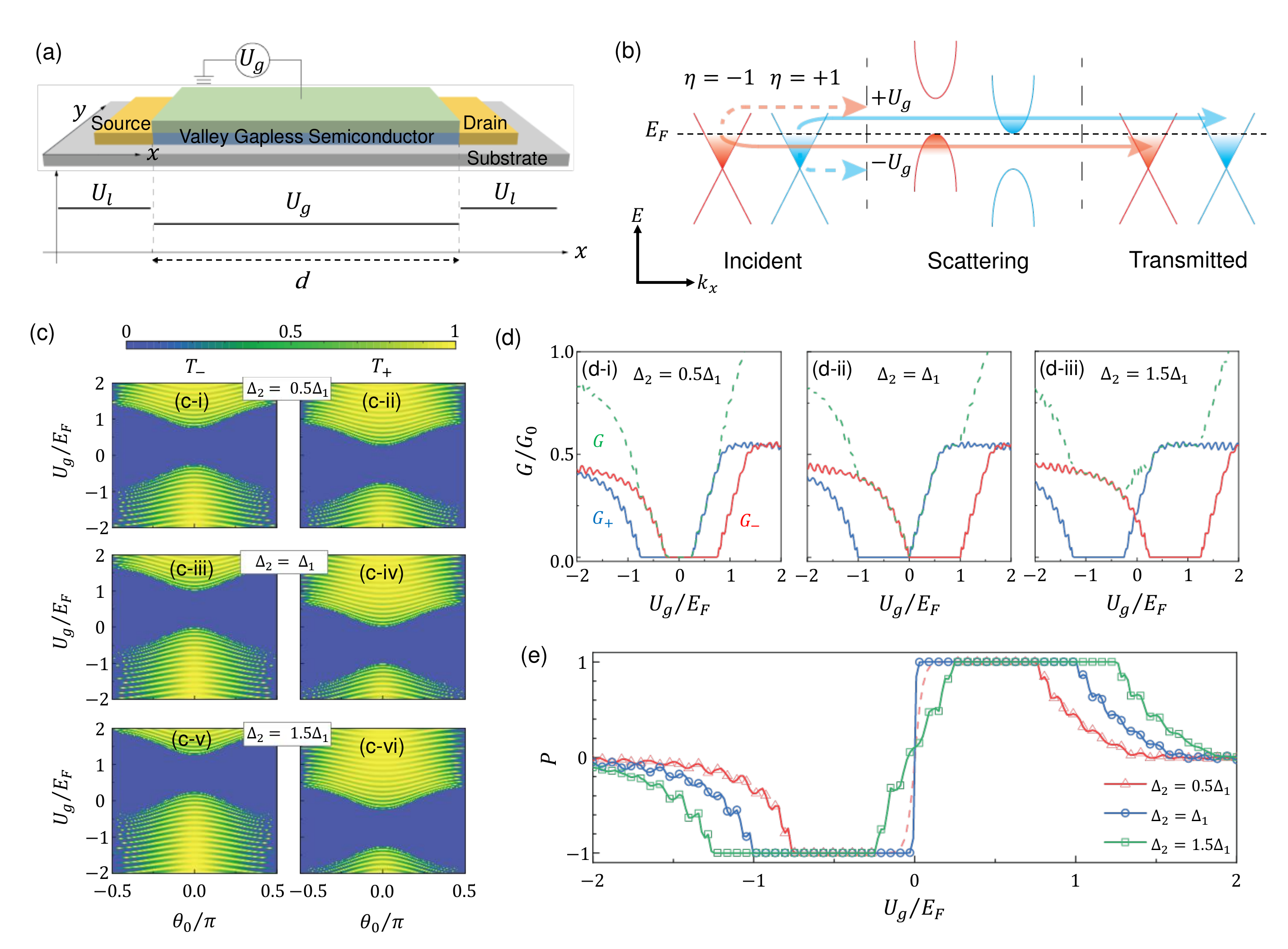}
    \caption{(a) Schematic of the gate-controlled VGS-based valley filter where the gate voltage, $U_g$ is applied to the central region with length, $d$.
    (b) The mechanism of the gate-controlled valley-filter effect. 
    For a positive (negative) gate voltage,
    only the electrons (holes) in the $\eta = +1$ ($-1$) valley are transmitted because of the VGS.
    (c) The valley-dependent transmission probability, $T_\eta$ with respect to the gate voltage, $U_g$ (in the unit of the Fermi level, $E_{F}$) and the incident direction of the electrons, $\theta_{0}$. The left (right) panels depict the transmission probability of the $\eta = -1$ $(+1)$ valley for various values of $\Delta_{1}$ and $\Delta_{2}$. The top, middle and bottom panels refer to the indirect-gap semiconducting, VGS and metallic phases with $\Delta_2 = 0.5\Delta_1$, $\Delta_2 = \Delta_1$ and $\Delta_2 = 1.5\Delta_1$ respectively. The gate-controlled (d) conductance, $G$ and (e) valley polarization efficiency, $P$ of the three phases. The conductance is shown in the unit of $G_0 = e^2/W\hbar$ with sample width, $W$. The red dashed line in (e) indicates the ill-defined $P$ due to the vanishing conductance.
    $E_{F} = 1$ is chosen as the energy unit and the system is isotropic with $\hbar v_{F_{x}}k_{F} = \hbar v_{F_{y}}k_{F} = 1$. 
    In the central region, $\Delta_{1} = 0.5E_{F}$ whereas the length is $k_{F}d = 20$.} \label{filter}
\end{figure*}

\section{Valley Filter} \label{device}
In this section, we investigate the valley-filter effect of the VGS, semiconducting and metallic phases as a means to generate valley-polarized carriers.
%
%We now investigate the transport properties of the VGS, semiconducting and metallic phases to show \textcolor{blue}{their} potential application in valleytronics.
%The first step is to generate valley-polarized carriers via .
%
In particular, since the electrons and holes in the VGS are bound to different valleys, naturally such a VGS-based valley filter is electrically manipulable. 
The device [Fig. \ref{filter}(a)] can be constructed by sandwiching a VGS between two normal metallic leads acting as the source and drain without any valley nonequivalence. 
The central region is controlled by a top gate. Figure \ref{filter}(b) illustrates the all-gate tunable feature of the VGS-based valley filter device. 

Considering that the system is translationally invariant in the $y$-direction, i.e., $k_{y}$ is a good quantum number, the corresponding Hamiltonian is 
\begin{equation}
 \mathcal{H}_{1}^{\eta }\left( x,k_{y}\right) = \mathcal{H}_{\eta }\left( -i\partial
_{x},k_{y}\right) -U\left( x\right) \tau _{0}\text{,}  \label{hvf}
\end{equation}%
with $\Delta _{1,2}\left( x\right) =\Delta _{1,2}\Theta \left(dx-x^{2}\right) $ and $U\left( x\right)=U_{l}\Theta \left( -x\right) +U_{g}\Theta \left( dx-x^{2}\right) +U_{l}\Theta \left( x-d\right) $, where $U_{l}$ is the electrostatic potential in the leads and the VGS is confined within the central region of length, $d$ and manipulated by the gate voltage, $U_{g}$.

The valley-filter effect manifests as the valley-dependent transmission probability, $T_{\eta} = \left\vert t_{\eta }\right\vert^{2}$ obtained by matching the wavefunctions (Appendix \ref{tunneling}) \cite{10.1063/1.3473725, katsnelson2006chiral, PhysRevA.96.013813, RevModPhys.80.1337, RevModPhys.81.109, allain2011klein}
\begin{equation}
\psi_{\eta}\left( x, y\right) =\left\{ 
\begin{array}{lc}
\psi_{\eta}^{+}\left( x, y\right) +r_{\eta }\psi_{\eta}^{-}\left(
x, y\right) \text{,} & x<0 \\ 
a\psi_{\eta}^{+}\left( x, y\right) + b\psi _{\eta}^{-}\left(x, y\right) 
\text{,} & 0\leq x\leq d \\ 
t_{\eta }\psi_{\eta}^{+}\left( x, y\right) \text{,} & x>d%
\end{array}%
\right. \text{,}
\end{equation}
at $x = 0$ and $x = d$ where
\begin{equation}
\psi_{\eta }^{\pm}\left( x, y\right) = 
\left[ 
\begin{array}{c}
\pm \eta e^{\mp i\eta \theta \left( x\right) } \\ 
S\left( x\right) \gamma \left( x\right)%
\end{array}%
\right] e^{\pm ik\left( x\right) x}e^{ik_{y}y}\text{,}
\end{equation}
is the wavefunction of the three regions characterized by $\hbar v_{F_{x}}k\left( x\right) =\pm \sqrt{Z_{+}\left( x\right) Z_{-}\left( x\right) -\left( \hbar v_{F_{y}}k_{y}\right) ^{2}}$, $\gamma \left( x\right) = \sqrt{1-2\Delta _{1}\left( x\right)/Z_{+}\left( x\right) }$, $S\left( x\right) = sgn\left[Z_{-}(x)\right] $,
\begin{equation}
    \theta \left(x\right) = \arctan \left[ \frac{v_{F_{y}}k_{y}}{v_{F_{x}}k\left( x\right)}\right]
    \text{,}
\end{equation}
and $Z_{\pm }\left( x\right) = E + U\left( x\right) +\eta \Delta _{2}\left( x\right) \pm \Delta _{1}\left( x\right) $ where $E$ is the energy of the incident electrons (Appendix \ref{spinor}).

At $T = 0$, the valley-dependent conductance is related to the transmission probability (Appendix \ref{transmission}) as
\begin{equation}
G_{\eta }\left( E_{F}\right) /G_{0}=\int_{-\pi /2}^{\pi /2} T_{\eta }\left( E_{F},\theta _{0}\right) \cos \theta _{0} d\theta_{0} \text{,}
\end{equation}
where $\theta _{0} = \arcsin \left( \hbar v_{F}k_{y}/E_{F}^{1/2}\right) $ and $E_{F}=U_{l}+E$\ are the direction and Fermi level of the incident electrons respectively. The conductance is normalized by the conductance unit, $G_{0} = e^{2}/W\hbar $ with sample width, $W$. 

Hereafter, $E_{F} = 1$ is chosen as the energy unit and the system is isotropic with $\hbar v_{F_{x}}k_{F}=\hbar v_{F_{y}}k_{F} = 1$. 
In the central region, $\Delta _{1} = 0.5E_{F}$ whereas the length is $k_{F}d = 20$. 
We argue that the results below are qualitatively similar for different combinations of the parameters.
Moreover, the results remain valid under the anisotropic case, i.e., $v_{F_{x}} \neq v_{F_{y}}$.

To demonstrate the valley-filter effect, we first obtain the valley-dependent transmission probability,  $T_{\eta }$ with respect to the incident direction of the electrons, $\theta _{0}$ and gate voltage, $U_{g}$ of the indirect-gap semiconducting ($\Delta _{2} = 0.5\Delta _{1}$), VGS ($\Delta _{2} = \Delta _{1}$)
and metallic ($\Delta _{2} =  1.5\Delta _{1}$) phases [Fig. \ref{filter}(c)]. 
For all three phases, the transmission coefficient vanishes beyond a critical angle, $\theta _{\eta }^{c} = \arcsin \left[ \sqrt{Z_+(d)Z_-(d)}/E_{F}\right] $. 
Electrons with incident angle beyond $\theta_{\eta }^{c}$ are reflected completely due to the evanescent wave in the central region. 
Moreover, the transmission probability vanishes when the gate voltage within the band gap of each valley, i.e., $-\left( \eta \Delta _{2} + \Delta _{1}\right) <U_{g} < -\left( \eta \Delta _{2}-\Delta _{1}\right)$, is swept. 
However, this voltage window is valley-dependent due to the energy shifting of the Dirac cones by $\Delta _{2}$ which results in carriers being valley-selectively transmitted. 
By tuning the gate voltage, only carriers from one valley are allowed to be transmitted whereas the transmission of carriers from the other valley is forbidden as illustrated schematically in Fig. \ref{filter}(b). 
Therefore, the total conductance, $G = G_{+} + G_{-}$ within this regime is solely contributed by carriers from one valley only, i.e., either $G = G_{+}$ or $G = G_{-}$ as shown in Fig. \ref{filter}(d). 
Beyond this voltage window, the conductance contribution from the other valley increases gradually and the total conductance oscillates due to Fabry-P\'{e}rot interference.
For 2D materials such as graphene and TMDC monolayers, the de Broglie wavelength is approximately $100$ nm or $1000$ \text{\AA} 
\cite{10.1063/5.0213720, yin2014graphene}, which is much larger than the characteristic length scale of the interfacial potential change (on the order of $a$ where $a = 2.46$ \text{\AA} is the graphene lattice constant). Under this condition, the interfaces of the square electrostatic potential are effectively smooth or long-range on the atomic lattice scale \cite{PhysRevB.75.115318, shon1998quantum, RevModPhys.80.1337, allain2011klein}. The interfaces will not be able to provide sufficient momentum transfer for a charge carrier to overcome the large intervalley momentum separation. Thus, intervalley scattering is negligible and the results of the valley conductance are valid. Nevertheless, it has been demonstrated that a smoothing of the potential step at the interfaces improves the flatness of the valley conductance plateaus \cite{rycerz2007valley}.
%
%\begin{figure*}
    %\centering
    %\includegraphics[width=0.8\textwidth]{fig4.pdf}
    %\caption{Caption} \textcolor{red}{[YS: can draw vertical lines from bottom to top, so to clearly illsutrate the (0,0), (0,1), etc states.]}
    %\label{valve}
%\end{figure*}

%This valley-selective nature manifests as

The valley-filter effect is characterized in terms of the valley polarization efficiency defined as 
\begin{equation}
    P = \frac{G_{+} - G_{-}}{G_{+} + G_{-}} 
    \text{.}
\end{equation}
Figure \ref{filter}(e) depicts the valley polarization efficiency, $P$ with respect to the gate voltage, $U_{g}$ of the three phases.
For each of them, a full valley polarization efficiency of $P = -1$ ($+1$) is achieved when the gate voltage satisfies 
$-\left(\Delta _{2}+\Delta _{1}\right) 
<U_{g}<\Delta _{1}-\Delta _{2}$ 
($\Delta_{2} - \Delta _{1} <U_{g}< \Delta _{2} + \Delta _{1}$).
The valley polarization efficiency is destroyed gradually when the gate voltage exceeds the aforementioned range of values.
In particular, the VGS ($\Delta_{2} = \Delta_{1}$) experiences a sharp flip of the valley polarization efficiency
between $P = -1$ and $P = +1$ at $U_{g} = 0$ [Fig. \ref{filter}(e)] owing to the valley-carrier coupling.
Hence, its conductance is solely contributed by electrons from the $\eta = -1$ ($+1$) valley when $U_{g} < 0 $ ($U_{g} > 0$).
On the other hand, the valley polarization efficiency of the semiconducting phase ($\Delta_{2} < \Delta _{1}$) is ill-defined for $\left\vert U_{g}\right\vert <\Delta _{1}-\Delta _{2}$ [red dashed line in Fig. \ref{filter}(e)] due to the vanishing conductance [Fig. \ref{filter}(d-i)]. 
For the metallic phase ($\Delta _{2}>\Delta _{1}$), its valley polarization efficiency is unstable and sensitive to the gate voltage.
The results demonstrate that the VGS phase is well suited for a valley filter device owing to the valley-carrier coupling.
The valley polarization efficiency, $P$ of the indirect-gap semiconductor, valley gapless semiconductor and metal phases can be experimentally determined by using helicity-resolved photoluminescence spectroscopy\cite{mak2012control, zeng2012valley, cao2012valley}.

\section{Conclusion} \label{conclusion}

%\ph{[PH: discuss experimental realization. The thing is how to realize the Haldane terms.]}

In summary, we have theoretically investigated a VGS-based valleytronic device where its valley-carrier coupling feature enables a fully valley-polarized current with all-electrical manipulation.
A possible realization scheme of the VGS phase is demonstrated in the honeycomb lattice subject to the Haldane and modified Haldane terms. 
The Haldane terms have been experimentally realized in many systems including
light-wave-controlled monolayer hexagonal boron nitride \cite{mitra2024light}, 
optically driven monolayer graphene \cite{PhysRevB.79.081406, mciver2020light},
topolectrical and quantum circuits \cite{yang2021observation,PhysRevLett.122.247702, PhysRevLett.129.140502} and
magnetic Weyl photonic crystal \cite{xi2023topological,wang2023experimental}.
Furthermore, the valley-filter effect of the VGS phase is studied in a single gate-controlled valley filter device where it manifests a sharp flip of the valley polarization efficiency between $P = -1$ and $P = +1$ at $U_{g} = 0$ due to the valley-carrier coupling. 
This highlights the potential application of the VGS phase in valleytronics including valley-based quantum computing and improving classical information storage and processing \cite{https://doi.org/10.1002/smll.201801483}.
It is possible to realize the VGS phase in other lattice systems \cite{PhysRevB.90.085103} such as the square \cite{PhysRevA.82.013608}, $\alpha$-$T_{3}$ \cite{PhysRevLett.112.026402, PhysRevB.109.235105} and dice \cite{PhysRevLett.81.5888, PhysRevB.64.155306} lattices, subject not only to the Haldane and modified Haldane terms, but also other effects such as Floquet driving \cite{annurev:/content/journals/10.1146/annurev-conmatphys-031218-013423,PhysRevB.111.045406, PhysRevX.4.031027, PhysRevA.68.013820,PhysRevB.95.045102,PhysRevLett.121.237401,PhysRevB.106.235405,qin2024light,stegmaier2024topologicaledgestatenucleation}, strain engineering 
\cite{PhysRevB.109.045403},
altermagnetism \cite{PhysRevB.110.L220402}
and spin-orbit coupling effects \cite{PhysRevB.84.195430, PhysRevResearch.6.043108, Lin_Fu, PhysRevB.110.195409}.

%We illustrate the concept of the VGS via a low-energy effective Hamiltonian which describes the states near the $K$ and $K^\prime$ valleys of the honeycomb lattice with the Haldane and modified Haldane terms. 
%Two valleytronic devices are discussed. Due to the valley-carrier coupling, a valley-polarized current is generated in a valley filter, where the current polarization is controlled by the gate voltage.  
%When two valley filters are serially arranged, one acting as a polarizer and the other as an analyzer, a valley valve is constructed. 

\begin{acknowledgments}
K. W. Lee is supported by the SUTD PhD Scholarship.
P.-H. Fu \& Y. S. Ang are supported by the Singapore Ministry of Education (MOE) Academic Research Fund (AcRF) Tier 2 Grant (MOE-T2EP50221-0019). 
J.-F. L. is supported by the National Natural Science Foundation of China (Grant No. 12174077).
C. H. L. is supported by Singapore’s NRF Quantum Engineering grant NRF2021-QEP2-02-P09 and Singapore’s MOE Tier-II grant Proposal ID: T2EP50222-0003.
\end{acknowledgments}

\appendix
\onecolumngrid

\section{Quantum Tunneling Problem}
\label{tunneling}

In studying the valley-filter effect of our system, we consider a quantum tunneling problem involving a lead-barrier-lead setup with 
$\Delta _{1,2}\left( x\right) =\Delta_{1,2}\Theta \left(dx - x^{2}\right) $ 
and $U\left( x\right) = U_{l}\Theta \left( -x\right) + U_{g}\Theta \left( dx-x^{2}\right) + U_{l}\Theta \left(x - d\right) $. 
The wavefunction describing the charge carrier, $\psi_{\eta}(x ,y)$ obeys the following time-independent Dirac equation \cite{dirac1928quantum, Dirac1930-DIRTPO}:
\begin{equation}
\mathcal{H}_{1}^{\eta }\psi_{\eta }(x ,y) = E\psi _{\eta }(x ,y) \text{.} 
\label{time-independent Dirac equation}
\end{equation}
Since the system is translationally invariant in the $y$-direction, the Hamiltonian becomes  
\begin{align}
\mathcal{H}_{1}^{\eta }(x, k_{y}) &= \mathcal{H}_{\eta}(-i \partial_{x}, k_{y}) - U(x)\tau_{0} \\\nonumber
&= \eta \hbar v_{F_{x}}(-i \partial_{x})\tau
_{x} + \hbar v_{F_{y}}k_{y}\tau _{y}
+ \Delta_{1}(x) \tau_{z} 
- \eta \Delta _{2}(x)\tau
_{0} - U(x)\tau_{0} 
\text{,} \nonumber  
\label{hamiltonian} 
\end{align}
%
%Since $k_{y}$ is a good quantum number, 
and
$\psi_{\eta }(x ,y) = \tilde{\psi}_{\eta}(x)e^{ik_{y}y}$ where $\tilde{\psi }_{\eta }(x)$ is a two-component spinor wavefunction with components $[\tilde{\psi }_{\eta }^{A}(x) ,\tilde{\psi }_{\eta }^{B}(x)]^{\top}$. Effectively, Eq. (\ref{time-independent Dirac equation}) becomes one-dimensional, i.e., 
\begin{equation}
\mathcal{H}_{1}^{\eta} \tilde{\psi}_{\eta }(x) = E\tilde{\psi }_{\eta }(x) \text{,}
\label{one-dimensional equation}
\end{equation}
which has plane-wave solutions describing the incident, reflected and transmitted waves \cite{10.1063/1.3473725, katsnelson2006chiral, PhysRevA.96.013813, RevModPhys.80.1337, RevModPhys.81.109, allain2011klein}. 
Assuming that the charge carrier propagates from left to right, 
for Region \uppercase\expandafter{\romannumeral 1\relax} ($x < 0$),
\begin{equation}
    \tilde{\psi}_{\eta}^{\mathit{I}}(x) = 
    \left(\begin{array}{c}
    \tilde{\psi }_{\eta}^{A_{\mathit{I}}} \\ 
    \tilde{\psi }_{\eta}^{B_{\mathit{I}}}
    \end{array} \right) e^{+ik_{x}x} + 
    \left(\begin{array}{c}
    \tilde{\psi }_{\eta}^{A_{\mathit{I}}} \\ 
    \tilde{\psi }_{\eta}^{B_{\mathit{I}}}
    \end{array} \right) r_{\eta}e^{-ik_{x}x} 
    \text{,} 
    \label{region 1}
\end{equation}
where $v_{F_{x}}k_{x} = v_{F}k_{F}\cos\theta_{0}$,
$v_{F_{y}}k_{y} = v_{F}k_{F}\sin\theta_{0}$ and 
\begin{equation}
    \theta_{0} = \arctan
    \left(\frac{v_{F_{y}}k_{y}}{v_{F_{x}}k_x}\right)
    \text{.}    
\end{equation}
For Region \uppercase\expandafter{\romannumeral 2\relax} ($0\leq x\leq d$),
\begin{equation}
    \tilde{\psi}_{\eta}^{\mathit{II}}(x) = 
    a \left(\begin{array}{c}
    \tilde{\psi }_{\eta}^{A_{\mathit{II}}} \\ 
    \tilde{\psi }_{\eta}^{B_{\mathit{II}}}
    \end{array} \right) e^{+iq_{x}x}
    + b \left(\begin{array}{c}
    \tilde{\psi }_{\eta}^{A_{\mathit{II}}} \\ 
    \tilde{\psi }_{\eta}^{B_{\mathit{II}}}
    \end{array} \right) e^{-iq_{x}x}
    \text{,} 
    \label{region 2}
\end{equation}
where $v_{F_{x}}q_{x} = v_{F}k_{F}\cos\theta$,
$v_{F_{y}}k_{y} = v_{F}k_{F}\sin\theta$ and 
\begin{equation}
    \theta =\arctan 
    \left(\frac{
    v_{F_{y}}k_{y}}{v_{F_{x}}q_x}\right) 
    \text{.}    
\end{equation}
For Region \uppercase\expandafter{\romannumeral 3\relax}
($x > d$),
\begin{equation}
    \tilde{\psi}_{\eta}^{\mathit{III}}(x) = t_{\eta}
    \left(\begin{array}{c}
    \tilde{\psi }_{\eta}^{A_{\mathit{III}}} \\ 
    \tilde{\psi }_{\eta}^{B_{\mathit{III}}}
    \end{array} \right)
    e^{+ik_{x}x}
    \text{.} 
    \label{region 3}
\end{equation}
From here onward, we introduce the notation 
\begin{equation}
    \theta(x) = \arctan \left[\frac {v_{F_{y}}k_{y}}{v_{F_{x}}k\left( x\right)}\right]
    \text{,}
\end{equation}
where 
$k(x) = k_{x}$ for $x < 0$ and $x > d$ whereas $k(x) = q_{x}$ for $0 \leq x \leq d$.

\section{Spinor Components}
\label{spinor}
We rewrite the Hamiltonian of our system as  
\begin{equation}
     \mathcal{H}_{1}^{\eta }[k(x) ,k_{y}] =\eta \hbar v_{F_{x}}k(x)\tau _{x} +\hbar v_{F_{y}}k_{y}\tau _{y} 
     + \Delta_{1}(x) \tau _{z} 
     -[\eta \Delta_{2}(x) +U(x)]\tau _{0}
     \text{.}
    \label{spinor hamiltonian}
\end{equation}
The secular equation of Eq. (\ref{spinor hamiltonian}) is expressed as $\det[\mathcal{H}_{1}^{\eta} - E] = 0$ where we obtain
%
%$\det [\eta \hbar v_{F_{x}}k(x)\tau _{x} +\hbar v_{F_{y}}k_{y}\tau _{y} -\Delta _{1}(x)\tau _{z} -[\eta \Delta _{2}(x) +U(x) +E]\tau _{0}] =0$
%
\begin{equation}
    [\hbar v_{F_{x}}k(x)]^{2} =[E +U(x) +\eta \Delta _{2}(x) +\Delta _{1}(x)][E +U(x) +\eta \Delta _{2}(x) -\Delta _{1}(x)] -(\hbar v_{F_{y}}k_{y})^{2}
     \text{.}
\end{equation}
By defining $Z_{\pm}(x) = E + U(x) +\eta \Delta_{2}(x) \pm\Delta _{1}(x)$, $\hbar v_{F_{x}}k(x) = \pm \sqrt{Z_{ +}(x)Z_{ -}(x) -(\hbar v_{F_{y}}k_{y})^{2}}$.
Next, we simplify the calculations by focusing on the
right-moving wavefunction and expand Eq. (\ref{one-dimensional equation}) into its matrix form as follows: 
\begin{equation}
    \left[\begin{array}{cc} -Z_{-}(x) & \eta \hbar v_{F_{x}}k_{}(x) -i\hbar v_{F_{y}}k_{y} \\
    \eta \hbar v_{F_{x}}k_{}(x) +i\hbar v_{F_{y}}k_{y} &  - Z_{+}(x)\end{array}\right]
    \left(\begin{array}{c}\tilde{\psi }_{\eta }^{A} \\
    \tilde{\psi}_{\eta}^{B}\end{array}\right) = 0
    \text{.}
\end{equation}
By multiplying the top row with the spinor components, we have 
\begin{equation}
    -Z_{-}(x)\tilde{\psi}_{\eta}^{A} + [\eta \hbar v_{F_{x}}k(x) -i\hbar v_{F_{y}}k_{y}]\tilde{\psi }_{\eta}^{B} = 0
    \text{.}
\end{equation}
Letting $\eta \hbar v_{F_{x}}k(x) -i\hbar v_{F_{y}}k_{y} =\eta \hbar v_{F}k_{F}e^{-i\eta \theta (x)}$ and $\hbar v_{F_{}}k_{F} = \pm \sqrt{[\hbar v_{F_{x}}k(x)]^{2} +(\hbar v_{F_{y}}k_{y})^{2}} = \pm \sqrt{Z_{ +}(x)Z_{ -}(x)}$,
%
%\begin{equation}
    %Z_{+}(x)\tilde{\psi }_{\eta }^{A}_{} =\eta \hbar v_{F_{}}k_{F}\exp [ -i\eta \theta (x)]\tilde{\psi }_{\eta }^{B}
%\end{equation}
%
\begin{equation}
    Z_{-}(x)\tilde{\psi }_{\eta }^{A} = \pm \eta \sqrt{Z_{ +}(x)Z_{ -}(x)}e^{-i\eta \theta (x)}\tilde{\psi}_{\eta}^{B}
    \label{almost}
    \text{.}
\end{equation}
Rearranging Eq. (\ref{almost}), 
%
%\begin{equation}
    %\frac{\tilde{\psi }_{\eta }^{A}}{\tilde{\psi }_{\eta }^{B}} =\frac{ \pm \eta \sqrt{Z_{ +}(x)Z_{ -}(x)}\exp [ -i\eta \theta (x)]}{Z_{ +}(x)}    
%\end{equation}
%
\begin{align}
    \frac{\tilde{\psi}_{\eta}^{A}}{\tilde{\psi }_{\eta }^{B}} 
    &= \frac{\pm \eta e^{ -i\eta \theta (x)}}
    {sgn[Z_{-}(x)]\sqrt{\frac{Z_{-}(x)}{Z_{+}(x)}}} \\\nonumber
    &= \frac{ \pm \eta e^{ -i\eta \theta (x)}}{S(x)\gamma (x)} \nonumber
    \text{,}
\end{align}
where $\gamma \left( x\right) = \sqrt{1 - 2\Delta_{1}\left( x\right)/Z_{+}\left( x\right) }$ and
$S\left( x\right) = sgn\left[Z_{-}(x)\right] $. 
Finally, we obtain the spinor components accounting for the right-moving wavefunction as
\begin{equation}
    \left(\begin{array}{c}\tilde{\psi }_{\eta }^{A} \\
    \tilde{\psi}_{\eta}^{B}\end{array}\right) = 
    \left[\begin{array}{c} \pm \eta e^{ -i\eta \theta (x)} \\ S(x)\gamma (x)\end{array}\right] 
    \text{,}
    \label{right-moving}
\end{equation}
and the spinor components of the left-moving wavefunction have a similar form, i.e., 
\begin{equation}
    \left(\begin{array}{c}\tilde{\psi }_{\eta }^{A} \\
    \tilde{\psi}_{\eta}^{B}\end{array}\right) = 
    \left[\begin{array}{c} \pm \eta e^{ +i\eta \theta (x)} \\ S(x)\gamma (x)\end{array}\right] 
    \text{.}
    \label{left-moving}
\end{equation}
Finally, we introduce the following compact form to account for the full solution of our quantum tunneling problem:
\begin{equation}
    \psi_{\eta}\left(x, y\right) =\left\{ 
    \begin{array}{lc}
    \psi_{\eta}^{+}\left( x, y\right) + r_{\eta }\psi_{\eta }^{-}\left(
    x, y\right) \text{,} & x < 0 \\ 
    a\psi_{\eta}^{+}\left( x, y\right) +b\psi _{\eta }^{-}\left( x, y\right) 
    \text{,} & 0\leq x\leq d \\ 
    t_{\eta}\psi_{\eta}^{+}\left( x, y\right) \text{,} & x>d%
    \end{array}%
    \right. 
    \text{,}
\end{equation}%
where
\begin{equation}
    \psi_{\eta}^{\pm}\left( x, y\right) =\left[
    \begin{array}{c}
    \pm \eta e^{\mp i\eta \theta \left( x\right) } \\ 
    S\left( x\right) \gamma \left( x\right)%
    \end{array}%
    \right] e^{\pm ik\left( x\right) x}e^{ik_{y}y}\text{.}
\end{equation}

\section{Valley-Dependent Scattering Coefficients}
\label{transmission}
The valley-dependent transmission and reflection amplitudes, $t_{\eta}$ and $r_{\eta}$ can be determined from the wavefunction continuity, i.e., the wavefunction has to satisfy the conditions $\psi^{\mathit{I}}_{\eta}(x = 0,y) = \psi^{\mathit{II}}_{\eta}(x = 0,y)$ and
$\psi^{\mathit{II}}_{\eta}(x = d,y) = \psi^{\mathit{III}}_{\eta}(x = d,y)$.
Consequently, we obtain the following:

\begin{equation}
    t_{\eta} = \frac{2i\gamma_{c}S_{c}\cos \theta_{0}\cos\theta_{c}e^{-idk_{0}}}
    {\sin dk_{c}(1 + \gamma_{c}^2 -2\gamma_{c}S_{c} \sin\theta_{0} \sin \theta_{c}) + 2i\gamma_{c}S_{c}\cos dk_{c}\cos\theta_{0}\cos\theta_{c}}
    \text{,}
    \label{transmission probability amplitude}
\end{equation}
and
\begin{equation}
r_{\eta} = \frac{-\sin d k_{c}e^{-i\eta (2 \theta_{0} + \theta_{c})}
    [\gamma_{c} S_{c} + e^{i \eta (\theta_{0} + \theta_{c})}] [i \eta (\sin \theta_{0} - \gamma_{c} S_{c} \sin \theta_{c}) - \gamma_{c} S_{c} \cos \theta_{c} + \cos \theta_{0}]}
    {\sin d k_{c} (1 + \gamma_{c}^2 -2 \gamma_{c} S_{c} \sin \theta_{0}  \sin  \theta_{c}) +2i\gamma_{c} S_{c} \cos dk_{c} \cos\theta_{0} \cos\theta_{c}}
    \text{,}
    \label{reflection probability amplitude}
\end{equation}
where 
$Z_\pm^{(g)} = E + U_{g} + \eta\Delta_{2} \pm \Delta_{1}$, 
$\hbar v_{F_{x}} k_c = \sqrt{Z_+^{(g)}Z_-^{(g)}-(\hbar v_{F_{y}} k_y)^2}$,
\begin{equation}
    \theta_c = \arctan\left(\frac{{v_{F_{y}}k_y}}{{v_{F_{x}}k_c}}\right)
    \text{,}
\end{equation} 
$\gamma_c = \sqrt{1 - {2\Delta_1}/{Z_+^{(g)}}}$ and  
$S_c = \operatorname{sgn} \left[Z_-^{(g)}\right]$
for the central region
whereas 
${\hbar v_{F_{x}}}  k_{0} = {\sqrt{(E + U_{l})^2 - (\hbar v_{F_{y}} k_y)^2}}$ and 
\begin{equation}
    \theta_{0} = \arctan\left(\frac{v_{F_{y}}
    k_y}{v_{F_{x}}k_0}\right) 
    \text{,}
\end{equation}
for the leads.

\twocolumngrid

\bibliography{references.bib}

%apsrev4-2.bst 2019-01-14 (MD) hand-edited version of apsrev4-1.bst
%Control: key (0)
%Control: author (8) initials jnrlst
%Control: editor formatted (1) identically to author
%Control: production of article title (0) allowed
%Control: page (0) single
%Control: year (1) truncated
%Control: production of eprint (0) enabled
\providecommand{\noopsort}[1]{}\providecommand{\singleletter}[1]{#1}%
\begin{thebibliography}{134}%
\makeatletter
\providecommand \@ifxundefined [1]{%
 \@ifx{#1\undefined}
}%
\providecommand \@ifnum [1]{%
 \ifnum #1\expandafter \@firstoftwo
 \else \expandafter \@secondoftwo
 \fi
}%
\providecommand \@ifx [1]{%
 \ifx #1\expandafter \@firstoftwo
 \else \expandafter \@secondoftwo
 \fi
}%
\providecommand \natexlab [1]{#1}%
\providecommand \enquote  [1]{``#1''}%
\providecommand \bibnamefont  [1]{#1}%
\providecommand \bibfnamefont [1]{#1}%
\providecommand \citenamefont [1]{#1}%
\providecommand \href@noop [0]{\@secondoftwo}%
\providecommand \href [0]{\begingroup \@sanitize@url \@href}%
\providecommand \@href[1]{\@@startlink{#1}\@@href}%
\providecommand \@@href[1]{\endgroup#1\@@endlink}%
\providecommand \@sanitize@url [0]{\catcode `\\12\catcode `\$12\catcode `\&12\catcode `\#12\catcode `\^12\catcode `\_12\catcode `\%12\relax}%
\providecommand \@@startlink[1]{}%
\providecommand \@@endlink[0]{}%
\providecommand \url  [0]{\begingroup\@sanitize@url \@url }%
\providecommand \@url [1]{\endgroup\@href {#1}{\urlprefix }}%
\providecommand \urlprefix  [0]{URL }%
\providecommand \Eprint [0]{\href }%
\providecommand \doibase [0]{https://doi.org/}%
\providecommand \selectlanguage [0]{\@gobble}%
\providecommand \bibinfo  [0]{\@secondoftwo}%
\providecommand \bibfield  [0]{\@secondoftwo}%
\providecommand \translation [1]{[#1]}%
\providecommand \BibitemOpen [0]{}%
\providecommand \bibitemStop [0]{}%
\providecommand \bibitemNoStop [0]{.\EOS\space}%
\providecommand \EOS [0]{\spacefactor3000\relax}%
\providecommand \BibitemShut  [1]{\csname bibitem#1\endcsname}%
\let\auto@bib@innerbib\@empty
%</preamble>
\bibitem [{\citenamefont {Zheng}\ \emph {et~al.}(2022)\citenamefont {Zheng}, \citenamefont {Zhao}, \citenamefont {Tan}, \citenamefont {Guan}, \citenamefont {Zhong}, \citenamefont {Yue}, \citenamefont {Xiang},\ and\ \citenamefont {Duan}}]{10.1063/5.0112893}%
  \BibitemOpen
  \bibfield  {author} {\bibinfo {author} {\bibfnamefont {J.-D.}\ \bibnamefont {Zheng}}, \bibinfo {author} {\bibfnamefont {Y.-F.}\ \bibnamefont {Zhao}}, \bibinfo {author} {\bibfnamefont {Y.-F.}\ \bibnamefont {Tan}}, \bibinfo {author} {\bibfnamefont {Z.}~\bibnamefont {Guan}}, \bibinfo {author} {\bibfnamefont {N.}~\bibnamefont {Zhong}}, \bibinfo {author} {\bibfnamefont {F.-Y.}\ \bibnamefont {Yue}}, \bibinfo {author} {\bibfnamefont {P.-H.}\ \bibnamefont {Xiang}},\ and\ \bibinfo {author} {\bibfnamefont {C.-G.}\ \bibnamefont {Duan}},\ }\bibfield  {title} {\bibinfo {title} {{Coupling of ferroelectric and valley properties in 2D materials}},\ }\href {https://doi.org/10.1063/5.0112893} {\bibfield  {journal} {\bibinfo  {journal} {J. Appl. Phys.}\ }\textbf {\bibinfo {volume} {132}},\ \bibinfo {pages} {120902} (\bibinfo {year} {2022})}\BibitemShut {NoStop}%
\bibitem [{\citenamefont {Liu}\ \emph {et~al.}(2019)\citenamefont {Liu}, \citenamefont {Gao}, \citenamefont {Zhang}, \citenamefont {He}, \citenamefont {Yu},\ and\ \citenamefont {Liu}}]{liu2019valleytronics}%
  \BibitemOpen
  \bibfield  {author} {\bibinfo {author} {\bibfnamefont {Y.}~\bibnamefont {Liu}}, \bibinfo {author} {\bibfnamefont {Y.}~\bibnamefont {Gao}}, \bibinfo {author} {\bibfnamefont {S.}~\bibnamefont {Zhang}}, \bibinfo {author} {\bibfnamefont {J.}~\bibnamefont {He}}, \bibinfo {author} {\bibfnamefont {J.}~\bibnamefont {Yu}},\ and\ \bibinfo {author} {\bibfnamefont {Z.}~\bibnamefont {Liu}},\ }\bibfield  {title} {\bibinfo {title} {{Valleytronics in transition metal dichalcogenides materials}},\ }\href {https://doi.org/https://doi.org/10.1007/s12274-019-2497-2} {\bibfield  {journal} {\bibinfo  {journal} {Nano Res.}\ }\textbf {\bibinfo {volume} {12}},\ \bibinfo {pages} {2695} (\bibinfo {year} {2019})}\BibitemShut {NoStop}%
\bibitem [{\citenamefont {Schaibley}\ \emph {et~al.}(2016)\citenamefont {Schaibley}, \citenamefont {Yu}, \citenamefont {Clark}, \citenamefont {Rivera}, \citenamefont {Ross}, \citenamefont {Seyler}, \citenamefont {Yao},\ and\ \citenamefont {Xu}}]{schaibley2016valleytronics}%
  \BibitemOpen
  \bibfield  {author} {\bibinfo {author} {\bibfnamefont {J.~R.}\ \bibnamefont {Schaibley}}, \bibinfo {author} {\bibfnamefont {H.}~\bibnamefont {Yu}}, \bibinfo {author} {\bibfnamefont {G.}~\bibnamefont {Clark}}, \bibinfo {author} {\bibfnamefont {P.}~\bibnamefont {Rivera}}, \bibinfo {author} {\bibfnamefont {J.~S.}\ \bibnamefont {Ross}}, \bibinfo {author} {\bibfnamefont {K.~L.}\ \bibnamefont {Seyler}}, \bibinfo {author} {\bibfnamefont {W.}~\bibnamefont {Yao}},\ and\ \bibinfo {author} {\bibfnamefont {X.}~\bibnamefont {Xu}},\ }\bibfield  {title} {\bibinfo {title} {{Valleytronics in 2D materials}},\ }\href {https://doi.org/https://doi.org/10.1038/natrevmats.2016.55} {\bibfield  {journal} {\bibinfo  {journal} {Nat Rev Mater}\ }\textbf {\bibinfo {volume} {1}},\ \bibinfo {pages} {16055} (\bibinfo {year} {2016})}\BibitemShut {NoStop}%
\bibitem [{\citenamefont {Yu}\ \emph {et~al.}(2015)\citenamefont {Yu}, \citenamefont {Cui}, \citenamefont {Xu},\ and\ \citenamefont {Yao}}]{yu2015valley}%
  \BibitemOpen
  \bibfield  {author} {\bibinfo {author} {\bibfnamefont {H.}~\bibnamefont {Yu}}, \bibinfo {author} {\bibfnamefont {X.}~\bibnamefont {Cui}}, \bibinfo {author} {\bibfnamefont {X.}~\bibnamefont {Xu}},\ and\ \bibinfo {author} {\bibfnamefont {W.}~\bibnamefont {Yao}},\ }\bibfield  {title} {\bibinfo {title} {{Valley excitons in two-dimensional semiconductors}},\ }\href {https://doi.org/https://doi.org/10.1093/nsr/nwu078} {\bibfield  {journal} {\bibinfo  {journal} {Natl. Sci. Rev.}\ }\textbf {\bibinfo {volume} {2}},\ \bibinfo {pages} {57} (\bibinfo {year} {2015})}\BibitemShut {NoStop}%
\bibitem [{\citenamefont {Xu}\ \emph {et~al.}(2014)\citenamefont {Xu}, \citenamefont {Yao}, \citenamefont {Xiao},\ and\ \citenamefont {Heinz}}]{xu2014spin}%
  \BibitemOpen
  \bibfield  {author} {\bibinfo {author} {\bibfnamefont {X.}~\bibnamefont {Xu}}, \bibinfo {author} {\bibfnamefont {W.}~\bibnamefont {Yao}}, \bibinfo {author} {\bibfnamefont {D.}~\bibnamefont {Xiao}},\ and\ \bibinfo {author} {\bibfnamefont {T.~F.}\ \bibnamefont {Heinz}},\ }\bibfield  {title} {\bibinfo {title} {{Spin and pseudospins in layered transition metal dichalcogenides}},\ }\href {https://doi.org/https://doi.org/10.1038/nphys2942} {\bibfield  {journal} {\bibinfo  {journal} {Nature Phys}\ }\textbf {\bibinfo {volume} {10}},\ \bibinfo {pages} {343} (\bibinfo {year} {2014})}\BibitemShut {NoStop}%
\bibitem [{\citenamefont {Luo}\ \emph {et~al.}(2024)\citenamefont {Luo}, \citenamefont {Huang}, \citenamefont {Qiao}, \citenamefont {Qi},\ and\ \citenamefont {Peng}}]{luo2024valleytronics}%
  \BibitemOpen
  \bibfield  {author} {\bibinfo {author} {\bibfnamefont {C.}~\bibnamefont {Luo}}, \bibinfo {author} {\bibfnamefont {Z.}~\bibnamefont {Huang}}, \bibinfo {author} {\bibfnamefont {H.}~\bibnamefont {Qiao}}, \bibinfo {author} {\bibfnamefont {X.}~\bibnamefont {Qi}},\ and\ \bibinfo {author} {\bibfnamefont {X.}~\bibnamefont {Peng}},\ }\bibfield  {title} {\bibinfo {title} {{Valleytronics in two-dimensional magnetic materials}},\ }\href {https://doi.org/10.1088/2515-7639/ad3b6e} {\bibfield  {journal} {\bibinfo  {journal} {J. Phys. Mater.}\ }\textbf {\bibinfo {volume} {7}},\ \bibinfo {pages} {022006} (\bibinfo {year} {2024})}\BibitemShut {NoStop}%
\bibitem [{\citenamefont {Yang}\ \emph {et~al.}(2024)\citenamefont {Yang}, \citenamefont {Long}, \citenamefont {Chen}, \citenamefont {Sa}, \citenamefont {Guo}, \citenamefont {Zheng}, \citenamefont {Pei}, \citenamefont {Zhan},\ and\ \citenamefont {Lu}}]{https://doi.org/10.1002/adom.202302900}%
  \BibitemOpen
  \bibfield  {author} {\bibinfo {author} {\bibfnamefont {S.}~\bibnamefont {Yang}}, \bibinfo {author} {\bibfnamefont {H.}~\bibnamefont {Long}}, \bibinfo {author} {\bibfnamefont {W.}~\bibnamefont {Chen}}, \bibinfo {author} {\bibfnamefont {B.}~\bibnamefont {Sa}}, \bibinfo {author} {\bibfnamefont {Z.}~\bibnamefont {Guo}}, \bibinfo {author} {\bibfnamefont {J.}~\bibnamefont {Zheng}}, \bibinfo {author} {\bibfnamefont {J.}~\bibnamefont {Pei}}, \bibinfo {author} {\bibfnamefont {H.}~\bibnamefont {Zhan}},\ and\ \bibinfo {author} {\bibfnamefont {Y.}~\bibnamefont {Lu}},\ }\bibfield  {title} {\bibinfo {title} {{Valleytronics Meets Straintronics: Valley Fine Structure Engineering of 2D Transition Metal Dichalcogenides}},\ }\href {https://doi.org/https://doi.org/10.1002/adom.202302900} {\bibfield  {journal} {\bibinfo  {journal} {Adv. Opt. Mater.}\ }\textbf {\bibinfo {volume} {12}},\ \bibinfo {pages} {2302900} (\bibinfo {year} {2024})}\BibitemShut {NoStop}%
\bibitem [{\citenamefont {Tyulnev}\ \emph {et~al.}(2024)\citenamefont {Tyulnev}, \citenamefont {Jim{\'e}nez-Gal{\'a}n}, \citenamefont {Poborska}, \citenamefont {Vamos}, \citenamefont {Russell}, \citenamefont {Tani}, \citenamefont {Smirnova}, \citenamefont {Ivanov}, \citenamefont {Silva},\ and\ \citenamefont {Biegert}}]{tyulnev2024valleytronics}%
  \BibitemOpen
  \bibfield  {author} {\bibinfo {author} {\bibfnamefont {I.}~\bibnamefont {Tyulnev}}, \bibinfo {author} {\bibfnamefont {{\'A}.}~\bibnamefont {Jim{\'e}nez-Gal{\'a}n}}, \bibinfo {author} {\bibfnamefont {J.}~\bibnamefont {Poborska}}, \bibinfo {author} {\bibfnamefont {L.}~\bibnamefont {Vamos}}, \bibinfo {author} {\bibfnamefont {P.~S.~J.}\ \bibnamefont {Russell}}, \bibinfo {author} {\bibfnamefont {F.}~\bibnamefont {Tani}}, \bibinfo {author} {\bibfnamefont {O.}~\bibnamefont {Smirnova}}, \bibinfo {author} {\bibfnamefont {M.}~\bibnamefont {Ivanov}}, \bibinfo {author} {\bibfnamefont {R.~E.}\ \bibnamefont {Silva}},\ and\ \bibinfo {author} {\bibfnamefont {J.}~\bibnamefont {Biegert}},\ }\bibfield  {title} {\bibinfo {title} {{Valleytronics in bulk ${\mathrm{MoS}}_{2}$ with a topologic optical field}},\ }\href {https://doi.org/https://doi.org/10.1038/s41586-024-07156-y} {\bibfield  {journal} {\bibinfo  {journal} {Nature}\ }\textbf {\bibinfo {volume} {628}},\ \bibinfo {pages} {746} (\bibinfo {year} {2024})}\BibitemShut
  {NoStop}%
\bibitem [{\citenamefont {Nirjhar}\ and\ \citenamefont {Ahmed}(2025)}]{doi:10.1021/acsaelm.4c02276}%
  \BibitemOpen
  \bibfield  {author} {\bibinfo {author} {\bibfnamefont {A.~R.}\ \bibnamefont {Nirjhar}}\ and\ \bibinfo {author} {\bibfnamefont {S.}~\bibnamefont {Ahmed}},\ }\bibfield  {title} {\bibinfo {title} {{Magnetic Proximity-Induced Colossal Valley Splitting in ${\mathrm{WTe}}_{2}$ for Room Temperature Valleytronics}},\ }\href {https://doi.org/10.1021/acsaelm.4c02276} {\bibfield  {journal} {\bibinfo  {journal} {ACS Appl. Electron. Mater.}\ }\textbf {\bibinfo {volume} {7}},\ \bibinfo {pages} {2012} (\bibinfo {year} {2025})}\BibitemShut {NoStop}%
\bibitem [{\citenamefont {Huang}\ \emph {et~al.}(2025)\citenamefont {Huang}, \citenamefont {Samanta}, \citenamefont {Shao},\ and\ \citenamefont {Tsymbal}}]{doi:10.1021/acsnano.4c12812}%
  \BibitemOpen
  \bibfield  {author} {\bibinfo {author} {\bibfnamefont {K.}~\bibnamefont {Huang}}, \bibinfo {author} {\bibfnamefont {K.}~\bibnamefont {Samanta}}, \bibinfo {author} {\bibfnamefont {D.-F.}\ \bibnamefont {Shao}},\ and\ \bibinfo {author} {\bibfnamefont {E.~Y.}\ \bibnamefont {Tsymbal}},\ }\bibfield  {title} {\bibinfo {title} {{Two-Dimensional Nonvolatile Valley Spin Valve}},\ }\href {https://doi.org/10.1021/acsnano.4c12812} {\bibfield  {journal} {\bibinfo  {journal} {ACS Nano}\ }\textbf {\bibinfo {volume} {19}},\ \bibinfo {pages} {3448} (\bibinfo {year} {2025})}\BibitemShut {NoStop}%
\bibitem [{\citenamefont {Jiang}\ \emph {et~al.}(2025{\natexlab{a}})\citenamefont {Jiang}, \citenamefont {Zhang}, \citenamefont {An}, \citenamefont {Tan}, \citenamefont {Dai}, \citenamefont {Chen}, \citenamefont {Chen}, \citenamefont {Cai}, \citenamefont {Fu}, \citenamefont {Z{\'u}{\~n}iga-P{\'e}rez} \emph {et~al.}}]{jiang2025chiral}%
  \BibitemOpen
  \bibfield  {author} {\bibinfo {author} {\bibfnamefont {H.}~\bibnamefont {Jiang}}, \bibinfo {author} {\bibfnamefont {Y.}~\bibnamefont {Zhang}}, \bibinfo {author} {\bibfnamefont {L.}~\bibnamefont {An}}, \bibinfo {author} {\bibfnamefont {Q.}~\bibnamefont {Tan}}, \bibinfo {author} {\bibfnamefont {X.}~\bibnamefont {Dai}}, \bibinfo {author} {\bibfnamefont {Y.}~\bibnamefont {Chen}}, \bibinfo {author} {\bibfnamefont {W.}~\bibnamefont {Chen}}, \bibinfo {author} {\bibfnamefont {H.}~\bibnamefont {Cai}}, \bibinfo {author} {\bibfnamefont {J.}~\bibnamefont {Fu}}, \bibinfo {author} {\bibfnamefont {J.}~\bibnamefont {Z{\'u}{\~n}iga-P{\'e}rez}}, \emph {et~al.},\ }\bibfield  {title} {\bibinfo {title} {{Chiral light detection with centrosymmetric-metamaterial-assisted valleytronics}},\ }\href {https://doi.org/https://doi.org/10.1038/s41563-025-02155-4} {\bibfield  {journal} {\bibinfo  {journal} {Nat. Mater.}\ ,\ \bibinfo {pages} {1}} (\bibinfo {year} {2025}{\natexlab{a}})}\BibitemShut {NoStop}%
\bibitem [{\citenamefont {Guo}\ \emph {et~al.}(2025)\citenamefont {Guo}, \citenamefont {Li},\ and\ \citenamefont {Wang}}]{PhysRevB.111.L140404}%
  \BibitemOpen
  \bibfield  {author} {\bibinfo {author} {\bibfnamefont {S.-D.}\ \bibnamefont {Guo}}, \bibinfo {author} {\bibfnamefont {P.}~\bibnamefont {Li}},\ and\ \bibinfo {author} {\bibfnamefont {G.}~\bibnamefont {Wang}},\ }\bibfield  {title} {\bibinfo {title} {{First-principles calculations study of valley polarization in antiferromagnetic bilayer systems}},\ }\href {https://doi.org/10.1103/PhysRevB.111.L140404} {\bibfield  {journal} {\bibinfo  {journal} {Phys. Rev. B}\ }\textbf {\bibinfo {volume} {111}},\ \bibinfo {pages} {L140404} (\bibinfo {year} {2025})}\BibitemShut {NoStop}%
\bibitem [{\citenamefont {Shirai}\ \emph {et~al.}(2025)\citenamefont {Shirai}, \citenamefont {Murotani}, \citenamefont {Fujimoto}, \citenamefont {Kanda}, \citenamefont {Yoshinobu},\ and\ \citenamefont {Matsunaga}}]{PhysRevB.111.L121201}%
  \BibitemOpen
  \bibfield  {author} {\bibinfo {author} {\bibfnamefont {A.~M.}\ \bibnamefont {Shirai}}, \bibinfo {author} {\bibfnamefont {Y.}~\bibnamefont {Murotani}}, \bibinfo {author} {\bibfnamefont {T.}~\bibnamefont {Fujimoto}}, \bibinfo {author} {\bibfnamefont {N.}~\bibnamefont {Kanda}}, \bibinfo {author} {\bibfnamefont {J.}~\bibnamefont {Yoshinobu}},\ and\ \bibinfo {author} {\bibfnamefont {R.}~\bibnamefont {Matsunaga}},\ }\bibfield  {title} {\bibinfo {title} {{Valley polarization dynamics of photoinjected carriers at the band edge in room-temperature silicon studied by terahertz polarimetry}},\ }\href {https://doi.org/10.1103/PhysRevB.111.L121201} {\bibfield  {journal} {\bibinfo  {journal} {Phys. Rev. B}\ }\textbf {\bibinfo {volume} {111}},\ \bibinfo {pages} {L121201} (\bibinfo {year} {2025})}\BibitemShut {NoStop}%
\bibitem [{\citenamefont {Shen}\ \emph {et~al.}(2025)\citenamefont {Shen}, \citenamefont {Zheng}, \citenamefont {Tong}, \citenamefont {Bao}, \citenamefont {Wan},\ and\ \citenamefont {Duan}}]{PhysRevB.111.075421}%
  \BibitemOpen
  \bibfield  {author} {\bibinfo {author} {\bibfnamefont {Y.-H.}\ \bibnamefont {Shen}}, \bibinfo {author} {\bibfnamefont {J.-D.}\ \bibnamefont {Zheng}}, \bibinfo {author} {\bibfnamefont {W.-Y.}\ \bibnamefont {Tong}}, \bibinfo {author} {\bibfnamefont {Z.-Q.}\ \bibnamefont {Bao}}, \bibinfo {author} {\bibfnamefont {X.-G.}\ \bibnamefont {Wan}},\ and\ \bibinfo {author} {\bibfnamefont {C.-G.}\ \bibnamefont {Duan}},\ }\bibfield  {title} {\bibinfo {title} {{Twist-induced quantum valley Hall states in bilayer germanene}},\ }\href {https://doi.org/10.1103/PhysRevB.111.075421} {\bibfield  {journal} {\bibinfo  {journal} {Phys. Rev. B}\ }\textbf {\bibinfo {volume} {111}},\ \bibinfo {pages} {075421} (\bibinfo {year} {2025})}\BibitemShut {NoStop}%
\bibitem [{\citenamefont {Herrmann}\ \emph {et~al.}(2025)\citenamefont {Herrmann}, \citenamefont {Klimmer}, \citenamefont {Lettau}, \citenamefont {Weickhardt}, \citenamefont {Papavasileiou}, \citenamefont {Mosina}, \citenamefont {Sofer}, \citenamefont {Paradisanos}, \citenamefont {Kartashov}, \citenamefont {Wilhelm} \emph {et~al.}}]{herrmann2025nonlinear}%
  \BibitemOpen
  \bibfield  {author} {\bibinfo {author} {\bibfnamefont {P.}~\bibnamefont {Herrmann}}, \bibinfo {author} {\bibfnamefont {S.}~\bibnamefont {Klimmer}}, \bibinfo {author} {\bibfnamefont {T.}~\bibnamefont {Lettau}}, \bibinfo {author} {\bibfnamefont {T.}~\bibnamefont {Weickhardt}}, \bibinfo {author} {\bibfnamefont {A.}~\bibnamefont {Papavasileiou}}, \bibinfo {author} {\bibfnamefont {K.}~\bibnamefont {Mosina}}, \bibinfo {author} {\bibfnamefont {Z.}~\bibnamefont {Sofer}}, \bibinfo {author} {\bibfnamefont {I.}~\bibnamefont {Paradisanos}}, \bibinfo {author} {\bibfnamefont {D.}~\bibnamefont {Kartashov}}, \bibinfo {author} {\bibfnamefont {J.}~\bibnamefont {Wilhelm}}, \emph {et~al.},\ }\bibfield  {title} {\bibinfo {title} {{Nonlinear valley selection rules and all-optical probe of broken time-reversal symmetry in monolayer ${\mathrm{WSe}}_{2}$}},\ }\href {https://doi.org/https://doi.org/10.1038/s41566-024-01591-z} {\bibfield  {journal} {\bibinfo  {journal} {Nat. Photon.}\ }\textbf {\bibinfo {volume} {19}},\ \bibinfo
  {pages} {300} (\bibinfo {year} {2025})}\BibitemShut {NoStop}%
\bibitem [{\citenamefont {Wang}\ \emph {et~al.}(2025)\citenamefont {Wang}, \citenamefont {Chang}, \citenamefont {Duan}, \citenamefont {Xu},\ and\ \citenamefont {Tang}}]{PhysRevLett.134.026904}%
  \BibitemOpen
  \bibfield  {author} {\bibinfo {author} {\bibfnamefont {R.}~\bibnamefont {Wang}}, \bibinfo {author} {\bibfnamefont {K.}~\bibnamefont {Chang}}, \bibinfo {author} {\bibfnamefont {W.}~\bibnamefont {Duan}}, \bibinfo {author} {\bibfnamefont {Y.}~\bibnamefont {Xu}},\ and\ \bibinfo {author} {\bibfnamefont {P.}~\bibnamefont {Tang}},\ }\bibfield  {title} {\bibinfo {title} {{Twist-Angle-Dependent Valley Polarization of Intralayer Moir\'e Excitons in van der Waals Superlattices}},\ }\href {https://doi.org/10.1103/PhysRevLett.134.026904} {\bibfield  {journal} {\bibinfo  {journal} {Phys. Rev. Lett.}\ }\textbf {\bibinfo {volume} {134}},\ \bibinfo {pages} {026904} (\bibinfo {year} {2025})}\BibitemShut {NoStop}%
\bibitem [{\citenamefont {Zhang}\ \emph {et~al.}(2025)\citenamefont {Zhang}, \citenamefont {Zhou},\ and\ \citenamefont {Cheng}}]{PhysRevB.111.165307}%
  \BibitemOpen
  \bibfield  {author} {\bibinfo {author} {\bibfnamefont {W.}~\bibnamefont {Zhang}}, \bibinfo {author} {\bibfnamefont {J.}~\bibnamefont {Zhou}},\ and\ \bibinfo {author} {\bibfnamefont {S.}~\bibnamefont {Cheng}},\ }\bibfield  {title} {\bibinfo {title} {{Valley transport in breathing kagome lattices with chiral spin textures}},\ }\href {https://doi.org/10.1103/PhysRevB.111.165307} {\bibfield  {journal} {\bibinfo  {journal} {Phys. Rev. B}\ }\textbf {\bibinfo {volume} {111}},\ \bibinfo {pages} {165307} (\bibinfo {year} {2025})}\BibitemShut {NoStop}%
\bibitem [{\citenamefont {Jiang}\ \emph {et~al.}(2025{\natexlab{b}})\citenamefont {Jiang}, \citenamefont {Qiao},\ and\ \citenamefont {Zhang}}]{PhysRevB.111.L140416}%
  \BibitemOpen
  \bibfield  {author} {\bibinfo {author} {\bibfnamefont {X.-C.}\ \bibnamefont {Jiang}}, \bibinfo {author} {\bibfnamefont {L.-Y.}\ \bibnamefont {Qiao}},\ and\ \bibinfo {author} {\bibfnamefont {Y.-Z.}\ \bibnamefont {Zhang}},\ }\bibfield  {title} {\bibinfo {title} {{Anomalous valley Hall effect in monolayer chromium-based triple-$Q$ magnets}},\ }\href {https://doi.org/10.1103/PhysRevB.111.L140416} {\bibfield  {journal} {\bibinfo  {journal} {Phys. Rev. B}\ }\textbf {\bibinfo {volume} {111}},\ \bibinfo {pages} {L140416} (\bibinfo {year} {2025}{\natexlab{b}})}\BibitemShut {NoStop}%
\bibitem [{\citenamefont {\ifmmode \check{Z}\else \v{Z}\fi{}uti\ifmmode~\acute{c}\else \'{c}\fi{}}\ \emph {et~al.}(2004)\citenamefont {\ifmmode \check{Z}\else \v{Z}\fi{}uti\ifmmode~\acute{c}\else \'{c}\fi{}}, \citenamefont {Fabian},\ and\ \citenamefont {Das~Sarma}}]{RevModPhys.76.323}%
  \BibitemOpen
  \bibfield  {author} {\bibinfo {author} {\bibfnamefont {I.}~\bibnamefont {\ifmmode \check{Z}\else \v{Z}\fi{}uti\ifmmode~\acute{c}\else \'{c}\fi{}}}, \bibinfo {author} {\bibfnamefont {J.}~\bibnamefont {Fabian}},\ and\ \bibinfo {author} {\bibfnamefont {S.}~\bibnamefont {Das~Sarma}},\ }\bibfield  {title} {\bibinfo {title} {{Spintronics: Fundamentals and applications}},\ }\href {https://doi.org/10.1103/RevModPhys.76.323} {\bibfield  {journal} {\bibinfo  {journal} {Rev. Mod. Phys.}\ }\textbf {\bibinfo {volume} {76}},\ \bibinfo {pages} {323} (\bibinfo {year} {2004})}\BibitemShut {NoStop}%
\bibitem [{\citenamefont {Gunawan}\ \emph {et~al.}(2006)\citenamefont {Gunawan}, \citenamefont {Shkolnikov}, \citenamefont {Vakili}, \citenamefont {Gokmen}, \citenamefont {De~Poortere},\ and\ \citenamefont {Shayegan}}]{PhysRevLett.97.186404}%
  \BibitemOpen
  \bibfield  {author} {\bibinfo {author} {\bibfnamefont {O.}~\bibnamefont {Gunawan}}, \bibinfo {author} {\bibfnamefont {Y.~P.}\ \bibnamefont {Shkolnikov}}, \bibinfo {author} {\bibfnamefont {K.}~\bibnamefont {Vakili}}, \bibinfo {author} {\bibfnamefont {T.}~\bibnamefont {Gokmen}}, \bibinfo {author} {\bibfnamefont {E.~P.}\ \bibnamefont {De~Poortere}},\ and\ \bibinfo {author} {\bibfnamefont {M.}~\bibnamefont {Shayegan}},\ }\bibfield  {title} {\bibinfo {title} {{Valley Susceptibility of an Interacting Two-Dimensional Electron System}},\ }\href {https://doi.org/10.1103/PhysRevLett.97.186404} {\bibfield  {journal} {\bibinfo  {journal} {Phys. Rev. Lett.}\ }\textbf {\bibinfo {volume} {97}},\ \bibinfo {pages} {186404} (\bibinfo {year} {2006})}\BibitemShut {NoStop}%
\bibitem [{\citenamefont {Gokmen}\ \emph {et~al.}(2008)\citenamefont {Gokmen}, \citenamefont {Padmanabhan}, \citenamefont {Gunawan}, \citenamefont {Shkolnikov}, \citenamefont {Vakili}, \citenamefont {De~Poortere},\ and\ \citenamefont {Shayegan}}]{PhysRevB.78.233306}%
  \BibitemOpen
  \bibfield  {author} {\bibinfo {author} {\bibfnamefont {T.}~\bibnamefont {Gokmen}}, \bibinfo {author} {\bibfnamefont {M.}~\bibnamefont {Padmanabhan}}, \bibinfo {author} {\bibfnamefont {O.}~\bibnamefont {Gunawan}}, \bibinfo {author} {\bibfnamefont {Y.~P.}\ \bibnamefont {Shkolnikov}}, \bibinfo {author} {\bibfnamefont {K.}~\bibnamefont {Vakili}}, \bibinfo {author} {\bibfnamefont {E.~P.}\ \bibnamefont {De~Poortere}},\ and\ \bibinfo {author} {\bibfnamefont {M.}~\bibnamefont {Shayegan}},\ }\bibfield  {title} {\bibinfo {title} {{Parallel magnetic-field tuning of valley splitting in AlAs two-dimensional electrons}},\ }\href {https://doi.org/10.1103/PhysRevB.78.233306} {\bibfield  {journal} {\bibinfo  {journal} {Phys. Rev. B}\ }\textbf {\bibinfo {volume} {78}},\ \bibinfo {pages} {233306} (\bibinfo {year} {2008})}\BibitemShut {NoStop}%
\bibitem [{\citenamefont {Mueed}\ \emph {et~al.}(2018)\citenamefont {Mueed}, \citenamefont {Hossain}, \citenamefont {Jo}, \citenamefont {Pfeiffer}, \citenamefont {West}, \citenamefont {Baldwin},\ and\ \citenamefont {Shayegan}}]{PhysRevLett.121.036802}%
  \BibitemOpen
  \bibfield  {author} {\bibinfo {author} {\bibfnamefont {M.~A.}\ \bibnamefont {Mueed}}, \bibinfo {author} {\bibfnamefont {M.~S.}\ \bibnamefont {Hossain}}, \bibinfo {author} {\bibfnamefont {I.}~\bibnamefont {Jo}}, \bibinfo {author} {\bibfnamefont {L.~N.}\ \bibnamefont {Pfeiffer}}, \bibinfo {author} {\bibfnamefont {K.~W.}\ \bibnamefont {West}}, \bibinfo {author} {\bibfnamefont {K.~W.}\ \bibnamefont {Baldwin}},\ and\ \bibinfo {author} {\bibfnamefont {M.}~\bibnamefont {Shayegan}},\ }\bibfield  {title} {\bibinfo {title} {{Realization of a Valley Superlattice}},\ }\href {https://doi.org/10.1103/PhysRevLett.121.036802} {\bibfield  {journal} {\bibinfo  {journal} {Phys. Rev. Lett.}\ }\textbf {\bibinfo {volume} {121}},\ \bibinfo {pages} {036802} (\bibinfo {year} {2018})}\BibitemShut {NoStop}%
\bibitem [{\citenamefont {Tong}\ \emph {et~al.}(2016)\citenamefont {Tong}, \citenamefont {Gong}, \citenamefont {Wan},\ and\ \citenamefont {Duan}}]{tong2016concepts}%
  \BibitemOpen
  \bibfield  {author} {\bibinfo {author} {\bibfnamefont {W.-Y.}\ \bibnamefont {Tong}}, \bibinfo {author} {\bibfnamefont {S.-J.}\ \bibnamefont {Gong}}, \bibinfo {author} {\bibfnamefont {X.}~\bibnamefont {Wan}},\ and\ \bibinfo {author} {\bibfnamefont {C.-G.}\ \bibnamefont {Duan}},\ }\bibfield  {title} {\bibinfo {title} {{Concepts of ferrovalley material and anomalous valley Hall effect}},\ }\href {https://doi.org/https://doi.org/10.1038/ncomms13612} {\bibfield  {journal} {\bibinfo  {journal} {Nat Commun}\ }\textbf {\bibinfo {volume} {7}},\ \bibinfo {pages} {13612} (\bibinfo {year} {2016})}\BibitemShut {NoStop}%
\bibitem [{\citenamefont {Shen}\ \emph {et~al.}(2017)\citenamefont {Shen}, \citenamefont {Tong}, \citenamefont {Gong},\ and\ \citenamefont {Duan}}]{Shen_2018}%
  \BibitemOpen
  \bibfield  {author} {\bibinfo {author} {\bibfnamefont {X.-W.}\ \bibnamefont {Shen}}, \bibinfo {author} {\bibfnamefont {W.-Y.}\ \bibnamefont {Tong}}, \bibinfo {author} {\bibfnamefont {S.-J.}\ \bibnamefont {Gong}},\ and\ \bibinfo {author} {\bibfnamefont {C.-G.}\ \bibnamefont {Duan}},\ }\bibfield  {title} {\bibinfo {title} {{Electrically tunable polarizer based on 2D orthorhombic ferrovalley materials}},\ }\href {https://doi.org/10.1088/2053-1583/aa8d3b} {\bibfield  {journal} {\bibinfo  {journal} {2D Mater.}\ }\textbf {\bibinfo {volume} {5}},\ \bibinfo {pages} {011001} (\bibinfo {year} {2017})}\BibitemShut {NoStop}%
\bibitem [{\citenamefont {Lai}\ \emph {et~al.}(2019)\citenamefont {Lai}, \citenamefont {Song}, \citenamefont {Wan}, \citenamefont {Xue}, \citenamefont {Wang}, \citenamefont {Ye}, \citenamefont {Dai}, \citenamefont {Zhang}, \citenamefont {Yang}, \citenamefont {Du} \emph {et~al.}}]{lai2019two}%
  \BibitemOpen
  \bibfield  {author} {\bibinfo {author} {\bibfnamefont {Y.}~\bibnamefont {Lai}}, \bibinfo {author} {\bibfnamefont {Z.}~\bibnamefont {Song}}, \bibinfo {author} {\bibfnamefont {Y.}~\bibnamefont {Wan}}, \bibinfo {author} {\bibfnamefont {M.}~\bibnamefont {Xue}}, \bibinfo {author} {\bibfnamefont {C.}~\bibnamefont {Wang}}, \bibinfo {author} {\bibfnamefont {Y.}~\bibnamefont {Ye}}, \bibinfo {author} {\bibfnamefont {L.}~\bibnamefont {Dai}}, \bibinfo {author} {\bibfnamefont {Z.}~\bibnamefont {Zhang}}, \bibinfo {author} {\bibfnamefont {W.}~\bibnamefont {Yang}}, \bibinfo {author} {\bibfnamefont {H.}~\bibnamefont {Du}}, \emph {et~al.},\ }\bibfield  {title} {\bibinfo {title} {{Two-dimensional ferromagnetism and driven ferroelectricity in van der Waals \ch{CuCrP2S6}}},\ }\href {https://doi.org/https://doi.org/10.1039/C9NR00738E} {\bibfield  {journal} {\bibinfo  {journal} {Nanoscale}\ }\textbf {\bibinfo {volume} {11}},\ \bibinfo {pages} {5163} (\bibinfo {year} {2019})}\BibitemShut {NoStop}%
\bibitem [{\citenamefont {Hu}\ \emph {et~al.}(2020)\citenamefont {Hu}, \citenamefont {Tong}, \citenamefont {Shen}, \citenamefont {Wan},\ and\ \citenamefont {Duan}}]{hu2020concepts}%
  \BibitemOpen
  \bibfield  {author} {\bibinfo {author} {\bibfnamefont {H.}~\bibnamefont {Hu}}, \bibinfo {author} {\bibfnamefont {W.-Y.}\ \bibnamefont {Tong}}, \bibinfo {author} {\bibfnamefont {Y.-H.}\ \bibnamefont {Shen}}, \bibinfo {author} {\bibfnamefont {X.}~\bibnamefont {Wan}},\ and\ \bibinfo {author} {\bibfnamefont {C.-G.}\ \bibnamefont {Duan}},\ }\bibfield  {title} {\bibinfo {title} {{Concepts of the half-valley-metal and quantum anomalous valley Hall effect}},\ }\href {https://doi.org/https://doi.org/10.1038/s41524-020-00397-1} {\bibfield  {journal} {\bibinfo  {journal} {npj Comput Mater}\ }\textbf {\bibinfo {volume} {6}},\ \bibinfo {pages} {129} (\bibinfo {year} {2020})}\BibitemShut {NoStop}%
\bibitem [{\citenamefont {Wan}\ \emph {et~al.}(2025)\citenamefont {Wan}, \citenamefont {Zhou}, \citenamefont {Zheng}, \citenamefont {Duan}, \citenamefont {Chen},\ and\ \citenamefont {Ouyang}}]{doi:10.1021/acsanm.4c06703}%
  \BibitemOpen
  \bibfield  {author} {\bibinfo {author} {\bibfnamefont {Z.}~\bibnamefont {Wan}}, \bibinfo {author} {\bibfnamefont {W.}~\bibnamefont {Zhou}}, \bibinfo {author} {\bibfnamefont {G.}~\bibnamefont {Zheng}}, \bibinfo {author} {\bibfnamefont {H.}~\bibnamefont {Duan}}, \bibinfo {author} {\bibfnamefont {Y.}~\bibnamefont {Chen}},\ and\ \bibinfo {author} {\bibfnamefont {F.}~\bibnamefont {Ouyang}},\ }\bibfield  {title} {\bibinfo {title} {{Tuning the Magnetization Direction and Valley Splitting of Monolayer Ferrovalley CrISe: Implications for Spintronics and Valleytronics Devices}},\ }\href {https://doi.org/10.1021/acsanm.4c06703} {\bibfield  {journal} {\bibinfo  {journal} {ACS Appl. Nano Mater.}\ }\textbf {\bibinfo {volume} {8}},\ \bibinfo {pages} {2466} (\bibinfo {year} {2025})}\BibitemShut {NoStop}%
\bibitem [{\citenamefont {Cao}\ \emph {et~al.}(2012)\citenamefont {Cao}, \citenamefont {Wang}, \citenamefont {Han}, \citenamefont {Ye}, \citenamefont {Zhu}, \citenamefont {Shi}, \citenamefont {Niu}, \citenamefont {Tan}, \citenamefont {Wang}, \citenamefont {Liu} \emph {et~al.}}]{cao2012valley}%
  \BibitemOpen
  \bibfield  {author} {\bibinfo {author} {\bibfnamefont {T.}~\bibnamefont {Cao}}, \bibinfo {author} {\bibfnamefont {G.}~\bibnamefont {Wang}}, \bibinfo {author} {\bibfnamefont {W.}~\bibnamefont {Han}}, \bibinfo {author} {\bibfnamefont {H.}~\bibnamefont {Ye}}, \bibinfo {author} {\bibfnamefont {C.}~\bibnamefont {Zhu}}, \bibinfo {author} {\bibfnamefont {J.}~\bibnamefont {Shi}}, \bibinfo {author} {\bibfnamefont {Q.}~\bibnamefont {Niu}}, \bibinfo {author} {\bibfnamefont {P.}~\bibnamefont {Tan}}, \bibinfo {author} {\bibfnamefont {E.}~\bibnamefont {Wang}}, \bibinfo {author} {\bibfnamefont {B.}~\bibnamefont {Liu}}, \emph {et~al.},\ }\bibfield  {title} {\bibinfo {title} {{Valley-selective circular dichroism of monolayer molybdenum disulphide}},\ }\href {https://doi.org/https://doi.org/10.1038/ncomms1882} {\bibfield  {journal} {\bibinfo  {journal} {Nat Commun}\ }\textbf {\bibinfo {volume} {3}},\ \bibinfo {pages} {887} (\bibinfo {year} {2012})}\BibitemShut {NoStop}%
\bibitem [{\citenamefont {Zeng}\ \emph {et~al.}(2012)\citenamefont {Zeng}, \citenamefont {Dai}, \citenamefont {Yao}, \citenamefont {Xiao},\ and\ \citenamefont {Cui}}]{zeng2012valley}%
  \BibitemOpen
  \bibfield  {author} {\bibinfo {author} {\bibfnamefont {H.}~\bibnamefont {Zeng}}, \bibinfo {author} {\bibfnamefont {J.}~\bibnamefont {Dai}}, \bibinfo {author} {\bibfnamefont {W.}~\bibnamefont {Yao}}, \bibinfo {author} {\bibfnamefont {D.}~\bibnamefont {Xiao}},\ and\ \bibinfo {author} {\bibfnamefont {X.}~\bibnamefont {Cui}},\ }\bibfield  {title} {\bibinfo {title} {{Valley polarization in ${\mathrm{MoS}}_{2}$ monolayers by optical pumping}},\ }\href {https://doi.org/https://doi.org/10.1038/nnano.2012.95} {\bibfield  {journal} {\bibinfo  {journal} {Nature Nanotech}\ }\textbf {\bibinfo {volume} {7}},\ \bibinfo {pages} {490} (\bibinfo {year} {2012})}\BibitemShut {NoStop}%
\bibitem [{\citenamefont {Jones}\ \emph {et~al.}(2013)\citenamefont {Jones}, \citenamefont {Yu}, \citenamefont {Ghimire}, \citenamefont {Wu}, \citenamefont {Aivazian}, \citenamefont {Ross}, \citenamefont {Zhao}, \citenamefont {Yan}, \citenamefont {Mandrus}, \citenamefont {Xiao} \emph {et~al.}}]{jones2013optical}%
  \BibitemOpen
  \bibfield  {author} {\bibinfo {author} {\bibfnamefont {A.~M.}\ \bibnamefont {Jones}}, \bibinfo {author} {\bibfnamefont {H.}~\bibnamefont {Yu}}, \bibinfo {author} {\bibfnamefont {N.~J.}\ \bibnamefont {Ghimire}}, \bibinfo {author} {\bibfnamefont {S.}~\bibnamefont {Wu}}, \bibinfo {author} {\bibfnamefont {G.}~\bibnamefont {Aivazian}}, \bibinfo {author} {\bibfnamefont {J.~S.}\ \bibnamefont {Ross}}, \bibinfo {author} {\bibfnamefont {B.}~\bibnamefont {Zhao}}, \bibinfo {author} {\bibfnamefont {J.}~\bibnamefont {Yan}}, \bibinfo {author} {\bibfnamefont {D.~G.}\ \bibnamefont {Mandrus}}, \bibinfo {author} {\bibfnamefont {D.}~\bibnamefont {Xiao}}, \emph {et~al.},\ }\bibfield  {title} {\bibinfo {title} {{Optical generation of excitonic valley coherence in monolayer ${\mathrm{WSe}}_{2}$}},\ }\href {https://doi.org/https://doi.org/10.1038/nnano.2013.151} {\bibfield  {journal} {\bibinfo  {journal} {Nature Nanotech}\ }\textbf {\bibinfo {volume} {8}},\ \bibinfo {pages} {634} (\bibinfo {year} {2013})}\BibitemShut {NoStop}%
\bibitem [{\citenamefont {Suzuki}\ \emph {et~al.}(2014)\citenamefont {Suzuki}, \citenamefont {Sakano}, \citenamefont {Zhang}, \citenamefont {Akashi}, \citenamefont {Morikawa}, \citenamefont {Harasawa}, \citenamefont {Yaji}, \citenamefont {Kuroda}, \citenamefont {Miyamoto}, \citenamefont {Okuda} \emph {et~al.}}]{suzuki2014valley}%
  \BibitemOpen
  \bibfield  {author} {\bibinfo {author} {\bibfnamefont {R.}~\bibnamefont {Suzuki}}, \bibinfo {author} {\bibfnamefont {M.}~\bibnamefont {Sakano}}, \bibinfo {author} {\bibfnamefont {Y.}~\bibnamefont {Zhang}}, \bibinfo {author} {\bibfnamefont {R.}~\bibnamefont {Akashi}}, \bibinfo {author} {\bibfnamefont {D.}~\bibnamefont {Morikawa}}, \bibinfo {author} {\bibfnamefont {A.}~\bibnamefont {Harasawa}}, \bibinfo {author} {\bibfnamefont {K.}~\bibnamefont {Yaji}}, \bibinfo {author} {\bibfnamefont {K.}~\bibnamefont {Kuroda}}, \bibinfo {author} {\bibfnamefont {K.}~\bibnamefont {Miyamoto}}, \bibinfo {author} {\bibfnamefont {T.}~\bibnamefont {Okuda}}, \emph {et~al.},\ }\bibfield  {title} {\bibinfo {title} {{Valley-dependent spin polarization in bulk ${\mathrm{MoS}}_{2}$ with broken inversion symmetry}},\ }\href {https://doi.org/https://doi.org/10.1038/nnano.2014.148} {\bibfield  {journal} {\bibinfo  {journal} {Nature Nanotech}\ }\textbf {\bibinfo {volume} {9}},\ \bibinfo {pages} {611} (\bibinfo {year} {2014})}\BibitemShut
  {NoStop}%
\bibitem [{\citenamefont {Aivazian}\ \emph {et~al.}(2015)\citenamefont {Aivazian}, \citenamefont {Gong}, \citenamefont {Jones}, \citenamefont {Chu}, \citenamefont {Yan}, \citenamefont {Mandrus}, \citenamefont {Zhang}, \citenamefont {Cobden}, \citenamefont {Yao},\ and\ \citenamefont {Xu}}]{aivazian2015magnetic}%
  \BibitemOpen
  \bibfield  {author} {\bibinfo {author} {\bibfnamefont {G.}~\bibnamefont {Aivazian}}, \bibinfo {author} {\bibfnamefont {Z.}~\bibnamefont {Gong}}, \bibinfo {author} {\bibfnamefont {A.~M.}\ \bibnamefont {Jones}}, \bibinfo {author} {\bibfnamefont {R.-L.}\ \bibnamefont {Chu}}, \bibinfo {author} {\bibfnamefont {J.}~\bibnamefont {Yan}}, \bibinfo {author} {\bibfnamefont {D.~G.}\ \bibnamefont {Mandrus}}, \bibinfo {author} {\bibfnamefont {C.}~\bibnamefont {Zhang}}, \bibinfo {author} {\bibfnamefont {D.}~\bibnamefont {Cobden}}, \bibinfo {author} {\bibfnamefont {W.}~\bibnamefont {Yao}},\ and\ \bibinfo {author} {\bibfnamefont {X.}~\bibnamefont {Xu}},\ }\bibfield  {title} {\bibinfo {title} {{Magnetic control of valley pseudospin in monolayer ${\mathrm{WSe}}_{2}$}},\ }\href {https://doi.org/https://doi.org/10.1038/nphys3201} {\bibfield  {journal} {\bibinfo  {journal} {Nature Phys}\ }\textbf {\bibinfo {volume} {11}},\ \bibinfo {pages} {148} (\bibinfo {year} {2015})}\BibitemShut {NoStop}%
\bibitem [{\citenamefont {Wang}\ \emph {et~al.}(2017)\citenamefont {Wang}, \citenamefont {Shan},\ and\ \citenamefont {Mak}}]{wang2017valley}%
  \BibitemOpen
  \bibfield  {author} {\bibinfo {author} {\bibfnamefont {Z.}~\bibnamefont {Wang}}, \bibinfo {author} {\bibfnamefont {J.}~\bibnamefont {Shan}},\ and\ \bibinfo {author} {\bibfnamefont {K.~F.}\ \bibnamefont {Mak}},\ }\bibfield  {title} {\bibinfo {title} {{Valley-and spin-polarized Landau levels in monolayer ${\mathrm{WSe}}_{2}$}},\ }\href {https://doi.org/https://doi.org/10.1038/nnano.2016.213} {\bibfield  {journal} {\bibinfo  {journal} {Nature Nanotech}\ }\textbf {\bibinfo {volume} {12}},\ \bibinfo {pages} {144} (\bibinfo {year} {2017})}\BibitemShut {NoStop}%
\bibitem [{\citenamefont {Zhong}\ \emph {et~al.}(2017)\citenamefont {Zhong}, \citenamefont {Seyler}, \citenamefont {Linpeng}, \citenamefont {Cheng}, \citenamefont {Sivadas}, \citenamefont {Huang}, \citenamefont {Schmidgall}, \citenamefont {Taniguchi}, \citenamefont {Watanabe}, \citenamefont {McGuire}, \citenamefont {Yao}, \citenamefont {Xiao}, \citenamefont {Fu},\ and\ \citenamefont {Xu}}]{doi:10.1126/sciadv.1603113}%
  \BibitemOpen
  \bibfield  {author} {\bibinfo {author} {\bibfnamefont {D.}~\bibnamefont {Zhong}}, \bibinfo {author} {\bibfnamefont {K.~L.}\ \bibnamefont {Seyler}}, \bibinfo {author} {\bibfnamefont {X.}~\bibnamefont {Linpeng}}, \bibinfo {author} {\bibfnamefont {R.}~\bibnamefont {Cheng}}, \bibinfo {author} {\bibfnamefont {N.}~\bibnamefont {Sivadas}}, \bibinfo {author} {\bibfnamefont {B.}~\bibnamefont {Huang}}, \bibinfo {author} {\bibfnamefont {E.}~\bibnamefont {Schmidgall}}, \bibinfo {author} {\bibfnamefont {T.}~\bibnamefont {Taniguchi}}, \bibinfo {author} {\bibfnamefont {K.}~\bibnamefont {Watanabe}}, \bibinfo {author} {\bibfnamefont {M.~A.}\ \bibnamefont {McGuire}}, \bibinfo {author} {\bibfnamefont {W.}~\bibnamefont {Yao}}, \bibinfo {author} {\bibfnamefont {D.}~\bibnamefont {Xiao}}, \bibinfo {author} {\bibfnamefont {K.-M.~C.}\ \bibnamefont {Fu}},\ and\ \bibinfo {author} {\bibfnamefont {X.}~\bibnamefont {Xu}},\ }\bibfield  {title} {\bibinfo {title} {{Van der Waals engineering of ferromagnetic semiconductor heterostructures
  for spin and valleytronics}},\ }\href {https://doi.org/10.1126/sciadv.1603113} {\bibfield  {journal} {\bibinfo  {journal} {Sci. Adv.}\ }\textbf {\bibinfo {volume} {3}},\ \bibinfo {pages} {e1603113} (\bibinfo {year} {2017})}\BibitemShut {NoStop}%
\bibitem [{\citenamefont {Dey}\ \emph {et~al.}(2017)\citenamefont {Dey}, \citenamefont {Yang}, \citenamefont {Robert}, \citenamefont {Wang}, \citenamefont {Urbaszek}, \citenamefont {Marie},\ and\ \citenamefont {Crooker}}]{PhysRevLett.119.137401}%
  \BibitemOpen
  \bibfield  {author} {\bibinfo {author} {\bibfnamefont {P.}~\bibnamefont {Dey}}, \bibinfo {author} {\bibfnamefont {L.}~\bibnamefont {Yang}}, \bibinfo {author} {\bibfnamefont {C.}~\bibnamefont {Robert}}, \bibinfo {author} {\bibfnamefont {G.}~\bibnamefont {Wang}}, \bibinfo {author} {\bibfnamefont {B.}~\bibnamefont {Urbaszek}}, \bibinfo {author} {\bibfnamefont {X.}~\bibnamefont {Marie}},\ and\ \bibinfo {author} {\bibfnamefont {S.~A.}\ \bibnamefont {Crooker}},\ }\bibfield  {title} {\bibinfo {title} {{Gate-Controlled Spin-Valley Locking of Resident Carriers in ${\mathrm{WSe}}_{2}$ Monolayers}},\ }\href {https://doi.org/10.1103/PhysRevLett.119.137401} {\bibfield  {journal} {\bibinfo  {journal} {Phys. Rev. Lett.}\ }\textbf {\bibinfo {volume} {119}},\ \bibinfo {pages} {137401} (\bibinfo {year} {2017})}\BibitemShut {NoStop}%
\bibitem [{\citenamefont {Wu}\ \emph {et~al.}(2019)\citenamefont {Wu}, \citenamefont {Zhou}, \citenamefont {Cai}, \citenamefont {Cheung}, \citenamefont {Liu}, \citenamefont {Huang}, \citenamefont {Lin}, \citenamefont {Han}, \citenamefont {An}, \citenamefont {Wang} \emph {et~al.}}]{wu2019intrinsic}%
  \BibitemOpen
  \bibfield  {author} {\bibinfo {author} {\bibfnamefont {Z.}~\bibnamefont {Wu}}, \bibinfo {author} {\bibfnamefont {B.~T.}\ \bibnamefont {Zhou}}, \bibinfo {author} {\bibfnamefont {X.}~\bibnamefont {Cai}}, \bibinfo {author} {\bibfnamefont {P.}~\bibnamefont {Cheung}}, \bibinfo {author} {\bibfnamefont {G.-B.}\ \bibnamefont {Liu}}, \bibinfo {author} {\bibfnamefont {M.}~\bibnamefont {Huang}}, \bibinfo {author} {\bibfnamefont {J.}~\bibnamefont {Lin}}, \bibinfo {author} {\bibfnamefont {T.}~\bibnamefont {Han}}, \bibinfo {author} {\bibfnamefont {L.}~\bibnamefont {An}}, \bibinfo {author} {\bibfnamefont {Y.}~\bibnamefont {Wang}}, \emph {et~al.},\ }\bibfield  {title} {\bibinfo {title} {{Intrinsic valley Hall transport in atomically thin ${\mathrm{MoS}}_{2}$}},\ }\href {https://doi.org/https://doi.org/10.1038/s41467-019-08629-9} {\bibfield  {journal} {\bibinfo  {journal} {Nat Commun}\ }\textbf {\bibinfo {volume} {10}},\ \bibinfo {pages} {611} (\bibinfo {year} {2019})}\BibitemShut {NoStop}%
\bibitem [{\citenamefont {Li}\ \emph {et~al.}(2020)\citenamefont {Li}, \citenamefont {Shao}, \citenamefont {Liu}, \citenamefont {Gao}, \citenamefont {Wang}, \citenamefont {Zheng}, \citenamefont {Hou}, \citenamefont {Shehzad}, \citenamefont {Yu}, \citenamefont {Miao} \emph {et~al.}}]{li2020room}%
  \BibitemOpen
  \bibfield  {author} {\bibinfo {author} {\bibfnamefont {L.}~\bibnamefont {Li}}, \bibinfo {author} {\bibfnamefont {L.}~\bibnamefont {Shao}}, \bibinfo {author} {\bibfnamefont {X.}~\bibnamefont {Liu}}, \bibinfo {author} {\bibfnamefont {A.}~\bibnamefont {Gao}}, \bibinfo {author} {\bibfnamefont {H.}~\bibnamefont {Wang}}, \bibinfo {author} {\bibfnamefont {B.}~\bibnamefont {Zheng}}, \bibinfo {author} {\bibfnamefont {G.}~\bibnamefont {Hou}}, \bibinfo {author} {\bibfnamefont {K.}~\bibnamefont {Shehzad}}, \bibinfo {author} {\bibfnamefont {L.}~\bibnamefont {Yu}}, \bibinfo {author} {\bibfnamefont {F.}~\bibnamefont {Miao}}, \emph {et~al.},\ }\bibfield  {title} {\bibinfo {title} {{Room-temperature valleytronic transistor}},\ }\href {https://doi.org/https://doi.org/10.1038/s41565-020-0727-0} {\bibfield  {journal} {\bibinfo  {journal} {Nat. Nanotechnol.}\ }\textbf {\bibinfo {volume} {15}},\ \bibinfo {pages} {743} (\bibinfo {year} {2020})}\BibitemShut {NoStop}%
\bibitem [{\citenamefont {Xiao}\ \emph {et~al.}(2012)\citenamefont {Xiao}, \citenamefont {Liu}, \citenamefont {Feng}, \citenamefont {Xu},\ and\ \citenamefont {Yao}}]{PhysRevLett.108.196802}%
  \BibitemOpen
  \bibfield  {author} {\bibinfo {author} {\bibfnamefont {D.}~\bibnamefont {Xiao}}, \bibinfo {author} {\bibfnamefont {G.-B.}\ \bibnamefont {Liu}}, \bibinfo {author} {\bibfnamefont {W.}~\bibnamefont {Feng}}, \bibinfo {author} {\bibfnamefont {X.}~\bibnamefont {Xu}},\ and\ \bibinfo {author} {\bibfnamefont {W.}~\bibnamefont {Yao}},\ }\bibfield  {title} {\bibinfo {title} {{Coupled Spin and Valley Physics in Monolayers of ${\mathrm{MoS}}_{2}$ and Other Group-VI Dichalcogenides}},\ }\href {https://doi.org/10.1103/PhysRevLett.108.196802} {\bibfield  {journal} {\bibinfo  {journal} {Phys. Rev. Lett.}\ }\textbf {\bibinfo {volume} {108}},\ \bibinfo {pages} {196802} (\bibinfo {year} {2012})}\BibitemShut {NoStop}%
\bibitem [{\citenamefont {Xu}\ \emph {et~al.}(2024)\citenamefont {Xu}, \citenamefont {Zhang}, \citenamefont {Liang},\ and\ \citenamefont {Zhu}}]{https://doi.org/10.1002/smll.202402139}%
  \BibitemOpen
  \bibfield  {author} {\bibinfo {author} {\bibfnamefont {R.}~\bibnamefont {Xu}}, \bibinfo {author} {\bibfnamefont {Z.}~\bibnamefont {Zhang}}, \bibinfo {author} {\bibfnamefont {J.}~\bibnamefont {Liang}},\ and\ \bibinfo {author} {\bibfnamefont {H.}~\bibnamefont {Zhu}},\ }\bibfield  {title} {\bibinfo {title} {{Valleytronics: Fundamental Challenges and Materials Beyond Transition Metal Chalcogenides}},\ }\href {https://doi.org/https://doi.org/10.1002/smll.202402139} {\bibfield  {journal} {\bibinfo  {journal} {Small}\ ,\ \bibinfo {pages} {2402139}} (\bibinfo {year} {2024})}\BibitemShut {NoStop}%
\bibitem [{\citenamefont {Gong}\ \emph {et~al.}(2013)\citenamefont {Gong}, \citenamefont {Liu}, \citenamefont {Yu}, \citenamefont {Xiao}, \citenamefont {Cui}, \citenamefont {Xu},\ and\ \citenamefont {Yao}}]{gong2013magnetoelectric}%
  \BibitemOpen
  \bibfield  {author} {\bibinfo {author} {\bibfnamefont {Z.}~\bibnamefont {Gong}}, \bibinfo {author} {\bibfnamefont {G.-B.}\ \bibnamefont {Liu}}, \bibinfo {author} {\bibfnamefont {H.}~\bibnamefont {Yu}}, \bibinfo {author} {\bibfnamefont {D.}~\bibnamefont {Xiao}}, \bibinfo {author} {\bibfnamefont {X.}~\bibnamefont {Cui}}, \bibinfo {author} {\bibfnamefont {X.}~\bibnamefont {Xu}},\ and\ \bibinfo {author} {\bibfnamefont {W.}~\bibnamefont {Yao}},\ }\bibfield  {title} {\bibinfo {title} {{Magnetoelectric effects and valley-controlled spin quantum gates in transition metal dichalcogenide bilayers}},\ }\href {https://doi.org/https://doi.org/10.1038/ncomms3053} {\bibfield  {journal} {\bibinfo  {journal} {Nat Commun}\ }\textbf {\bibinfo {volume} {4}},\ \bibinfo {pages} {2053} (\bibinfo {year} {2013})}\BibitemShut {NoStop}%
\bibitem [{\citenamefont {Wu}\ \emph {et~al.}(2013)\citenamefont {Wu}, \citenamefont {Ross}, \citenamefont {Liu}, \citenamefont {Aivazian}, \citenamefont {Jones}, \citenamefont {Fei}, \citenamefont {Zhu}, \citenamefont {Xiao}, \citenamefont {Yao}, \citenamefont {Cobden} \emph {et~al.}}]{wu2013electrical}%
  \BibitemOpen
  \bibfield  {author} {\bibinfo {author} {\bibfnamefont {S.}~\bibnamefont {Wu}}, \bibinfo {author} {\bibfnamefont {J.~S.}\ \bibnamefont {Ross}}, \bibinfo {author} {\bibfnamefont {G.-B.}\ \bibnamefont {Liu}}, \bibinfo {author} {\bibfnamefont {G.}~\bibnamefont {Aivazian}}, \bibinfo {author} {\bibfnamefont {A.}~\bibnamefont {Jones}}, \bibinfo {author} {\bibfnamefont {Z.}~\bibnamefont {Fei}}, \bibinfo {author} {\bibfnamefont {W.}~\bibnamefont {Zhu}}, \bibinfo {author} {\bibfnamefont {D.}~\bibnamefont {Xiao}}, \bibinfo {author} {\bibfnamefont {W.}~\bibnamefont {Yao}}, \bibinfo {author} {\bibfnamefont {D.}~\bibnamefont {Cobden}}, \emph {et~al.},\ }\bibfield  {title} {\bibinfo {title} {{Electrical tuning of valley magnetic moment through symmetry control in bilayer ${\mathrm{MoS}}_{2}$}},\ }\href {https://doi.org/https://doi.org/10.1038/nphys2524} {\bibfield  {journal} {\bibinfo  {journal} {Nature Phys}\ }\textbf {\bibinfo {volume} {9}},\ \bibinfo {pages} {149} (\bibinfo {year} {2013})}\BibitemShut {NoStop}%
\bibitem [{\citenamefont {Jones}\ \emph {et~al.}(2014)\citenamefont {Jones}, \citenamefont {Yu}, \citenamefont {Ross}, \citenamefont {Klement}, \citenamefont {Ghimire}, \citenamefont {Yan}, \citenamefont {Mandrus}, \citenamefont {Yao},\ and\ \citenamefont {Xu}}]{jones2014spin}%
  \BibitemOpen
  \bibfield  {author} {\bibinfo {author} {\bibfnamefont {A.~M.}\ \bibnamefont {Jones}}, \bibinfo {author} {\bibfnamefont {H.}~\bibnamefont {Yu}}, \bibinfo {author} {\bibfnamefont {J.~S.}\ \bibnamefont {Ross}}, \bibinfo {author} {\bibfnamefont {P.}~\bibnamefont {Klement}}, \bibinfo {author} {\bibfnamefont {N.~J.}\ \bibnamefont {Ghimire}}, \bibinfo {author} {\bibfnamefont {J.}~\bibnamefont {Yan}}, \bibinfo {author} {\bibfnamefont {D.~G.}\ \bibnamefont {Mandrus}}, \bibinfo {author} {\bibfnamefont {W.}~\bibnamefont {Yao}},\ and\ \bibinfo {author} {\bibfnamefont {X.}~\bibnamefont {Xu}},\ }\bibfield  {title} {\bibinfo {title} {{Spin--layer locking effects in optical orientation of exciton spin in bilayer ${\mathrm{WSe}}_{2}$}},\ }\href {https://doi.org/https://doi.org/10.1038/nphys2848} {\bibfield  {journal} {\bibinfo  {journal} {Nature Phys}\ }\textbf {\bibinfo {volume} {10}},\ \bibinfo {pages} {130} (\bibinfo {year} {2014})}\BibitemShut {NoStop}%
\bibitem [{\citenamefont {Sui}\ \emph {et~al.}(2015)\citenamefont {Sui}, \citenamefont {Chen}, \citenamefont {Ma}, \citenamefont {Shan}, \citenamefont {Tian}, \citenamefont {Watanabe}, \citenamefont {Taniguchi}, \citenamefont {Jin}, \citenamefont {Yao}, \citenamefont {Xiao} \emph {et~al.}}]{sui2015gate}%
  \BibitemOpen
  \bibfield  {author} {\bibinfo {author} {\bibfnamefont {M.}~\bibnamefont {Sui}}, \bibinfo {author} {\bibfnamefont {G.}~\bibnamefont {Chen}}, \bibinfo {author} {\bibfnamefont {L.}~\bibnamefont {Ma}}, \bibinfo {author} {\bibfnamefont {W.-Y.}\ \bibnamefont {Shan}}, \bibinfo {author} {\bibfnamefont {D.}~\bibnamefont {Tian}}, \bibinfo {author} {\bibfnamefont {K.}~\bibnamefont {Watanabe}}, \bibinfo {author} {\bibfnamefont {T.}~\bibnamefont {Taniguchi}}, \bibinfo {author} {\bibfnamefont {X.}~\bibnamefont {Jin}}, \bibinfo {author} {\bibfnamefont {W.}~\bibnamefont {Yao}}, \bibinfo {author} {\bibfnamefont {D.}~\bibnamefont {Xiao}}, \emph {et~al.},\ }\bibfield  {title} {\bibinfo {title} {{Gate-tunable topological valley transport in bilayer graphene}},\ }\href {https://doi.org/https://doi.org/10.1038/nphys3485} {\bibfield  {journal} {\bibinfo  {journal} {Nature Phys}\ }\textbf {\bibinfo {volume} {11}},\ \bibinfo {pages} {1027} (\bibinfo {year} {2015})}\BibitemShut {NoStop}%
\bibitem [{\citenamefont {Shimazaki}\ \emph {et~al.}(2015)\citenamefont {Shimazaki}, \citenamefont {Yamamoto}, \citenamefont {Borzenets}, \citenamefont {Watanabe}, \citenamefont {Taniguchi},\ and\ \citenamefont {Tarucha}}]{shimazaki2015generation}%
  \BibitemOpen
  \bibfield  {author} {\bibinfo {author} {\bibfnamefont {Y.}~\bibnamefont {Shimazaki}}, \bibinfo {author} {\bibfnamefont {M.}~\bibnamefont {Yamamoto}}, \bibinfo {author} {\bibfnamefont {I.~V.}\ \bibnamefont {Borzenets}}, \bibinfo {author} {\bibfnamefont {K.}~\bibnamefont {Watanabe}}, \bibinfo {author} {\bibfnamefont {T.}~\bibnamefont {Taniguchi}},\ and\ \bibinfo {author} {\bibfnamefont {S.}~\bibnamefont {Tarucha}},\ }\bibfield  {title} {\bibinfo {title} {{Generation and detection of pure valley current by electrically induced Berry curvature in bilayer graphene}},\ }\href {https://doi.org/https://doi.org/10.1038/nphys3551} {\bibfield  {journal} {\bibinfo  {journal} {Nature Phys}\ }\textbf {\bibinfo {volume} {11}},\ \bibinfo {pages} {1032} (\bibinfo {year} {2015})}\BibitemShut {NoStop}%
\bibitem [{\citenamefont {Lee}\ \emph {et~al.}(2016)\citenamefont {Lee}, \citenamefont {Mak},\ and\ \citenamefont {Shan}}]{lee2016electrical}%
  \BibitemOpen
  \bibfield  {author} {\bibinfo {author} {\bibfnamefont {J.}~\bibnamefont {Lee}}, \bibinfo {author} {\bibfnamefont {K.~F.}\ \bibnamefont {Mak}},\ and\ \bibinfo {author} {\bibfnamefont {J.}~\bibnamefont {Shan}},\ }\bibfield  {title} {\bibinfo {title} {{Electrical control of the valley Hall effect in bilayer ${\mathrm{MoS}}_{2}$ transistors}},\ }\href {https://doi.org/https://doi.org/10.1038/nnano.2015.337} {\bibfield  {journal} {\bibinfo  {journal} {Nature Nanotech}\ }\textbf {\bibinfo {volume} {11}},\ \bibinfo {pages} {421} (\bibinfo {year} {2016})}\BibitemShut {NoStop}%
\bibitem [{\citenamefont {Scuri}\ \emph {et~al.}(2020)\citenamefont {Scuri}, \citenamefont {Andersen}, \citenamefont {Zhou}, \citenamefont {Wild}, \citenamefont {Sung}, \citenamefont {Gelly}, \citenamefont {B\'erub\'e}, \citenamefont {Heo}, \citenamefont {Shao}, \citenamefont {Joe}, \citenamefont {Mier~Valdivia}, \citenamefont {Taniguchi}, \citenamefont {Watanabe}, \citenamefont {Lon\ifmmode~\check{c}\else \v{c}\fi{}ar}, \citenamefont {Kim}, \citenamefont {Lukin},\ and\ \citenamefont {Park}}]{PhysRevLett.124.217403}%
  \BibitemOpen
  \bibfield  {author} {\bibinfo {author} {\bibfnamefont {G.}~\bibnamefont {Scuri}}, \bibinfo {author} {\bibfnamefont {T.~I.}\ \bibnamefont {Andersen}}, \bibinfo {author} {\bibfnamefont {Y.}~\bibnamefont {Zhou}}, \bibinfo {author} {\bibfnamefont {D.~S.}\ \bibnamefont {Wild}}, \bibinfo {author} {\bibfnamefont {J.}~\bibnamefont {Sung}}, \bibinfo {author} {\bibfnamefont {R.~J.}\ \bibnamefont {Gelly}}, \bibinfo {author} {\bibfnamefont {D.}~\bibnamefont {B\'erub\'e}}, \bibinfo {author} {\bibfnamefont {H.}~\bibnamefont {Heo}}, \bibinfo {author} {\bibfnamefont {L.}~\bibnamefont {Shao}}, \bibinfo {author} {\bibfnamefont {A.~Y.}\ \bibnamefont {Joe}}, \bibinfo {author} {\bibfnamefont {A.~M.}\ \bibnamefont {Mier~Valdivia}}, \bibinfo {author} {\bibfnamefont {T.}~\bibnamefont {Taniguchi}}, \bibinfo {author} {\bibfnamefont {K.}~\bibnamefont {Watanabe}}, \bibinfo {author} {\bibfnamefont {M.}~\bibnamefont {Lon\ifmmode~\check{c}\else \v{c}\fi{}ar}}, \bibinfo {author} {\bibfnamefont {P.}~\bibnamefont {Kim}}, \bibinfo {author}
  {\bibfnamefont {M.~D.}\ \bibnamefont {Lukin}},\ and\ \bibinfo {author} {\bibfnamefont {H.}~\bibnamefont {Park}},\ }\bibfield  {title} {\bibinfo {title} {{Electrically Tunable Valley Dynamics in Twisted ${\mathrm{WSe}}_{2}/{\mathrm{WSe}}_{2}$ Bilayers}},\ }\href {https://doi.org/10.1103/PhysRevLett.124.217403} {\bibfield  {journal} {\bibinfo  {journal} {Phys. Rev. Lett.}\ }\textbf {\bibinfo {volume} {124}},\ \bibinfo {pages} {217403} (\bibinfo {year} {2020})}\BibitemShut {NoStop}%
\bibitem [{\citenamefont {Yu}\ \emph {et~al.}(2020)\citenamefont {Yu}, \citenamefont {Guan}, \citenamefont {Sheng}, \citenamefont {Gao},\ and\ \citenamefont {Yang}}]{PhysRevLett.124.037701}%
  \BibitemOpen
  \bibfield  {author} {\bibinfo {author} {\bibfnamefont {Z.-M.}\ \bibnamefont {Yu}}, \bibinfo {author} {\bibfnamefont {S.}~\bibnamefont {Guan}}, \bibinfo {author} {\bibfnamefont {X.-L.}\ \bibnamefont {Sheng}}, \bibinfo {author} {\bibfnamefont {W.}~\bibnamefont {Gao}},\ and\ \bibinfo {author} {\bibfnamefont {S.~A.}\ \bibnamefont {Yang}},\ }\bibfield  {title} {\bibinfo {title} {{Valley-Layer Coupling: A New Design Principle for Valleytronics}},\ }\href {https://doi.org/10.1103/PhysRevLett.124.037701} {\bibfield  {journal} {\bibinfo  {journal} {Phys. Rev. Lett.}\ }\textbf {\bibinfo {volume} {124}},\ \bibinfo {pages} {037701} (\bibinfo {year} {2020})}\BibitemShut {NoStop}%
\bibitem [{\citenamefont {Zhang}\ \emph {et~al.}(2023{\natexlab{a}})\citenamefont {Zhang}, \citenamefont {Xu}, \citenamefont {Huang}, \citenamefont {Dai}, \citenamefont {Kou},\ and\ \citenamefont {Ma}}]{zhang2023layer}%
  \BibitemOpen
  \bibfield  {author} {\bibinfo {author} {\bibfnamefont {T.}~\bibnamefont {Zhang}}, \bibinfo {author} {\bibfnamefont {X.}~\bibnamefont {Xu}}, \bibinfo {author} {\bibfnamefont {B.}~\bibnamefont {Huang}}, \bibinfo {author} {\bibfnamefont {Y.}~\bibnamefont {Dai}}, \bibinfo {author} {\bibfnamefont {L.}~\bibnamefont {Kou}},\ and\ \bibinfo {author} {\bibfnamefont {Y.}~\bibnamefont {Ma}},\ }\bibfield  {title} {\bibinfo {title} {{Layer-polarized anomalous Hall effects in valleytronic van der Waals bilayers}},\ }\href {https://doi.org/https://doi.org/10.1039/D2MH00906D} {\bibfield  {journal} {\bibinfo  {journal} {Mater. Horiz.}\ }\textbf {\bibinfo {volume} {10}},\ \bibinfo {pages} {483} (\bibinfo {year} {2023}{\natexlab{a}})}\BibitemShut {NoStop}%
\bibitem [{\citenamefont {McCann}\ \emph {et~al.}(2006)\citenamefont {McCann}, \citenamefont {Kechedzhi}, \citenamefont {Fal'ko}, \citenamefont {Suzuura}, \citenamefont {Ando},\ and\ \citenamefont {Altshuler}}]{PhysRevLett.97.146805}%
  \BibitemOpen
  \bibfield  {author} {\bibinfo {author} {\bibfnamefont {E.}~\bibnamefont {McCann}}, \bibinfo {author} {\bibfnamefont {K.}~\bibnamefont {Kechedzhi}}, \bibinfo {author} {\bibfnamefont {V.~I.}\ \bibnamefont {Fal'ko}}, \bibinfo {author} {\bibfnamefont {H.}~\bibnamefont {Suzuura}}, \bibinfo {author} {\bibfnamefont {T.}~\bibnamefont {Ando}},\ and\ \bibinfo {author} {\bibfnamefont {B.~L.}\ \bibnamefont {Altshuler}},\ }\bibfield  {title} {\bibinfo {title} {{Weak-Localization Magnetoresistance and Valley Symmetry in Graphene}},\ }\href {https://doi.org/10.1103/PhysRevLett.97.146805} {\bibfield  {journal} {\bibinfo  {journal} {Phys. Rev. Lett.}\ }\textbf {\bibinfo {volume} {97}},\ \bibinfo {pages} {146805} (\bibinfo {year} {2006})}\BibitemShut {NoStop}%
\bibitem [{\citenamefont {Morozov}\ \emph {et~al.}(2006)\citenamefont {Morozov}, \citenamefont {Novoselov}, \citenamefont {Katsnelson}, \citenamefont {Schedin}, \citenamefont {Ponomarenko}, \citenamefont {Jiang},\ and\ \citenamefont {Geim}}]{PhysRevLett.97.016801}%
  \BibitemOpen
  \bibfield  {author} {\bibinfo {author} {\bibfnamefont {S.~V.}\ \bibnamefont {Morozov}}, \bibinfo {author} {\bibfnamefont {K.~S.}\ \bibnamefont {Novoselov}}, \bibinfo {author} {\bibfnamefont {M.~I.}\ \bibnamefont {Katsnelson}}, \bibinfo {author} {\bibfnamefont {F.}~\bibnamefont {Schedin}}, \bibinfo {author} {\bibfnamefont {L.~A.}\ \bibnamefont {Ponomarenko}}, \bibinfo {author} {\bibfnamefont {D.}~\bibnamefont {Jiang}},\ and\ \bibinfo {author} {\bibfnamefont {A.~K.}\ \bibnamefont {Geim}},\ }\bibfield  {title} {\bibinfo {title} {{Strong Suppression of Weak Localization in Graphene}},\ }\href {https://doi.org/10.1103/PhysRevLett.97.016801} {\bibfield  {journal} {\bibinfo  {journal} {Phys. Rev. Lett.}\ }\textbf {\bibinfo {volume} {97}},\ \bibinfo {pages} {016801} (\bibinfo {year} {2006})}\BibitemShut {NoStop}%
\bibitem [{\citenamefont {Morpurgo}\ and\ \citenamefont {Guinea}(2006)}]{PhysRevLett.97.196804}%
  \BibitemOpen
  \bibfield  {author} {\bibinfo {author} {\bibfnamefont {A.~F.}\ \bibnamefont {Morpurgo}}\ and\ \bibinfo {author} {\bibfnamefont {F.}~\bibnamefont {Guinea}},\ }\bibfield  {title} {\bibinfo {title} {{Intervalley Scattering, Long-Range Disorder, and Effective Time-Reversal Symmetry Breaking in Graphene}},\ }\href {https://doi.org/10.1103/PhysRevLett.97.196804} {\bibfield  {journal} {\bibinfo  {journal} {Phys. Rev. Lett.}\ }\textbf {\bibinfo {volume} {97}},\ \bibinfo {pages} {196804} (\bibinfo {year} {2006})}\BibitemShut {NoStop}%
\bibitem [{\citenamefont {Ang}\ \emph {et~al.}(2017)\citenamefont {Ang}, \citenamefont {Yang}, \citenamefont {Zhang}, \citenamefont {Ma},\ and\ \citenamefont {Ang}}]{PhysRevB.96.245410}%
  \BibitemOpen
  \bibfield  {author} {\bibinfo {author} {\bibfnamefont {Y.~S.}\ \bibnamefont {Ang}}, \bibinfo {author} {\bibfnamefont {S.~A.}\ \bibnamefont {Yang}}, \bibinfo {author} {\bibfnamefont {C.}~\bibnamefont {Zhang}}, \bibinfo {author} {\bibfnamefont {Z.}~\bibnamefont {Ma}},\ and\ \bibinfo {author} {\bibfnamefont {L.~K.}\ \bibnamefont {Ang}},\ }\bibfield  {title} {\bibinfo {title} {{Valleytronics in merging Dirac cones: All-electric-controlled valley filter, valve, and universal reversible logic gate}},\ }\href {https://doi.org/10.1103/PhysRevB.96.245410} {\bibfield  {journal} {\bibinfo  {journal} {Phys. Rev. B}\ }\textbf {\bibinfo {volume} {96}},\ \bibinfo {pages} {245410} (\bibinfo {year} {2017})}\BibitemShut {NoStop}%
\bibitem [{\citenamefont {Chen}\ \emph {et~al.}(2022)\citenamefont {Chen}, \citenamefont {Zhou}, \citenamefont {Yan}, \citenamefont {Liu}, \citenamefont {Xu}, \citenamefont {Wang}, \citenamefont {Wan}, \citenamefont {He}, \citenamefont {Zhang},\ and\ \citenamefont {Chai}}]{chen2022room}%
  \BibitemOpen
  \bibfield  {author} {\bibinfo {author} {\bibfnamefont {J.}~\bibnamefont {Chen}}, \bibinfo {author} {\bibfnamefont {Y.}~\bibnamefont {Zhou}}, \bibinfo {author} {\bibfnamefont {J.}~\bibnamefont {Yan}}, \bibinfo {author} {\bibfnamefont {J.}~\bibnamefont {Liu}}, \bibinfo {author} {\bibfnamefont {L.}~\bibnamefont {Xu}}, \bibinfo {author} {\bibfnamefont {J.}~\bibnamefont {Wang}}, \bibinfo {author} {\bibfnamefont {T.}~\bibnamefont {Wan}}, \bibinfo {author} {\bibfnamefont {Y.}~\bibnamefont {He}}, \bibinfo {author} {\bibfnamefont {W.}~\bibnamefont {Zhang}},\ and\ \bibinfo {author} {\bibfnamefont {Y.}~\bibnamefont {Chai}},\ }\bibfield  {title} {\bibinfo {title} {{Room-temperature valley transistors for low-power neuromorphic computing}},\ }\href {https://doi.org/https://doi.org/10.1038/s41467-022-35396-x} {\bibfield  {journal} {\bibinfo  {journal} {Nat Commun}\ }\textbf {\bibinfo {volume} {13}},\ \bibinfo {pages} {7758} (\bibinfo {year} {2022})}\BibitemShut {NoStop}%
\bibitem [{\citenamefont {Xin}\ \emph {et~al.}(2018)\citenamefont {Xin}, \citenamefont {Tang}, \citenamefont {Liu}, \citenamefont {Zhao}, \citenamefont {Pan},\ and\ \citenamefont {Zhu}}]{xin2018valleytronics}%
  \BibitemOpen
  \bibfield  {author} {\bibinfo {author} {\bibfnamefont {J.}~\bibnamefont {Xin}}, \bibinfo {author} {\bibfnamefont {Y.}~\bibnamefont {Tang}}, \bibinfo {author} {\bibfnamefont {Y.}~\bibnamefont {Liu}}, \bibinfo {author} {\bibfnamefont {X.}~\bibnamefont {Zhao}}, \bibinfo {author} {\bibfnamefont {H.}~\bibnamefont {Pan}},\ and\ \bibinfo {author} {\bibfnamefont {T.}~\bibnamefont {Zhu}},\ }\bibfield  {title} {\bibinfo {title} {{Valleytronics in thermoelectric materials}},\ }\href {https://doi.org/https://doi.org/10.1038/s41535-018-0083-6} {\bibfield  {journal} {\bibinfo  {journal} {npj Quant Mater}\ }\textbf {\bibinfo {volume} {3}},\ \bibinfo {pages} {9} (\bibinfo {year} {2018})}\BibitemShut {NoStop}%
\bibitem [{\citenamefont {Xie}\ \emph {et~al.}(2023)\citenamefont {Xie}, \citenamefont {Efetov},\ and\ \citenamefont {Law}}]{PhysRevResearch.5.023029}%
  \BibitemOpen
  \bibfield  {author} {\bibinfo {author} {\bibfnamefont {Y.-M.}\ \bibnamefont {Xie}}, \bibinfo {author} {\bibfnamefont {D.~K.}\ \bibnamefont {Efetov}},\ and\ \bibinfo {author} {\bibfnamefont {K.~T.}\ \bibnamefont {Law}},\ }\bibfield  {title} {\bibinfo {title} {{${\ensuremath{\varphi}}_{0}$-Josephson junction in twisted bilayer graphene induced by a valley-polarized state}},\ }\href {https://doi.org/10.1103/PhysRevResearch.5.023029} {\bibfield  {journal} {\bibinfo  {journal} {Phys. Rev. Res.}\ }\textbf {\bibinfo {volume} {5}},\ \bibinfo {pages} {023029} (\bibinfo {year} {2023})}\BibitemShut {NoStop}%
\bibitem [{\citenamefont {Beenakker}\ \emph {et~al.}(2018)\citenamefont {Beenakker}, \citenamefont {Gnezdilov}, \citenamefont {Dresselhaus}, \citenamefont {Ostroukh}, \citenamefont {Herasymenko}, \citenamefont {Adagideli},\ and\ \citenamefont {Tworzyd\l{}o}}]{PhysRevB.97.241403}%
  \BibitemOpen
  \bibfield  {author} {\bibinfo {author} {\bibfnamefont {C.~W.~J.}\ \bibnamefont {Beenakker}}, \bibinfo {author} {\bibfnamefont {N.~V.}\ \bibnamefont {Gnezdilov}}, \bibinfo {author} {\bibfnamefont {E.}~\bibnamefont {Dresselhaus}}, \bibinfo {author} {\bibfnamefont {V.~P.}\ \bibnamefont {Ostroukh}}, \bibinfo {author} {\bibfnamefont {Y.}~\bibnamefont {Herasymenko}}, \bibinfo {author} {\bibfnamefont {i.~d.~I.}\ \bibnamefont {Adagideli}},\ and\ \bibinfo {author} {\bibfnamefont {J.}~\bibnamefont {Tworzyd\l{}o}},\ }\bibfield  {title} {\bibinfo {title} {{Valley switch in a graphene superlattice due to pseudo-Andreev reflection}},\ }\href {https://doi.org/10.1103/PhysRevB.97.241403} {\bibfield  {journal} {\bibinfo  {journal} {Phys. Rev. B}\ }\textbf {\bibinfo {volume} {97}},\ \bibinfo {pages} {241403} (\bibinfo {year} {2018})}\BibitemShut {NoStop}%
\bibitem [{\citenamefont {Wang}\ \emph {et~al.}(2020)\citenamefont {Wang}, \citenamefont {Liu}, \citenamefont {Wang},\ and\ \citenamefont {Liu}}]{PhysRevB.101.245428}%
  \BibitemOpen
  \bibfield  {author} {\bibinfo {author} {\bibfnamefont {J.~J.}\ \bibnamefont {Wang}}, \bibinfo {author} {\bibfnamefont {S.}~\bibnamefont {Liu}}, \bibinfo {author} {\bibfnamefont {J.}~\bibnamefont {Wang}},\ and\ \bibinfo {author} {\bibfnamefont {J.-F.}\ \bibnamefont {Liu}},\ }\bibfield  {title} {\bibinfo {title} {{Valley supercurrent in the Kekul\'e graphene superlattice heterojunction}},\ }\href {https://doi.org/10.1103/PhysRevB.101.245428} {\bibfield  {journal} {\bibinfo  {journal} {Phys. Rev. B}\ }\textbf {\bibinfo {volume} {101}},\ \bibinfo {pages} {245428} (\bibinfo {year} {2020})}\BibitemShut {NoStop}%
\bibitem [{\citenamefont {Wu}\ \emph {et~al.}(2020)\citenamefont {Wu}, \citenamefont {Meng}, \citenamefont {Kong}, \citenamefont {Zhang}, \citenamefont {Bai},\ and\ \citenamefont {Xu}}]{PhysRevB.101.125406}%
  \BibitemOpen
  \bibfield  {author} {\bibinfo {author} {\bibfnamefont {X.}~\bibnamefont {Wu}}, \bibinfo {author} {\bibfnamefont {H.}~\bibnamefont {Meng}}, \bibinfo {author} {\bibfnamefont {F.}~\bibnamefont {Kong}}, \bibinfo {author} {\bibfnamefont {H.}~\bibnamefont {Zhang}}, \bibinfo {author} {\bibfnamefont {Y.}~\bibnamefont {Bai}},\ and\ \bibinfo {author} {\bibfnamefont {N.}~\bibnamefont {Xu}},\ }\bibfield  {title} {\bibinfo {title} {{Tunable nonlocal valley-entangled Cooper pair splitter realized in bilayer-graphene van der Waals spin valves}},\ }\href {https://doi.org/10.1103/PhysRevB.101.125406} {\bibfield  {journal} {\bibinfo  {journal} {Phys. Rev. B}\ }\textbf {\bibinfo {volume} {101}},\ \bibinfo {pages} {125406} (\bibinfo {year} {2020})}\BibitemShut {NoStop}%
\bibitem [{\citenamefont {Chen}\ \emph {et~al.}(2015)\citenamefont {Chen}, \citenamefont {Zhang},\ and\ \citenamefont {Guo}}]{PhysRevB.92.155427}%
  \BibitemOpen
  \bibfield  {author} {\bibinfo {author} {\bibfnamefont {X.}~\bibnamefont {Chen}}, \bibinfo {author} {\bibfnamefont {L.}~\bibnamefont {Zhang}},\ and\ \bibinfo {author} {\bibfnamefont {H.}~\bibnamefont {Guo}},\ }\bibfield  {title} {\bibinfo {title} {{Valley caloritronics and its realization by graphene nanoribbons}},\ }\href {https://doi.org/10.1103/PhysRevB.92.155427} {\bibfield  {journal} {\bibinfo  {journal} {Phys. Rev. B}\ }\textbf {\bibinfo {volume} {92}},\ \bibinfo {pages} {155427} (\bibinfo {year} {2015})}\BibitemShut {NoStop}%
\bibitem [{\citenamefont {An}\ \emph {et~al.}(2017)\citenamefont {An}, \citenamefont {Xiao}, \citenamefont {Tu}, \citenamefont {Yu}, \citenamefont {Fal'ko},\ and\ \citenamefont {Yao}}]{PhysRevLett.118.096602}%
  \BibitemOpen
  \bibfield  {author} {\bibinfo {author} {\bibfnamefont {X.-T.}\ \bibnamefont {An}}, \bibinfo {author} {\bibfnamefont {J.}~\bibnamefont {Xiao}}, \bibinfo {author} {\bibfnamefont {M.~W.-Y.}\ \bibnamefont {Tu}}, \bibinfo {author} {\bibfnamefont {H.}~\bibnamefont {Yu}}, \bibinfo {author} {\bibfnamefont {V.~I.}\ \bibnamefont {Fal'ko}},\ and\ \bibinfo {author} {\bibfnamefont {W.}~\bibnamefont {Yao}},\ }\bibfield  {title} {\bibinfo {title} {{Realization of Valley and Spin Pumps by Scattering at Nonmagnetic Disorders}},\ }\href {https://doi.org/10.1103/PhysRevLett.118.096602} {\bibfield  {journal} {\bibinfo  {journal} {Phys. Rev. Lett.}\ }\textbf {\bibinfo {volume} {118}},\ \bibinfo {pages} {096602} (\bibinfo {year} {2017})}\BibitemShut {NoStop}%
\bibitem [{\citenamefont {Wu}\ \emph {et~al.}(2011)\citenamefont {Wu}, \citenamefont {Zhai}, \citenamefont {Peeters}, \citenamefont {Xu},\ and\ \citenamefont {Chang}}]{PhysRevLett.106.176802}%
  \BibitemOpen
  \bibfield  {author} {\bibinfo {author} {\bibfnamefont {Z.}~\bibnamefont {Wu}}, \bibinfo {author} {\bibfnamefont {F.}~\bibnamefont {Zhai}}, \bibinfo {author} {\bibfnamefont {F.~M.}\ \bibnamefont {Peeters}}, \bibinfo {author} {\bibfnamefont {H.~Q.}\ \bibnamefont {Xu}},\ and\ \bibinfo {author} {\bibfnamefont {K.}~\bibnamefont {Chang}},\ }\bibfield  {title} {\bibinfo {title} {{Valley-Dependent Brewster Angles and Goos-H\"anchen Effect in Strained Graphene}},\ }\href {https://doi.org/10.1103/PhysRevLett.106.176802} {\bibfield  {journal} {\bibinfo  {journal} {Phys. Rev. Lett.}\ }\textbf {\bibinfo {volume} {106}},\ \bibinfo {pages} {176802} (\bibinfo {year} {2011})}\BibitemShut {NoStop}%
\bibitem [{\citenamefont {Qiu}\ \emph {et~al.}(2019)\citenamefont {Qiu}, \citenamefont {Lv},\ and\ \citenamefont {Cao}}]{Qiu_2019}%
  \BibitemOpen
  \bibfield  {author} {\bibinfo {author} {\bibfnamefont {X.}~\bibnamefont {Qiu}}, \bibinfo {author} {\bibfnamefont {Q.}~\bibnamefont {Lv}},\ and\ \bibinfo {author} {\bibfnamefont {Z.}~\bibnamefont {Cao}},\ }\bibfield  {title} {\bibinfo {title} {{A high-quality spin and valley beam splitter in ${\mathrm{WSe}}_{2}$ tunnelling junction through the Goos–Hänchen shift}},\ }\href {https://doi.org/10.1088/1361-648X/ab0b07} {\bibfield  {journal} {\bibinfo  {journal} {J. Phys.: Condens. Matter}\ }\textbf {\bibinfo {volume} {31}},\ \bibinfo {pages} {225303} (\bibinfo {year} {2019})}\BibitemShut {NoStop}%
\bibitem [{\citenamefont {Liu}\ \emph {et~al.}(2023{\natexlab{a}})\citenamefont {Liu}, \citenamefont {Zhang}, \citenamefont {Zhou}, \citenamefont {Nie}, \citenamefont {Li},\ and\ \citenamefont {Zhang}}]{LIU2023108550}%
  \BibitemOpen
  \bibfield  {author} {\bibinfo {author} {\bibfnamefont {R.}~\bibnamefont {Liu}}, \bibinfo {author} {\bibfnamefont {Y.}~\bibnamefont {Zhang}}, \bibinfo {author} {\bibfnamefont {Y.}~\bibnamefont {Zhou}}, \bibinfo {author} {\bibfnamefont {J.}~\bibnamefont {Nie}}, \bibinfo {author} {\bibfnamefont {L.}~\bibnamefont {Li}},\ and\ \bibinfo {author} {\bibfnamefont {Y.}~\bibnamefont {Zhang}},\ }\bibfield  {title} {\bibinfo {title} {{Polarization-driven high Rabi frequency of piezotronic valley transistors}},\ }\href {https://doi.org/https://doi.org/10.1016/j.nanoen.2023.108550} {\bibfield  {journal} {\bibinfo  {journal} {Nano Energy}\ }\textbf {\bibinfo {volume} {113}},\ \bibinfo {pages} {108550} (\bibinfo {year} {2023}{\natexlab{a}})}\BibitemShut {NoStop}%
\bibitem [{\citenamefont {Lai}\ \emph {et~al.}(2023)\citenamefont {Lai}, \citenamefont {Zhang}, \citenamefont {Wang}, \citenamefont {Rasmita}, \citenamefont {Deng}, \citenamefont {Liu},\ and\ \citenamefont {Gao}}]{doi:10.1021/acs.nanolett.2c03947}%
  \BibitemOpen
  \bibfield  {author} {\bibinfo {author} {\bibfnamefont {S.}~\bibnamefont {Lai}}, \bibinfo {author} {\bibfnamefont {Z.}~\bibnamefont {Zhang}}, \bibinfo {author} {\bibfnamefont {N.}~\bibnamefont {Wang}}, \bibinfo {author} {\bibfnamefont {A.}~\bibnamefont {Rasmita}}, \bibinfo {author} {\bibfnamefont {Y.}~\bibnamefont {Deng}}, \bibinfo {author} {\bibfnamefont {Z.}~\bibnamefont {Liu}},\ and\ \bibinfo {author} {\bibfnamefont {W.-b.}\ \bibnamefont {Gao}},\ }\bibfield  {title} {\bibinfo {title} {{Dual-Gate All-Electrical Valleytronic Transistors}},\ }\href {https://doi.org/10.1021/acs.nanolett.2c03947} {\bibfield  {journal} {\bibinfo  {journal} {Nano Lett.}\ }\textbf {\bibinfo {volume} {23}},\ \bibinfo {pages} {192} (\bibinfo {year} {2023})}\BibitemShut {NoStop}%
\bibitem [{\citenamefont {Shao}\ \emph {et~al.}(2024)\citenamefont {Shao}, \citenamefont {Geng}, \citenamefont {Liu}, \citenamefont {Lado}, \citenamefont {Chen},\ and\ \citenamefont {Xing}}]{PhysRevLett.132.156301}%
  \BibitemOpen
  \bibfield  {author} {\bibinfo {author} {\bibfnamefont {K.}~\bibnamefont {Shao}}, \bibinfo {author} {\bibfnamefont {H.}~\bibnamefont {Geng}}, \bibinfo {author} {\bibfnamefont {E.}~\bibnamefont {Liu}}, \bibinfo {author} {\bibfnamefont {J.~L.}\ \bibnamefont {Lado}}, \bibinfo {author} {\bibfnamefont {W.}~\bibnamefont {Chen}},\ and\ \bibinfo {author} {\bibfnamefont {D.~Y.}\ \bibnamefont {Xing}},\ }\bibfield  {title} {\bibinfo {title} {{Non-Hermitian Moir\'e Valley Filter}},\ }\href {https://doi.org/10.1103/PhysRevLett.132.156301} {\bibfield  {journal} {\bibinfo  {journal} {Phys. Rev. Lett.}\ }\textbf {\bibinfo {volume} {132}},\ \bibinfo {pages} {156301} (\bibinfo {year} {2024})}\BibitemShut {NoStop}%
\bibitem [{\citenamefont {Xiao}\ \emph {et~al.}(2007)\citenamefont {Xiao}, \citenamefont {Yao},\ and\ \citenamefont {Niu}}]{PhysRevLett.99.236809}%
  \BibitemOpen
  \bibfield  {author} {\bibinfo {author} {\bibfnamefont {D.}~\bibnamefont {Xiao}}, \bibinfo {author} {\bibfnamefont {W.}~\bibnamefont {Yao}},\ and\ \bibinfo {author} {\bibfnamefont {Q.}~\bibnamefont {Niu}},\ }\bibfield  {title} {\bibinfo {title} {{Valley-Contrasting Physics in Graphene: Magnetic Moment and Topological Transport}},\ }\href {https://doi.org/10.1103/PhysRevLett.99.236809} {\bibfield  {journal} {\bibinfo  {journal} {Phys. Rev. Lett.}\ }\textbf {\bibinfo {volume} {99}},\ \bibinfo {pages} {236809} (\bibinfo {year} {2007})}\BibitemShut {NoStop}%
\bibitem [{\citenamefont {Das}\ \emph {et~al.}(2024)\citenamefont {Das}, \citenamefont {Ghorai}, \citenamefont {Culcer},\ and\ \citenamefont {Agarwal}}]{PhysRevLett.132.096302}%
  \BibitemOpen
  \bibfield  {author} {\bibinfo {author} {\bibfnamefont {K.}~\bibnamefont {Das}}, \bibinfo {author} {\bibfnamefont {K.}~\bibnamefont {Ghorai}}, \bibinfo {author} {\bibfnamefont {D.}~\bibnamefont {Culcer}},\ and\ \bibinfo {author} {\bibfnamefont {A.}~\bibnamefont {Agarwal}},\ }\bibfield  {title} {\bibinfo {title} {{Nonlinear Valley Hall Effect}},\ }\href {https://doi.org/10.1103/PhysRevLett.132.096302} {\bibfield  {journal} {\bibinfo  {journal} {Phys. Rev. Lett.}\ }\textbf {\bibinfo {volume} {132}},\ \bibinfo {pages} {096302} (\bibinfo {year} {2024})}\BibitemShut {NoStop}%
\bibitem [{\citenamefont {Ghadimi}\ \emph {et~al.}(2024)\citenamefont {Ghadimi}, \citenamefont {Mondal}, \citenamefont {Kim},\ and\ \citenamefont {Yang}}]{PhysRevLett.133.196603}%
  \BibitemOpen
  \bibfield  {author} {\bibinfo {author} {\bibfnamefont {R.}~\bibnamefont {Ghadimi}}, \bibinfo {author} {\bibfnamefont {C.}~\bibnamefont {Mondal}}, \bibinfo {author} {\bibfnamefont {S.}~\bibnamefont {Kim}},\ and\ \bibinfo {author} {\bibfnamefont {B.-J.}\ \bibnamefont {Yang}},\ }\bibfield  {title} {\bibinfo {title} {{Quantum Valley Hall Effect without Berry Curvature}},\ }\href {https://doi.org/10.1103/PhysRevLett.133.196603} {\bibfield  {journal} {\bibinfo  {journal} {Phys. Rev. Lett.}\ }\textbf {\bibinfo {volume} {133}},\ \bibinfo {pages} {196603} (\bibinfo {year} {2024})}\BibitemShut {NoStop}%
\bibitem [{\citenamefont {Zhang}\ \emph {et~al.}(2023{\natexlab{b}})\citenamefont {Zhang}, \citenamefont {Shao}, \citenamefont {Wang}, \citenamefont {Yang}, \citenamefont {Yang},\ and\ \citenamefont {Tsymbal}}]{PhysRevLett.131.246301}%
  \BibitemOpen
  \bibfield  {author} {\bibinfo {author} {\bibfnamefont {S.-H.}\ \bibnamefont {Zhang}}, \bibinfo {author} {\bibfnamefont {D.-F.}\ \bibnamefont {Shao}}, \bibinfo {author} {\bibfnamefont {Z.-A.}\ \bibnamefont {Wang}}, \bibinfo {author} {\bibfnamefont {J.}~\bibnamefont {Yang}}, \bibinfo {author} {\bibfnamefont {W.}~\bibnamefont {Yang}},\ and\ \bibinfo {author} {\bibfnamefont {E.~Y.}\ \bibnamefont {Tsymbal}},\ }\bibfield  {title} {\bibinfo {title} {{Tunneling Valley Hall Effect Driven by Tilted Dirac Fermions}},\ }\href {https://doi.org/10.1103/PhysRevLett.131.246301} {\bibfield  {journal} {\bibinfo  {journal} {Phys. Rev. Lett.}\ }\textbf {\bibinfo {volume} {131}},\ \bibinfo {pages} {246301} (\bibinfo {year} {2023}{\natexlab{b}})}\BibitemShut {NoStop}%
\bibitem [{\citenamefont {Zhou}\ \emph {et~al.}(2017)\citenamefont {Zhou}, \citenamefont {Sun},\ and\ \citenamefont {Jena}}]{PhysRevLett.119.046403}%
  \BibitemOpen
  \bibfield  {author} {\bibinfo {author} {\bibfnamefont {J.}~\bibnamefont {Zhou}}, \bibinfo {author} {\bibfnamefont {Q.}~\bibnamefont {Sun}},\ and\ \bibinfo {author} {\bibfnamefont {P.}~\bibnamefont {Jena}},\ }\bibfield  {title} {\bibinfo {title} {{Valley-Polarized Quantum Anomalous Hall Effect in Ferrimagnetic Honeycomb Lattices}},\ }\href {https://doi.org/10.1103/PhysRevLett.119.046403} {\bibfield  {journal} {\bibinfo  {journal} {Phys. Rev. Lett.}\ }\textbf {\bibinfo {volume} {119}},\ \bibinfo {pages} {046403} (\bibinfo {year} {2017})}\BibitemShut {NoStop}%
\bibitem [{\citenamefont {Pan}\ \emph {et~al.}(2014)\citenamefont {Pan}, \citenamefont {Li}, \citenamefont {Liu}, \citenamefont {Zhu}, \citenamefont {Qiao},\ and\ \citenamefont {Yao}}]{PhysRevLett.112.106802}%
  \BibitemOpen
  \bibfield  {author} {\bibinfo {author} {\bibfnamefont {H.}~\bibnamefont {Pan}}, \bibinfo {author} {\bibfnamefont {Z.}~\bibnamefont {Li}}, \bibinfo {author} {\bibfnamefont {C.-C.}\ \bibnamefont {Liu}}, \bibinfo {author} {\bibfnamefont {G.}~\bibnamefont {Zhu}}, \bibinfo {author} {\bibfnamefont {Z.}~\bibnamefont {Qiao}},\ and\ \bibinfo {author} {\bibfnamefont {Y.}~\bibnamefont {Yao}},\ }\bibfield  {title} {\bibinfo {title} {{Valley-Polarized Quantum Anomalous Hall Effect in Silicene}},\ }\href {https://doi.org/10.1103/PhysRevLett.112.106802} {\bibfield  {journal} {\bibinfo  {journal} {Phys. Rev. Lett.}\ }\textbf {\bibinfo {volume} {112}},\ \bibinfo {pages} {106802} (\bibinfo {year} {2014})}\BibitemShut {NoStop}%
\bibitem [{\citenamefont {Pan}\ \emph {et~al.}(2015{\natexlab{a}})\citenamefont {Pan}, \citenamefont {Li}, \citenamefont {Zhang},\ and\ \citenamefont {Yang}}]{PhysRevB.92.041404}%
  \BibitemOpen
  \bibfield  {author} {\bibinfo {author} {\bibfnamefont {H.}~\bibnamefont {Pan}}, \bibinfo {author} {\bibfnamefont {X.}~\bibnamefont {Li}}, \bibinfo {author} {\bibfnamefont {F.}~\bibnamefont {Zhang}},\ and\ \bibinfo {author} {\bibfnamefont {S.~A.}\ \bibnamefont {Yang}},\ }\bibfield  {title} {\bibinfo {title} {{Perfect valley filter in a topological domain wall}},\ }\href {https://doi.org/10.1103/PhysRevB.92.041404} {\bibfield  {journal} {\bibinfo  {journal} {Phys. Rev. B}\ }\textbf {\bibinfo {volume} {92}},\ \bibinfo {pages} {041404} (\bibinfo {year} {2015}{\natexlab{a}})}\BibitemShut {NoStop}%
\bibitem [{\citenamefont {Pan}\ \emph {et~al.}(2015{\natexlab{b}})\citenamefont {Pan}, \citenamefont {Li}, \citenamefont {Jiang}, \citenamefont {Yao},\ and\ \citenamefont {Yang}}]{PhysRevB.91.045404}%
  \BibitemOpen
  \bibfield  {author} {\bibinfo {author} {\bibfnamefont {H.}~\bibnamefont {Pan}}, \bibinfo {author} {\bibfnamefont {X.}~\bibnamefont {Li}}, \bibinfo {author} {\bibfnamefont {H.}~\bibnamefont {Jiang}}, \bibinfo {author} {\bibfnamefont {Y.}~\bibnamefont {Yao}},\ and\ \bibinfo {author} {\bibfnamefont {S.~A.}\ \bibnamefont {Yang}},\ }\bibfield  {title} {\bibinfo {title} {{Valley-polarized quantum anomalous Hall phase and disorder-induced valley-filtered chiral edge channels}},\ }\href {https://doi.org/10.1103/PhysRevB.91.045404} {\bibfield  {journal} {\bibinfo  {journal} {Phys. Rev. B}\ }\textbf {\bibinfo {volume} {91}},\ \bibinfo {pages} {045404} (\bibinfo {year} {2015}{\natexlab{b}})}\BibitemShut {NoStop}%
\bibitem [{\citenamefont {Rycerz}\ \emph {et~al.}(2007)\citenamefont {Rycerz}, \citenamefont {Tworzyd{\l}o},\ and\ \citenamefont {Beenakker}}]{rycerz2007valley}%
  \BibitemOpen
  \bibfield  {author} {\bibinfo {author} {\bibfnamefont {A.}~\bibnamefont {Rycerz}}, \bibinfo {author} {\bibfnamefont {J.}~\bibnamefont {Tworzyd{\l}o}},\ and\ \bibinfo {author} {\bibfnamefont {C.}~\bibnamefont {Beenakker}},\ }\bibfield  {title} {\bibinfo {title} {{Valley filter and valley valve in graphene}},\ }\href {https://doi.org/https://doi.org/10.1038/nphys547} {\bibfield  {journal} {\bibinfo  {journal} {Nature Phys}\ }\textbf {\bibinfo {volume} {3}},\ \bibinfo {pages} {172} (\bibinfo {year} {2007})}\BibitemShut {NoStop}%
\bibitem [{\citenamefont {Wang}(2008)}]{PhysRevLett.100.156404}%
  \BibitemOpen
  \bibfield  {author} {\bibinfo {author} {\bibfnamefont {X.~L.}\ \bibnamefont {Wang}},\ }\bibfield  {title} {\bibinfo {title} {{Proposal for a New Class of Materials: Spin Gapless Semiconductors}},\ }\href {https://doi.org/10.1103/PhysRevLett.100.156404} {\bibfield  {journal} {\bibinfo  {journal} {Phys. Rev. Lett.}\ }\textbf {\bibinfo {volume} {100}},\ \bibinfo {pages} {156404} (\bibinfo {year} {2008})}\BibitemShut {NoStop}%
\bibitem [{\citenamefont {Guo}\ \emph {et~al.}(2024{\natexlab{a}})\citenamefont {Guo}, \citenamefont {Tao}, \citenamefont {Wang}, \citenamefont {Chen}, \citenamefont {Huang},\ and\ \citenamefont {Ang}}]{guo2024proposal}%
  \BibitemOpen
  \bibfield  {author} {\bibinfo {author} {\bibfnamefont {S.-D.}\ \bibnamefont {Guo}}, \bibinfo {author} {\bibfnamefont {Y.-L.}\ \bibnamefont {Tao}}, \bibinfo {author} {\bibfnamefont {G.}~\bibnamefont {Wang}}, \bibinfo {author} {\bibfnamefont {S.}~\bibnamefont {Chen}}, \bibinfo {author} {\bibfnamefont {D.}~\bibnamefont {Huang}},\ and\ \bibinfo {author} {\bibfnamefont {Y.~S.}\ \bibnamefont {Ang}},\ }\bibfield  {title} {\bibinfo {title} {{Proposal for valleytronic materials: Ferrovalley metal and valley gapless semiconductor}},\ }\href {https://doi.org/https://doi.org/10.1007/s11467-023-1334-y} {\bibfield  {journal} {\bibinfo  {journal} {Front. Phys.}\ }\textbf {\bibinfo {volume} {19}},\ \bibinfo {pages} {23302} (\bibinfo {year} {2024}{\natexlab{a}})}\BibitemShut {NoStop}%
\bibitem [{\citenamefont {Krizman}\ \emph {et~al.}(2024)\citenamefont {Krizman}, \citenamefont {Bermejo-Ortiz}, \citenamefont {Zakusylo}, \citenamefont {Hajlaoui}, \citenamefont {Takashiro}, \citenamefont {Rosmus}, \citenamefont {Olszowska}, \citenamefont {Ko\l{}odziej}, \citenamefont {Bauer}, \citenamefont {Guldner}, \citenamefont {Springholz},\ and\ \citenamefont {de~Vaulchier}}]{PhysRevLett.132.166601}%
  \BibitemOpen
  \bibfield  {author} {\bibinfo {author} {\bibfnamefont {G.}~\bibnamefont {Krizman}}, \bibinfo {author} {\bibfnamefont {J.}~\bibnamefont {Bermejo-Ortiz}}, \bibinfo {author} {\bibfnamefont {T.}~\bibnamefont {Zakusylo}}, \bibinfo {author} {\bibfnamefont {M.}~\bibnamefont {Hajlaoui}}, \bibinfo {author} {\bibfnamefont {T.}~\bibnamefont {Takashiro}}, \bibinfo {author} {\bibfnamefont {M.}~\bibnamefont {Rosmus}}, \bibinfo {author} {\bibfnamefont {N.}~\bibnamefont {Olszowska}}, \bibinfo {author} {\bibfnamefont {J.~J.}\ \bibnamefont {Ko\l{}odziej}}, \bibinfo {author} {\bibfnamefont {G.}~\bibnamefont {Bauer}}, \bibinfo {author} {\bibfnamefont {Y.}~\bibnamefont {Guldner}}, \bibinfo {author} {\bibfnamefont {G.}~\bibnamefont {Springholz}},\ and\ \bibinfo {author} {\bibfnamefont {L.-A.}\ \bibnamefont {de~Vaulchier}},\ }\bibfield  {title} {\bibinfo {title} {{Valley-Polarized Quantum Hall Phase in a Strain-Controlled Dirac System}},\ }\href {https://doi.org/10.1103/PhysRevLett.132.166601} {\bibfield  {journal} {\bibinfo
  {journal} {Phys. Rev. Lett.}\ }\textbf {\bibinfo {volume} {132}},\ \bibinfo {pages} {166601} (\bibinfo {year} {2024})}\BibitemShut {NoStop}%
\bibitem [{\citenamefont {Haldane}(1988)}]{PhysRevLett.61.2015}%
  \BibitemOpen
  \bibfield  {author} {\bibinfo {author} {\bibfnamefont {F.~D.~M.}\ \bibnamefont {Haldane}},\ }\bibfield  {title} {\bibinfo {title} {{Model for a Quantum Hall Effect without Landau Levels: Condensed-Matter Realization of the "Parity Anomaly"}},\ }\href {https://doi.org/10.1103/PhysRevLett.61.2015} {\bibfield  {journal} {\bibinfo  {journal} {Phys. Rev. Lett.}\ }\textbf {\bibinfo {volume} {61}},\ \bibinfo {pages} {2015} (\bibinfo {year} {1988})}\BibitemShut {NoStop}%
\bibitem [{\citenamefont {Colom\'es}\ and\ \citenamefont {Franz}(2018)}]{PhysRevLett.120.086603}%
  \BibitemOpen
  \bibfield  {author} {\bibinfo {author} {\bibfnamefont {E.}~\bibnamefont {Colom\'es}}\ and\ \bibinfo {author} {\bibfnamefont {M.}~\bibnamefont {Franz}},\ }\bibfield  {title} {\bibinfo {title} {{Antichiral Edge States in a Modified Haldane Nanoribbon}},\ }\href {https://doi.org/10.1103/PhysRevLett.120.086603} {\bibfield  {journal} {\bibinfo  {journal} {Phys. Rev. Lett.}\ }\textbf {\bibinfo {volume} {120}},\ \bibinfo {pages} {086603} (\bibinfo {year} {2018})}\BibitemShut {NoStop}%
\bibitem [{\citenamefont {Qi}\ \emph {et~al.}(2006)\citenamefont {Qi}, \citenamefont {Wu},\ and\ \citenamefont {Zhang}}]{PhysRevB.74.085308}%
  \BibitemOpen
  \bibfield  {author} {\bibinfo {author} {\bibfnamefont {X.-L.}\ \bibnamefont {Qi}}, \bibinfo {author} {\bibfnamefont {Y.-S.}\ \bibnamefont {Wu}},\ and\ \bibinfo {author} {\bibfnamefont {S.-C.}\ \bibnamefont {Zhang}},\ }\bibfield  {title} {\bibinfo {title} {{Topological quantization of the spin Hall effect in two-dimensional paramagnetic semiconductors}},\ }\href {https://doi.org/10.1103/PhysRevB.74.085308} {\bibfield  {journal} {\bibinfo  {journal} {Phys. Rev. B}\ }\textbf {\bibinfo {volume} {74}},\ \bibinfo {pages} {085308} (\bibinfo {year} {2006})}\BibitemShut {NoStop}%
\bibitem [{\citenamefont {Ezawa}(2013{\natexlab{a}})}]{PhysRevB.87.155415}%
  \BibitemOpen
  \bibfield  {author} {\bibinfo {author} {\bibfnamefont {M.}~\bibnamefont {Ezawa}},\ }\bibfield  {title} {\bibinfo {title} {{Spin valleytronics in silicene: Quantum spin Hall--quantum anomalous Hall insulators and single-valley semimetals}},\ }\href {https://doi.org/10.1103/PhysRevB.87.155415} {\bibfield  {journal} {\bibinfo  {journal} {Phys. Rev. B}\ }\textbf {\bibinfo {volume} {87}},\ \bibinfo {pages} {155415} (\bibinfo {year} {2013}{\natexlab{a}})}\BibitemShut {NoStop}%
\bibitem [{\citenamefont {Jotzu}\ \emph {et~al.}(2014)\citenamefont {Jotzu}, \citenamefont {Messer}, \citenamefont {Desbuquois}, \citenamefont {Lebrat}, \citenamefont {Uehlinger}, \citenamefont {Greif},\ and\ \citenamefont {Esslinger}}]{jotzu2014experimental}%
  \BibitemOpen
  \bibfield  {author} {\bibinfo {author} {\bibfnamefont {G.}~\bibnamefont {Jotzu}}, \bibinfo {author} {\bibfnamefont {M.}~\bibnamefont {Messer}}, \bibinfo {author} {\bibfnamefont {R.}~\bibnamefont {Desbuquois}}, \bibinfo {author} {\bibfnamefont {M.}~\bibnamefont {Lebrat}}, \bibinfo {author} {\bibfnamefont {T.}~\bibnamefont {Uehlinger}}, \bibinfo {author} {\bibfnamefont {D.}~\bibnamefont {Greif}},\ and\ \bibinfo {author} {\bibfnamefont {T.}~\bibnamefont {Esslinger}},\ }\bibfield  {title} {\bibinfo {title} {{Experimental realization of the topological Haldane model with ultracold fermions}},\ }\href {https://doi.org/https://doi.org/10.1038/nature13915} {\bibfield  {journal} {\bibinfo  {journal} {Nature}\ }\textbf {\bibinfo {volume} {515}},\ \bibinfo {pages} {237} (\bibinfo {year} {2014})}\BibitemShut {NoStop}%
\bibitem [{\citenamefont {McIver}\ \emph {et~al.}(2020)\citenamefont {McIver}, \citenamefont {Schulte}, \citenamefont {Stein}, \citenamefont {Matsuyama}, \citenamefont {Jotzu}, \citenamefont {Meier},\ and\ \citenamefont {Cavalleri}}]{mciver2020light}%
  \BibitemOpen
  \bibfield  {author} {\bibinfo {author} {\bibfnamefont {J.~W.}\ \bibnamefont {McIver}}, \bibinfo {author} {\bibfnamefont {B.}~\bibnamefont {Schulte}}, \bibinfo {author} {\bibfnamefont {F.-U.}\ \bibnamefont {Stein}}, \bibinfo {author} {\bibfnamefont {T.}~\bibnamefont {Matsuyama}}, \bibinfo {author} {\bibfnamefont {G.}~\bibnamefont {Jotzu}}, \bibinfo {author} {\bibfnamefont {G.}~\bibnamefont {Meier}},\ and\ \bibinfo {author} {\bibfnamefont {A.}~\bibnamefont {Cavalleri}},\ }\bibfield  {title} {\bibinfo {title} {{Light-induced anomalous Hall effect in graphene}},\ }\href {https://doi.org/https://doi.org/10.1038/s41567-019-0698-y} {\bibfield  {journal} {\bibinfo  {journal} {Nat. Phys.}\ }\textbf {\bibinfo {volume} {16}},\ \bibinfo {pages} {38} (\bibinfo {year} {2020})}\BibitemShut {NoStop}%
\bibitem [{\citenamefont {Li}\ \emph {et~al.}(2021)\citenamefont {Li}, \citenamefont {Jiang}, \citenamefont {Shen}, \citenamefont {Zhang}, \citenamefont {Li}, \citenamefont {Tao}, \citenamefont {Devakul}, \citenamefont {Watanabe}, \citenamefont {Taniguchi}, \citenamefont {Fu} \emph {et~al.}}]{li2021quantum}%
  \BibitemOpen
  \bibfield  {author} {\bibinfo {author} {\bibfnamefont {T.}~\bibnamefont {Li}}, \bibinfo {author} {\bibfnamefont {S.}~\bibnamefont {Jiang}}, \bibinfo {author} {\bibfnamefont {B.}~\bibnamefont {Shen}}, \bibinfo {author} {\bibfnamefont {Y.}~\bibnamefont {Zhang}}, \bibinfo {author} {\bibfnamefont {L.}~\bibnamefont {Li}}, \bibinfo {author} {\bibfnamefont {Z.}~\bibnamefont {Tao}}, \bibinfo {author} {\bibfnamefont {T.}~\bibnamefont {Devakul}}, \bibinfo {author} {\bibfnamefont {K.}~\bibnamefont {Watanabe}}, \bibinfo {author} {\bibfnamefont {T.}~\bibnamefont {Taniguchi}}, \bibinfo {author} {\bibfnamefont {L.}~\bibnamefont {Fu}}, \emph {et~al.},\ }\bibfield  {title} {\bibinfo {title} {{Quantum anomalous Hall effect from intertwined moir{\'e} bands}},\ }\href {https://doi.org/https://doi.org/10.1038/s41586-021-04171-1} {\bibfield  {journal} {\bibinfo  {journal} {Nature}\ }\textbf {\bibinfo {volume} {600}},\ \bibinfo {pages} {641} (\bibinfo {year} {2021})}\BibitemShut {NoStop}%
\bibitem [{\citenamefont {Zhou}\ \emph {et~al.}(2020)\citenamefont {Zhou}, \citenamefont {Liu}, \citenamefont {Yang}, \citenamefont {Hu}, \citenamefont {Ma}, \citenamefont {Xue}, \citenamefont {Wang}, \citenamefont {Deng},\ and\ \citenamefont {Zhang}}]{PhysRevLett.125.263603}%
  \BibitemOpen
  \bibfield  {author} {\bibinfo {author} {\bibfnamefont {P.}~\bibnamefont {Zhou}}, \bibinfo {author} {\bibfnamefont {G.-G.}\ \bibnamefont {Liu}}, \bibinfo {author} {\bibfnamefont {Y.}~\bibnamefont {Yang}}, \bibinfo {author} {\bibfnamefont {Y.-H.}\ \bibnamefont {Hu}}, \bibinfo {author} {\bibfnamefont {S.}~\bibnamefont {Ma}}, \bibinfo {author} {\bibfnamefont {H.}~\bibnamefont {Xue}}, \bibinfo {author} {\bibfnamefont {Q.}~\bibnamefont {Wang}}, \bibinfo {author} {\bibfnamefont {L.}~\bibnamefont {Deng}},\ and\ \bibinfo {author} {\bibfnamefont {B.}~\bibnamefont {Zhang}},\ }\bibfield  {title} {\bibinfo {title} {{Observation of Photonic Antichiral Edge States}},\ }\href {https://doi.org/10.1103/PhysRevLett.125.263603} {\bibfield  {journal} {\bibinfo  {journal} {Phys. Rev. Lett.}\ }\textbf {\bibinfo {volume} {125}},\ \bibinfo {pages} {263603} (\bibinfo {year} {2020})}\BibitemShut {NoStop}%
\bibitem [{\citenamefont {Yang}\ \emph {et~al.}(2021)\citenamefont {Yang}, \citenamefont {Zhu}, \citenamefont {Hang},\ and\ \citenamefont {Chong}}]{yang2021observation}%
  \BibitemOpen
  \bibfield  {author} {\bibinfo {author} {\bibfnamefont {Y.}~\bibnamefont {Yang}}, \bibinfo {author} {\bibfnamefont {D.}~\bibnamefont {Zhu}}, \bibinfo {author} {\bibfnamefont {Z.}~\bibnamefont {Hang}},\ and\ \bibinfo {author} {\bibfnamefont {Y.}~\bibnamefont {Chong}},\ }\bibfield  {title} {\bibinfo {title} {{Observation of antichiral edge states in a circuit lattice}},\ }\href {https://doi.org/https://doi.org/10.1007/s11433-021-1675-0} {\bibfield  {journal} {\bibinfo  {journal} {Sci. China Phys. Mech. Astron.}\ }\textbf {\bibinfo {volume} {64}},\ \bibinfo {pages} {1} (\bibinfo {year} {2021})}\BibitemShut {NoStop}%
\bibitem [{\citenamefont {Vila}\ \emph {et~al.}(2019)\citenamefont {Vila}, \citenamefont {Hung}, \citenamefont {Roche},\ and\ \citenamefont {Saito}}]{PhysRevB.99.161404}%
  \BibitemOpen
  \bibfield  {author} {\bibinfo {author} {\bibfnamefont {M.}~\bibnamefont {Vila}}, \bibinfo {author} {\bibfnamefont {N.~T.}\ \bibnamefont {Hung}}, \bibinfo {author} {\bibfnamefont {S.}~\bibnamefont {Roche}},\ and\ \bibinfo {author} {\bibfnamefont {R.}~\bibnamefont {Saito}},\ }\bibfield  {title} {\bibinfo {title} {{Tunable circular dichroism and valley polarization in the modified Haldane model}},\ }\href {https://doi.org/10.1103/PhysRevB.99.161404} {\bibfield  {journal} {\bibinfo  {journal} {Phys. Rev. B}\ }\textbf {\bibinfo {volume} {99}},\ \bibinfo {pages} {161404} (\bibinfo {year} {2019})}\BibitemShut {NoStop}%
\bibitem [{\citenamefont {Ezawa}(2013{\natexlab{b}})}]{PhysRevLett.110.026603}%
  \BibitemOpen
  \bibfield  {author} {\bibinfo {author} {\bibfnamefont {M.}~\bibnamefont {Ezawa}},\ }\bibfield  {title} {\bibinfo {title} {{Photoinduced Topological Phase Transition and a Single Dirac-Cone State in Silicene}},\ }\href {https://doi.org/10.1103/PhysRevLett.110.026603} {\bibfield  {journal} {\bibinfo  {journal} {Phys. Rev. Lett.}\ }\textbf {\bibinfo {volume} {110}},\ \bibinfo {pages} {026603} (\bibinfo {year} {2013}{\natexlab{b}})}\BibitemShut {NoStop}%
\bibitem [{\citenamefont {Zhao}\ \emph {et~al.}(2024)\citenamefont {Zhao}, \citenamefont {Kang}, \citenamefont {Zhang}, \citenamefont {Kn{\"u}ppel}, \citenamefont {Tao}, \citenamefont {Li}, \citenamefont {Tschirhart}, \citenamefont {Redekop}, \citenamefont {Watanabe}, \citenamefont {Taniguchi} \emph {et~al.}}]{zhao2024realization}%
  \BibitemOpen
  \bibfield  {author} {\bibinfo {author} {\bibfnamefont {W.}~\bibnamefont {Zhao}}, \bibinfo {author} {\bibfnamefont {K.}~\bibnamefont {Kang}}, \bibinfo {author} {\bibfnamefont {Y.}~\bibnamefont {Zhang}}, \bibinfo {author} {\bibfnamefont {P.}~\bibnamefont {Kn{\"u}ppel}}, \bibinfo {author} {\bibfnamefont {Z.}~\bibnamefont {Tao}}, \bibinfo {author} {\bibfnamefont {L.}~\bibnamefont {Li}}, \bibinfo {author} {\bibfnamefont {C.~L.}\ \bibnamefont {Tschirhart}}, \bibinfo {author} {\bibfnamefont {E.}~\bibnamefont {Redekop}}, \bibinfo {author} {\bibfnamefont {K.}~\bibnamefont {Watanabe}}, \bibinfo {author} {\bibfnamefont {T.}~\bibnamefont {Taniguchi}}, \emph {et~al.},\ }\bibfield  {title} {\bibinfo {title} {{Realization of the Haldane Chern insulator in a moir{\'e} lattice}},\ }\href {https://doi.org/https://doi.org/10.1038/s41567-023-02284-0} {\bibfield  {journal} {\bibinfo  {journal} {Nat. Phys.}\ }\textbf {\bibinfo {volume} {20}},\ \bibinfo {pages} {275} (\bibinfo {year} {2024})}\BibitemShut {NoStop}%
\bibitem [{\citenamefont {Mitra}\ \emph {et~al.}(2024)\citenamefont {Mitra}, \citenamefont {Jim{\'e}nez-Gal{\'a}n}, \citenamefont {Aulich}, \citenamefont {Neuhaus}, \citenamefont {Silva}, \citenamefont {Pervak}, \citenamefont {Kling},\ and\ \citenamefont {Biswas}}]{mitra2024light}%
  \BibitemOpen
  \bibfield  {author} {\bibinfo {author} {\bibfnamefont {S.}~\bibnamefont {Mitra}}, \bibinfo {author} {\bibfnamefont {{\'A}.}~\bibnamefont {Jim{\'e}nez-Gal{\'a}n}}, \bibinfo {author} {\bibfnamefont {M.}~\bibnamefont {Aulich}}, \bibinfo {author} {\bibfnamefont {M.}~\bibnamefont {Neuhaus}}, \bibinfo {author} {\bibfnamefont {R.~E.}\ \bibnamefont {Silva}}, \bibinfo {author} {\bibfnamefont {V.}~\bibnamefont {Pervak}}, \bibinfo {author} {\bibfnamefont {M.~F.}\ \bibnamefont {Kling}},\ and\ \bibinfo {author} {\bibfnamefont {S.}~\bibnamefont {Biswas}},\ }\bibfield  {title} {\bibinfo {title} {{Light-wave-controlled Haldane model in monolayer hexagonal boron nitride}},\ }\href {https://doi.org/https://doi.org/10.1038/s41586-024-07244-z} {\bibfield  {journal} {\bibinfo  {journal} {Nature}\ ,\ \bibinfo {pages} {1}} (\bibinfo {year} {2024})}\BibitemShut {NoStop}%
\bibitem [{\citenamefont {Oka}\ and\ \citenamefont {Aoki}(2009)}]{PhysRevB.79.081406}%
  \BibitemOpen
  \bibfield  {author} {\bibinfo {author} {\bibfnamefont {T.}~\bibnamefont {Oka}}\ and\ \bibinfo {author} {\bibfnamefont {H.}~\bibnamefont {Aoki}},\ }\bibfield  {title} {\bibinfo {title} {{Photovoltaic Hall effect in graphene}},\ }\href {https://doi.org/10.1103/PhysRevB.79.081406} {\bibfield  {journal} {\bibinfo  {journal} {Phys. Rev. B}\ }\textbf {\bibinfo {volume} {79}},\ \bibinfo {pages} {081406} (\bibinfo {year} {2009})}\BibitemShut {NoStop}%
\bibitem [{\citenamefont {Hofmann}\ \emph {et~al.}(2019)\citenamefont {Hofmann}, \citenamefont {Helbig}, \citenamefont {Lee}, \citenamefont {Greiter},\ and\ \citenamefont {Thomale}}]{PhysRevLett.122.247702}%
  \BibitemOpen
  \bibfield  {author} {\bibinfo {author} {\bibfnamefont {T.}~\bibnamefont {Hofmann}}, \bibinfo {author} {\bibfnamefont {T.}~\bibnamefont {Helbig}}, \bibinfo {author} {\bibfnamefont {C.~H.}\ \bibnamefont {Lee}}, \bibinfo {author} {\bibfnamefont {M.}~\bibnamefont {Greiter}},\ and\ \bibinfo {author} {\bibfnamefont {R.}~\bibnamefont {Thomale}},\ }\bibfield  {title} {\bibinfo {title} {{Chiral Voltage Propagation and Calibration in a Topolectrical Chern Circuit}},\ }\href {https://doi.org/10.1103/PhysRevLett.122.247702} {\bibfield  {journal} {\bibinfo  {journal} {Phys. Rev. Lett.}\ }\textbf {\bibinfo {volume} {122}},\ \bibinfo {pages} {247702} (\bibinfo {year} {2019})}\BibitemShut {NoStop}%
\bibitem [{\citenamefont {Liu}\ \emph {et~al.}(2023{\natexlab{b}})\citenamefont {Liu}, \citenamefont {Shi}, \citenamefont {Shen}, \citenamefont {Chen}, \citenamefont {Chen}, \citenamefont {Chen},\ and\ \citenamefont {Dong}}]{liu2023antichiral}%
  \BibitemOpen
  \bibfield  {author} {\bibinfo {author} {\bibfnamefont {J.-W.}\ \bibnamefont {Liu}}, \bibinfo {author} {\bibfnamefont {F.-L.}\ \bibnamefont {Shi}}, \bibinfo {author} {\bibfnamefont {K.}~\bibnamefont {Shen}}, \bibinfo {author} {\bibfnamefont {X.-D.}\ \bibnamefont {Chen}}, \bibinfo {author} {\bibfnamefont {K.}~\bibnamefont {Chen}}, \bibinfo {author} {\bibfnamefont {W.-J.}\ \bibnamefont {Chen}},\ and\ \bibinfo {author} {\bibfnamefont {J.-W.}\ \bibnamefont {Dong}},\ }\bibfield  {title} {\bibinfo {title} {{Antichiral surface states in time-reversal-invariant photonic semimetals}},\ }\href {https://doi.org/https://doi.org/10.1038/s41467-023-37670-y} {\bibfield  {journal} {\bibinfo  {journal} {Nat Commun}\ }\textbf {\bibinfo {volume} {14}},\ \bibinfo {pages} {2027} (\bibinfo {year} {2023}{\natexlab{b}})}\BibitemShut {NoStop}%
\bibitem [{\citenamefont {Xi}\ \emph {et~al.}(2023)\citenamefont {Xi}, \citenamefont {Yan}, \citenamefont {Yang}, \citenamefont {Meng}, \citenamefont {Zhu}, \citenamefont {Chen}, \citenamefont {Wang}, \citenamefont {Zhou}, \citenamefont {Shum}, \citenamefont {Yang} \emph {et~al.}}]{xi2023topological}%
  \BibitemOpen
  \bibfield  {author} {\bibinfo {author} {\bibfnamefont {X.}~\bibnamefont {Xi}}, \bibinfo {author} {\bibfnamefont {B.}~\bibnamefont {Yan}}, \bibinfo {author} {\bibfnamefont {L.}~\bibnamefont {Yang}}, \bibinfo {author} {\bibfnamefont {Y.}~\bibnamefont {Meng}}, \bibinfo {author} {\bibfnamefont {Z.-X.}\ \bibnamefont {Zhu}}, \bibinfo {author} {\bibfnamefont {J.-M.}\ \bibnamefont {Chen}}, \bibinfo {author} {\bibfnamefont {Z.}~\bibnamefont {Wang}}, \bibinfo {author} {\bibfnamefont {P.}~\bibnamefont {Zhou}}, \bibinfo {author} {\bibfnamefont {P.~P.}\ \bibnamefont {Shum}}, \bibinfo {author} {\bibfnamefont {Y.}~\bibnamefont {Yang}}, \emph {et~al.},\ }\bibfield  {title} {\bibinfo {title} {{Topological antichiral surface states in a magnetic Weyl photonic crystal}},\ }\href {https://doi.org/https://doi.org/10.1038/s41467-023-37710-7} {\bibfield  {journal} {\bibinfo  {journal} {Nat Commun}\ }\textbf {\bibinfo {volume} {14}},\ \bibinfo {pages} {1991} (\bibinfo {year} {2023})}\BibitemShut {NoStop}%
\bibitem [{\citenamefont {Yuan}\ \emph {et~al.}(2013)\citenamefont {Yuan}, \citenamefont {Bahramy}, \citenamefont {Morimoto}, \citenamefont {Wu}, \citenamefont {Nomura}, \citenamefont {Yang}, \citenamefont {Shimotani}, \citenamefont {Suzuki}, \citenamefont {Toh}, \citenamefont {Kloc} \emph {et~al.}}]{yuan2013zeeman}%
  \BibitemOpen
  \bibfield  {author} {\bibinfo {author} {\bibfnamefont {H.}~\bibnamefont {Yuan}}, \bibinfo {author} {\bibfnamefont {M.~S.}\ \bibnamefont {Bahramy}}, \bibinfo {author} {\bibfnamefont {K.}~\bibnamefont {Morimoto}}, \bibinfo {author} {\bibfnamefont {S.}~\bibnamefont {Wu}}, \bibinfo {author} {\bibfnamefont {K.}~\bibnamefont {Nomura}}, \bibinfo {author} {\bibfnamefont {B.-J.}\ \bibnamefont {Yang}}, \bibinfo {author} {\bibfnamefont {H.}~\bibnamefont {Shimotani}}, \bibinfo {author} {\bibfnamefont {R.}~\bibnamefont {Suzuki}}, \bibinfo {author} {\bibfnamefont {M.}~\bibnamefont {Toh}}, \bibinfo {author} {\bibfnamefont {C.}~\bibnamefont {Kloc}}, \emph {et~al.},\ }\bibfield  {title} {\bibinfo {title} {{Zeeman-type spin splitting controlled by an electric field}},\ }\href {https://doi.org/https://doi.org/10.1038/nphys2691} {\bibfield  {journal} {\bibinfo  {journal} {Nature Phys}\ }\textbf {\bibinfo {volume} {9}},\ \bibinfo {pages} {563} (\bibinfo {year} {2013})}\BibitemShut {NoStop}%
\bibitem [{\citenamefont {Yang}\ \emph {et~al.}(2018)\citenamefont {Yang}, \citenamefont {Molina}, \citenamefont {Kim}, \citenamefont {Barroso}, \citenamefont {Lohmann}, \citenamefont {Liu}, \citenamefont {Xu}, \citenamefont {Wu}, \citenamefont {Bartels}, \citenamefont {Watanabe}, \citenamefont {Taniguchi},\ and\ \citenamefont {Shi}}]{doi:10.1021/acs.nanolett.8b00691}%
  \BibitemOpen
  \bibfield  {author} {\bibinfo {author} {\bibfnamefont {B.}~\bibnamefont {Yang}}, \bibinfo {author} {\bibfnamefont {E.}~\bibnamefont {Molina}}, \bibinfo {author} {\bibfnamefont {J.}~\bibnamefont {Kim}}, \bibinfo {author} {\bibfnamefont {D.}~\bibnamefont {Barroso}}, \bibinfo {author} {\bibfnamefont {M.}~\bibnamefont {Lohmann}}, \bibinfo {author} {\bibfnamefont {Y.}~\bibnamefont {Liu}}, \bibinfo {author} {\bibfnamefont {Y.}~\bibnamefont {Xu}}, \bibinfo {author} {\bibfnamefont {R.}~\bibnamefont {Wu}}, \bibinfo {author} {\bibfnamefont {L.}~\bibnamefont {Bartels}}, \bibinfo {author} {\bibfnamefont {K.}~\bibnamefont {Watanabe}}, \bibinfo {author} {\bibfnamefont {T.}~\bibnamefont {Taniguchi}},\ and\ \bibinfo {author} {\bibfnamefont {J.}~\bibnamefont {Shi}},\ }\bibfield  {title} {\bibinfo {title} {{Effect of Distance on Photoluminescence Quenching and Proximity-Induced Spin–Orbit Coupling in Graphene/${\mathrm{WSe}}_{2}$ Heterostructures}},\ }\href {https://doi.org/10.1021/acs.nanolett.8b00691} {\bibfield
  {journal} {\bibinfo  {journal} {Nano Lett.}\ }\textbf {\bibinfo {volume} {18}},\ \bibinfo {pages} {3580} (\bibinfo {year} {2018})}\BibitemShut {NoStop}%
\bibitem [{\citenamefont {Cheiwchanchamnangij}\ \emph {et~al.}(2013)\citenamefont {Cheiwchanchamnangij}, \citenamefont {Lambrecht}, \citenamefont {Song},\ and\ \citenamefont {Dery}}]{PhysRevB.88.155404}%
  \BibitemOpen
  \bibfield  {author} {\bibinfo {author} {\bibfnamefont {T.}~\bibnamefont {Cheiwchanchamnangij}}, \bibinfo {author} {\bibfnamefont {W.~R.~L.}\ \bibnamefont {Lambrecht}}, \bibinfo {author} {\bibfnamefont {Y.}~\bibnamefont {Song}},\ and\ \bibinfo {author} {\bibfnamefont {H.}~\bibnamefont {Dery}},\ }\bibfield  {title} {\bibinfo {title} {{Strain effects on the spin-orbit-induced band structure splittings in monolayer ${\mathrm{MoS}}_{2}$ and graphene}},\ }\href {https://doi.org/10.1103/PhysRevB.88.155404} {\bibfield  {journal} {\bibinfo  {journal} {Phys. Rev. B}\ }\textbf {\bibinfo {volume} {88}},\ \bibinfo {pages} {155404} (\bibinfo {year} {2013})}\BibitemShut {NoStop}%
\bibitem [{\citenamefont {Fujita}\ \emph {et~al.}(2010)\citenamefont {Fujita}, \citenamefont {Jalil},\ and\ \citenamefont {Tan}}]{10.1063/1.3473725}%
  \BibitemOpen
  \bibfield  {author} {\bibinfo {author} {\bibfnamefont {T.}~\bibnamefont {Fujita}}, \bibinfo {author} {\bibfnamefont {M.~B.~A.}\ \bibnamefont {Jalil}},\ and\ \bibinfo {author} {\bibfnamefont {S.~G.}\ \bibnamefont {Tan}},\ }\bibfield  {title} {\bibinfo {title} {{Valley filter in strain engineered graphene}},\ }\href {https://doi.org/10.1063/1.3473725} {\bibfield  {journal} {\bibinfo  {journal} {Appl. Phys. Lett.}\ }\textbf {\bibinfo {volume} {97}},\ \bibinfo {pages} {043508} (\bibinfo {year} {2010})}\BibitemShut {NoStop}%
\bibitem [{\citenamefont {Katsnelson}\ \emph {et~al.}(2006)\citenamefont {Katsnelson}, \citenamefont {Novoselov},\ and\ \citenamefont {Geim}}]{katsnelson2006chiral}%
  \BibitemOpen
  \bibfield  {author} {\bibinfo {author} {\bibfnamefont {M.~I.}\ \bibnamefont {Katsnelson}}, \bibinfo {author} {\bibfnamefont {K.~S.}\ \bibnamefont {Novoselov}},\ and\ \bibinfo {author} {\bibfnamefont {A.~K.}\ \bibnamefont {Geim}},\ }\bibfield  {title} {\bibinfo {title} {{Chiral tunnelling and the Klein paradox in graphene}},\ }\href {https://doi.org/https://doi.org/10.1038/nphys384} {\bibfield  {journal} {\bibinfo  {journal} {Nature Phys}\ }\textbf {\bibinfo {volume} {2}},\ \bibinfo {pages} {620} (\bibinfo {year} {2006})}\BibitemShut {NoStop}%
\bibitem [{\citenamefont {Ozawa}\ \emph {et~al.}(2017)\citenamefont {Ozawa}, \citenamefont {Amo}, \citenamefont {Bloch},\ and\ \citenamefont {Carusotto}}]{PhysRevA.96.013813}%
  \BibitemOpen
  \bibfield  {author} {\bibinfo {author} {\bibfnamefont {T.}~\bibnamefont {Ozawa}}, \bibinfo {author} {\bibfnamefont {A.}~\bibnamefont {Amo}}, \bibinfo {author} {\bibfnamefont {J.}~\bibnamefont {Bloch}},\ and\ \bibinfo {author} {\bibfnamefont {I.}~\bibnamefont {Carusotto}},\ }\bibfield  {title} {\bibinfo {title} {{Klein tunneling in driven-dissipative photonic graphene}},\ }\href {https://doi.org/10.1103/PhysRevA.96.013813} {\bibfield  {journal} {\bibinfo  {journal} {Phys. Rev. A}\ }\textbf {\bibinfo {volume} {96}},\ \bibinfo {pages} {013813} (\bibinfo {year} {2017})}\BibitemShut {NoStop}%
\bibitem [{\citenamefont {Beenakker}(2008)}]{RevModPhys.80.1337}%
  \BibitemOpen
  \bibfield  {author} {\bibinfo {author} {\bibfnamefont {C.~W.~J.}\ \bibnamefont {Beenakker}},\ }\bibfield  {title} {\bibinfo {title} {{Colloquium: Andreev reflection and Klein tunneling in graphene}},\ }\href {https://doi.org/10.1103/RevModPhys.80.1337} {\bibfield  {journal} {\bibinfo  {journal} {Rev. Mod. Phys.}\ }\textbf {\bibinfo {volume} {80}},\ \bibinfo {pages} {1337} (\bibinfo {year} {2008})}\BibitemShut {NoStop}%
\bibitem [{\citenamefont {Castro~Neto}\ \emph {et~al.}(2009)\citenamefont {Castro~Neto}, \citenamefont {Guinea}, \citenamefont {Peres}, \citenamefont {Novoselov},\ and\ \citenamefont {Geim}}]{RevModPhys.81.109}%
  \BibitemOpen
  \bibfield  {author} {\bibinfo {author} {\bibfnamefont {A.~H.}\ \bibnamefont {Castro~Neto}}, \bibinfo {author} {\bibfnamefont {F.}~\bibnamefont {Guinea}}, \bibinfo {author} {\bibfnamefont {N.~M.~R.}\ \bibnamefont {Peres}}, \bibinfo {author} {\bibfnamefont {K.~S.}\ \bibnamefont {Novoselov}},\ and\ \bibinfo {author} {\bibfnamefont {A.~K.}\ \bibnamefont {Geim}},\ }\bibfield  {title} {\bibinfo {title} {{The electronic properties of graphene}},\ }\href {https://doi.org/10.1103/RevModPhys.81.109} {\bibfield  {journal} {\bibinfo  {journal} {Rev. Mod. Phys.}\ }\textbf {\bibinfo {volume} {81}},\ \bibinfo {pages} {109} (\bibinfo {year} {2009})}\BibitemShut {NoStop}%
\bibitem [{\citenamefont {Allain}\ and\ \citenamefont {Fuchs}(2011)}]{allain2011klein}%
  \BibitemOpen
  \bibfield  {author} {\bibinfo {author} {\bibfnamefont {P.~E.}\ \bibnamefont {Allain}}\ and\ \bibinfo {author} {\bibfnamefont {J.-N.}\ \bibnamefont {Fuchs}},\ }\bibfield  {title} {\bibinfo {title} {{Klein tunneling in graphene: optics with massless electrons}},\ }\href {https://doi.org/https://doi.org/10.1140/epjb/e2011-20351-3} {\bibfield  {journal} {\bibinfo  {journal} {Eur. Phys. J. B}\ }\textbf {\bibinfo {volume} {83}},\ \bibinfo {pages} {301} (\bibinfo {year} {2011})}\BibitemShut {NoStop}%
\bibitem [{\citenamefont {Myers}\ \emph {et~al.}(2024)\citenamefont {Myers}, \citenamefont {Sulas-Kern}, \citenamefont {Fei}, \citenamefont {Ghoshal}, \citenamefont {Hermosilla-Palacios},\ and\ \citenamefont {Blackburn}}]{10.1063/5.0213720}%
  \BibitemOpen
  \bibfield  {author} {\bibinfo {author} {\bibfnamefont {A.~R.}\ \bibnamefont {Myers}}, \bibinfo {author} {\bibfnamefont {D.~B.}\ \bibnamefont {Sulas-Kern}}, \bibinfo {author} {\bibfnamefont {R.}~\bibnamefont {Fei}}, \bibinfo {author} {\bibfnamefont {D.}~\bibnamefont {Ghoshal}}, \bibinfo {author} {\bibfnamefont {M.~A.}\ \bibnamefont {Hermosilla-Palacios}},\ and\ \bibinfo {author} {\bibfnamefont {J.~L.}\ \bibnamefont {Blackburn}},\ }\bibfield  {title} {\bibinfo {title} {{Quantifying carrier density in monolayer ${\mathrm{MoS}}_{2}$ by optical spectroscopy}},\ }\href {https://doi.org/10.1063/5.0213720} {\bibfield  {journal} {\bibinfo  {journal} {J. Chem. Phys.}\ }\textbf {\bibinfo {volume} {161}},\ \bibinfo {pages} {044706} (\bibinfo {year} {2024})}\BibitemShut {NoStop}%
\bibitem [{\citenamefont {Yin}\ \emph {et~al.}(2014)\citenamefont {Yin}, \citenamefont {Cheng}, \citenamefont {Wang}, \citenamefont {Jin},\ and\ \citenamefont {Wang}}]{yin2014graphene}%
  \BibitemOpen
  \bibfield  {author} {\bibinfo {author} {\bibfnamefont {Y.}~\bibnamefont {Yin}}, \bibinfo {author} {\bibfnamefont {Z.}~\bibnamefont {Cheng}}, \bibinfo {author} {\bibfnamefont {L.}~\bibnamefont {Wang}}, \bibinfo {author} {\bibfnamefont {K.}~\bibnamefont {Jin}},\ and\ \bibinfo {author} {\bibfnamefont {W.}~\bibnamefont {Wang}},\ }\bibfield  {title} {\bibinfo {title} {{Graphene, a material for high temperature devices--intrinsic carrier density, carrier drift velocity and lattice energy}},\ }\href {https://doi.org/https://doi.org/10.1038/srep05758} {\bibfield  {journal} {\bibinfo  {journal} {Sci Rep}\ }\textbf {\bibinfo {volume} {4}},\ \bibinfo {pages} {5758} (\bibinfo {year} {2014})}\BibitemShut {NoStop}%
\bibitem [{\citenamefont {Friesen}\ \emph {et~al.}(2007)\citenamefont {Friesen}, \citenamefont {Chutia}, \citenamefont {Tahan},\ and\ \citenamefont {Coppersmith}}]{PhysRevB.75.115318}%
  \BibitemOpen
  \bibfield  {author} {\bibinfo {author} {\bibfnamefont {M.}~\bibnamefont {Friesen}}, \bibinfo {author} {\bibfnamefont {S.}~\bibnamefont {Chutia}}, \bibinfo {author} {\bibfnamefont {C.}~\bibnamefont {Tahan}},\ and\ \bibinfo {author} {\bibfnamefont {S.~N.}\ \bibnamefont {Coppersmith}},\ }\bibfield  {title} {\bibinfo {title} {{Valley splitting theory of $\mathrm{Si}\mathrm{Ge}/\mathrm{Si}/\mathrm{Si}\mathrm{Ge}$ quantum wells}},\ }\href {https://doi.org/10.1103/PhysRevB.75.115318} {\bibfield  {journal} {\bibinfo  {journal} {Phys. Rev. B}\ }\textbf {\bibinfo {volume} {75}},\ \bibinfo {pages} {115318} (\bibinfo {year} {2007})}\BibitemShut {NoStop}%
\bibitem [{\citenamefont {Shon}\ and\ \citenamefont {Ando}(1998)}]{shon1998quantum}%
  \BibitemOpen
  \bibfield  {author} {\bibinfo {author} {\bibfnamefont {N.~H.}\ \bibnamefont {Shon}}\ and\ \bibinfo {author} {\bibfnamefont {T.}~\bibnamefont {Ando}},\ }\bibfield  {title} {\bibinfo {title} {{Quantum transport in two-dimensional graphite system}},\ }\href {https://doi.org/https://doi.org/10.1143/JPSJ.67.2421} {\bibfield  {journal} {\bibinfo  {journal} {J. Phys. Soc. Jpn.}\ }\textbf {\bibinfo {volume} {67}},\ \bibinfo {pages} {2421} (\bibinfo {year} {1998})}\BibitemShut {NoStop}%
\bibitem [{\citenamefont {Mak}\ \emph {et~al.}(2012)\citenamefont {Mak}, \citenamefont {He}, \citenamefont {Shan},\ and\ \citenamefont {Heinz}}]{mak2012control}%
  \BibitemOpen
  \bibfield  {author} {\bibinfo {author} {\bibfnamefont {K.~F.}\ \bibnamefont {Mak}}, \bibinfo {author} {\bibfnamefont {K.}~\bibnamefont {He}}, \bibinfo {author} {\bibfnamefont {J.}~\bibnamefont {Shan}},\ and\ \bibinfo {author} {\bibfnamefont {T.~F.}\ \bibnamefont {Heinz}},\ }\bibfield  {title} {\bibinfo {title} {{Control of valley polarization in monolayer ${\mathrm{MoS}}_{2}$ by optical helicity}},\ }\href {https://doi.org/https://doi.org/10.1038/nnano.2012.96} {\bibfield  {journal} {\bibinfo  {journal} {Nature Nanotech}\ }\textbf {\bibinfo {volume} {7}},\ \bibinfo {pages} {494} (\bibinfo {year} {2012})}\BibitemShut {NoStop}%
\bibitem [{\citenamefont {Koh}\ \emph {et~al.}(2022)\citenamefont {Koh}, \citenamefont {Tai},\ and\ \citenamefont {Lee}}]{PhysRevLett.129.140502}%
  \BibitemOpen
  \bibfield  {author} {\bibinfo {author} {\bibfnamefont {J.~M.}\ \bibnamefont {Koh}}, \bibinfo {author} {\bibfnamefont {T.}~\bibnamefont {Tai}},\ and\ \bibinfo {author} {\bibfnamefont {C.~H.}\ \bibnamefont {Lee}},\ }\bibfield  {title} {\bibinfo {title} {{Simulation of Interaction-Induced Chiral Topological Dynamics on a Digital Quantum Computer}},\ }\href {https://doi.org/10.1103/PhysRevLett.129.140502} {\bibfield  {journal} {\bibinfo  {journal} {Phys. Rev. Lett.}\ }\textbf {\bibinfo {volume} {129}},\ \bibinfo {pages} {140502} (\bibinfo {year} {2022})}\BibitemShut {NoStop}%
\bibitem [{\citenamefont {Wang}\ \emph {et~al.}(2023)\citenamefont {Wang}, \citenamefont {Valligatla}, \citenamefont {Yin}, \citenamefont {Schwarz}, \citenamefont {Medina-S{\'a}nchez}, \citenamefont {Baunack}, \citenamefont {Lee}, \citenamefont {Thomale}, \citenamefont {Li}, \citenamefont {Fomin} \emph {et~al.}}]{wang2023experimental}%
  \BibitemOpen
  \bibfield  {author} {\bibinfo {author} {\bibfnamefont {J.}~\bibnamefont {Wang}}, \bibinfo {author} {\bibfnamefont {S.}~\bibnamefont {Valligatla}}, \bibinfo {author} {\bibfnamefont {Y.}~\bibnamefont {Yin}}, \bibinfo {author} {\bibfnamefont {L.}~\bibnamefont {Schwarz}}, \bibinfo {author} {\bibfnamefont {M.}~\bibnamefont {Medina-S{\'a}nchez}}, \bibinfo {author} {\bibfnamefont {S.}~\bibnamefont {Baunack}}, \bibinfo {author} {\bibfnamefont {C.~H.}\ \bibnamefont {Lee}}, \bibinfo {author} {\bibfnamefont {R.}~\bibnamefont {Thomale}}, \bibinfo {author} {\bibfnamefont {S.}~\bibnamefont {Li}}, \bibinfo {author} {\bibfnamefont {V.~M.}\ \bibnamefont {Fomin}}, \emph {et~al.},\ }\bibfield  {title} {\bibinfo {title} {{Experimental observation of Berry phases in optical M{\"o}bius-strip microcavities}},\ }\href {https://doi.org/https://doi.org/10.1038/s41566-022-01107-7} {\bibfield  {journal} {\bibinfo  {journal} {Nat. Photon.}\ }\textbf {\bibinfo {volume} {17}},\ \bibinfo {pages} {120} (\bibinfo {year} {2023})}\BibitemShut
  {NoStop}%
\bibitem [{\citenamefont {Vitale}\ \emph {et~al.}(2018)\citenamefont {Vitale}, \citenamefont {Nezich}, \citenamefont {Varghese}, \citenamefont {Kim}, \citenamefont {Gedik}, \citenamefont {Jarillo-Herrero}, \citenamefont {Xiao},\ and\ \citenamefont {Rothschild}}]{https://doi.org/10.1002/smll.201801483}%
  \BibitemOpen
  \bibfield  {author} {\bibinfo {author} {\bibfnamefont {S.~A.}\ \bibnamefont {Vitale}}, \bibinfo {author} {\bibfnamefont {D.}~\bibnamefont {Nezich}}, \bibinfo {author} {\bibfnamefont {J.~O.}\ \bibnamefont {Varghese}}, \bibinfo {author} {\bibfnamefont {P.}~\bibnamefont {Kim}}, \bibinfo {author} {\bibfnamefont {N.}~\bibnamefont {Gedik}}, \bibinfo {author} {\bibfnamefont {P.}~\bibnamefont {Jarillo-Herrero}}, \bibinfo {author} {\bibfnamefont {D.}~\bibnamefont {Xiao}},\ and\ \bibinfo {author} {\bibfnamefont {M.}~\bibnamefont {Rothschild}},\ }\bibfield  {title} {\bibinfo {title} {{Valleytronics: Opportunities, Challenges, and Paths Forward}},\ }\href {https://doi.org/https://doi.org/10.1002/smll.201801483} {\bibfield  {journal} {\bibinfo  {journal} {Small}\ }\textbf {\bibinfo {volume} {14}},\ \bibinfo {pages} {1801483} (\bibinfo {year} {2018})}\BibitemShut {NoStop}%
\bibitem [{\citenamefont {Lee}\ and\ \citenamefont {Qi}(2014)}]{PhysRevB.90.085103}%
  \BibitemOpen
  \bibfield  {author} {\bibinfo {author} {\bibfnamefont {C.~H.}\ \bibnamefont {Lee}}\ and\ \bibinfo {author} {\bibfnamefont {X.-L.}\ \bibnamefont {Qi}},\ }\bibfield  {title} {\bibinfo {title} {{Lattice construction of pseudopotential Hamiltonians for fractional Chern insulators}},\ }\href {https://doi.org/10.1103/PhysRevB.90.085103} {\bibfield  {journal} {\bibinfo  {journal} {Phys. Rev. B}\ }\textbf {\bibinfo {volume} {90}},\ \bibinfo {pages} {085103} (\bibinfo {year} {2014})}\BibitemShut {NoStop}%
\bibitem [{\citenamefont {Stanescu}\ \emph {et~al.}(2010)\citenamefont {Stanescu}, \citenamefont {Galitski},\ and\ \citenamefont {Das~Sarma}}]{PhysRevA.82.013608}%
  \BibitemOpen
  \bibfield  {author} {\bibinfo {author} {\bibfnamefont {T.~D.}\ \bibnamefont {Stanescu}}, \bibinfo {author} {\bibfnamefont {V.}~\bibnamefont {Galitski}},\ and\ \bibinfo {author} {\bibfnamefont {S.}~\bibnamefont {Das~Sarma}},\ }\bibfield  {title} {\bibinfo {title} {{Topological states in two-dimensional optical lattices}},\ }\href {https://doi.org/10.1103/PhysRevA.82.013608} {\bibfield  {journal} {\bibinfo  {journal} {Phys. Rev. A}\ }\textbf {\bibinfo {volume} {82}},\ \bibinfo {pages} {013608} (\bibinfo {year} {2010})}\BibitemShut {NoStop}%
\bibitem [{\citenamefont {Raoux}\ \emph {et~al.}(2014)\citenamefont {Raoux}, \citenamefont {Morigi}, \citenamefont {Fuchs}, \citenamefont {Pi\'echon},\ and\ \citenamefont {Montambaux}}]{PhysRevLett.112.026402}%
  \BibitemOpen
  \bibfield  {author} {\bibinfo {author} {\bibfnamefont {A.}~\bibnamefont {Raoux}}, \bibinfo {author} {\bibfnamefont {M.}~\bibnamefont {Morigi}}, \bibinfo {author} {\bibfnamefont {J.-N.}\ \bibnamefont {Fuchs}}, \bibinfo {author} {\bibfnamefont {F.}~\bibnamefont {Pi\'echon}},\ and\ \bibinfo {author} {\bibfnamefont {G.}~\bibnamefont {Montambaux}},\ }\bibfield  {title} {\bibinfo {title} {{From Dia- to Paramagnetic Orbital Susceptibility of Massless Fermions}},\ }\href {https://doi.org/10.1103/PhysRevLett.112.026402} {\bibfield  {journal} {\bibinfo  {journal} {Phys. Rev. Lett.}\ }\textbf {\bibinfo {volume} {112}},\ \bibinfo {pages} {026402} (\bibinfo {year} {2014})}\BibitemShut {NoStop}%
\bibitem [{\citenamefont {Lee}\ \emph {et~al.}(2024)\citenamefont {Lee}, \citenamefont {Fu},\ and\ \citenamefont {Ang}}]{PhysRevB.109.235105}%
  \BibitemOpen
  \bibfield  {author} {\bibinfo {author} {\bibfnamefont {K.~W.}\ \bibnamefont {Lee}}, \bibinfo {author} {\bibfnamefont {P.-H.}\ \bibnamefont {Fu}},\ and\ \bibinfo {author} {\bibfnamefont {Y.~S.}\ \bibnamefont {Ang}},\ }\bibfield  {title} {\bibinfo {title} {{Interplay between Haldane and modified Haldane models in $\ensuremath{\alpha}\text{\ensuremath{-}}{T}_{3}$ lattice: Band structures, phase diagrams, and edge states}},\ }\href {https://doi.org/10.1103/PhysRevB.109.235105} {\bibfield  {journal} {\bibinfo  {journal} {Phys. Rev. B}\ }\textbf {\bibinfo {volume} {109}},\ \bibinfo {pages} {235105} (\bibinfo {year} {2024})}\BibitemShut {NoStop}%
\bibitem [{\citenamefont {Vidal}\ \emph {et~al.}(1998)\citenamefont {Vidal}, \citenamefont {Mosseri},\ and\ \citenamefont {Dou\ifmmode~\mbox{\c{c}}\else \c{c}\fi{}ot}}]{PhysRevLett.81.5888}%
  \BibitemOpen
  \bibfield  {author} {\bibinfo {author} {\bibfnamefont {J.}~\bibnamefont {Vidal}}, \bibinfo {author} {\bibfnamefont {R.}~\bibnamefont {Mosseri}},\ and\ \bibinfo {author} {\bibfnamefont {B.}~\bibnamefont {Dou\ifmmode~\mbox{\c{c}}\else \c{c}\fi{}ot}},\ }\bibfield  {title} {\bibinfo {title} {{Aharonov-Bohm Cages in Two-Dimensional Structures}},\ }\href {https://doi.org/10.1103/PhysRevLett.81.5888} {\bibfield  {journal} {\bibinfo  {journal} {Phys. Rev. Lett.}\ }\textbf {\bibinfo {volume} {81}},\ \bibinfo {pages} {5888} (\bibinfo {year} {1998})}\BibitemShut {NoStop}%
\bibitem [{\citenamefont {Vidal}\ \emph {et~al.}(2001)\citenamefont {Vidal}, \citenamefont {Butaud}, \citenamefont {Dou\ifmmode~\mbox{\c{c}}\else \c{c}\fi{}ot},\ and\ \citenamefont {Mosseri}}]{PhysRevB.64.155306}%
  \BibitemOpen
  \bibfield  {author} {\bibinfo {author} {\bibfnamefont {J.}~\bibnamefont {Vidal}}, \bibinfo {author} {\bibfnamefont {P.}~\bibnamefont {Butaud}}, \bibinfo {author} {\bibfnamefont {B.}~\bibnamefont {Dou\ifmmode~\mbox{\c{c}}\else \c{c}\fi{}ot}},\ and\ \bibinfo {author} {\bibfnamefont {R.}~\bibnamefont {Mosseri}},\ }\bibfield  {title} {\bibinfo {title} {{Disorder and interactions in Aharonov-Bohm cages}},\ }\href {https://doi.org/10.1103/PhysRevB.64.155306} {\bibfield  {journal} {\bibinfo  {journal} {Phys. Rev. B}\ }\textbf {\bibinfo {volume} {64}},\ \bibinfo {pages} {155306} (\bibinfo {year} {2001})}\BibitemShut {NoStop}%
\bibitem [{\citenamefont {Oka}\ and\ \citenamefont {Kitamura}(2019)}]{annurev:/content/journals/10.1146/annurev-conmatphys-031218-013423}%
  \BibitemOpen
  \bibfield  {author} {\bibinfo {author} {\bibfnamefont {T.}~\bibnamefont {Oka}}\ and\ \bibinfo {author} {\bibfnamefont {S.}~\bibnamefont {Kitamura}},\ }\bibfield  {title} {\bibinfo {title} {{Floquet Engineering of Quantum Materials}},\ }\href {https://doi.org/https://doi.org/10.1146/annurev-conmatphys-031218-013423"} {\bibfield  {journal} {\bibinfo  {journal} {Annu. Rev. Condens. Matter Phys.}\ }\textbf {\bibinfo {volume} {10}},\ \bibinfo {pages} {387} (\bibinfo {year} {2019})}\BibitemShut {NoStop}%
\bibitem [{\citenamefont {Lee}\ \emph {et~al.}(2025)\citenamefont {Lee}, \citenamefont {Calderon}, \citenamefont {Yu}, \citenamefont {Lee}, \citenamefont {Ang},\ and\ \citenamefont {Fu}}]{PhysRevB.111.045406}%
  \BibitemOpen
  \bibfield  {author} {\bibinfo {author} {\bibfnamefont {K.~W.}\ \bibnamefont {Lee}}, \bibinfo {author} {\bibfnamefont {M.~J.~A.}\ \bibnamefont {Calderon}}, \bibinfo {author} {\bibfnamefont {X.-L.}\ \bibnamefont {Yu}}, \bibinfo {author} {\bibfnamefont {C.~H.}\ \bibnamefont {Lee}}, \bibinfo {author} {\bibfnamefont {Y.~S.}\ \bibnamefont {Ang}},\ and\ \bibinfo {author} {\bibfnamefont {P.-H.}\ \bibnamefont {Fu}},\ }\bibfield  {title} {\bibinfo {title} {{Floquet engineering of topological phase transitions in a quantum spin Hall $\ensuremath{\alpha}\text{\ensuremath{-}}{T}_{3}$ system}},\ }\href {https://doi.org/10.1103/PhysRevB.111.045406} {\bibfield  {journal} {\bibinfo  {journal} {Phys. Rev. B}\ }\textbf {\bibinfo {volume} {111}},\ \bibinfo {pages} {045406} (\bibinfo {year} {2025})}\BibitemShut {NoStop}%
\bibitem [{\citenamefont {Goldman}\ and\ \citenamefont {Dalibard}(2014)}]{PhysRevX.4.031027}%
  \BibitemOpen
  \bibfield  {author} {\bibinfo {author} {\bibfnamefont {N.}~\bibnamefont {Goldman}}\ and\ \bibinfo {author} {\bibfnamefont {J.}~\bibnamefont {Dalibard}},\ }\bibfield  {title} {\bibinfo {title} {{Periodically Driven Quantum Systems: Effective Hamiltonians and Engineered Gauge Fields}},\ }\href {https://doi.org/10.1103/PhysRevX.4.031027} {\bibfield  {journal} {\bibinfo  {journal} {Phys. Rev. X}\ }\textbf {\bibinfo {volume} {4}},\ \bibinfo {pages} {031027} (\bibinfo {year} {2014})}\BibitemShut {NoStop}%
\bibitem [{\citenamefont {Rahav}\ \emph {et~al.}(2003)\citenamefont {Rahav}, \citenamefont {Gilary},\ and\ \citenamefont {Fishman}}]{PhysRevA.68.013820}%
  \BibitemOpen
  \bibfield  {author} {\bibinfo {author} {\bibfnamefont {S.}~\bibnamefont {Rahav}}, \bibinfo {author} {\bibfnamefont {I.}~\bibnamefont {Gilary}},\ and\ \bibinfo {author} {\bibfnamefont {S.}~\bibnamefont {Fishman}},\ }\bibfield  {title} {\bibinfo {title} {{Effective Hamiltonians for periodically driven systems}},\ }\href {https://doi.org/10.1103/PhysRevA.68.013820} {\bibfield  {journal} {\bibinfo  {journal} {Phys. Rev. A}\ }\textbf {\bibinfo {volume} {68}},\ \bibinfo {pages} {013820} (\bibinfo {year} {2003})}\BibitemShut {NoStop}%
\bibitem [{\citenamefont {Plekhanov}\ \emph {et~al.}(2017)\citenamefont {Plekhanov}, \citenamefont {Roux},\ and\ \citenamefont {Le~Hur}}]{PhysRevB.95.045102}%
  \BibitemOpen
  \bibfield  {author} {\bibinfo {author} {\bibfnamefont {K.}~\bibnamefont {Plekhanov}}, \bibinfo {author} {\bibfnamefont {G.}~\bibnamefont {Roux}},\ and\ \bibinfo {author} {\bibfnamefont {K.}~\bibnamefont {Le~Hur}},\ }\bibfield  {title} {\bibinfo {title} {{Floquet engineering of Haldane Chern insulators and chiral bosonic phase transitions}},\ }\href {https://doi.org/10.1103/PhysRevB.95.045102} {\bibfield  {journal} {\bibinfo  {journal} {Phys. Rev. B}\ }\textbf {\bibinfo {volume} {95}},\ \bibinfo {pages} {045102} (\bibinfo {year} {2017})}\BibitemShut {NoStop}%
\bibitem [{\citenamefont {Lee}\ \emph {et~al.}(2018)\citenamefont {Lee}, \citenamefont {Ho}, \citenamefont {Yang}, \citenamefont {Gong},\ and\ \citenamefont {Papi\ifmmode~\acute{c}\else \'{c}\fi{}}}]{PhysRevLett.121.237401}%
  \BibitemOpen
  \bibfield  {author} {\bibinfo {author} {\bibfnamefont {C.~H.}\ \bibnamefont {Lee}}, \bibinfo {author} {\bibfnamefont {W.~W.}\ \bibnamefont {Ho}}, \bibinfo {author} {\bibfnamefont {B.}~\bibnamefont {Yang}}, \bibinfo {author} {\bibfnamefont {J.}~\bibnamefont {Gong}},\ and\ \bibinfo {author} {\bibfnamefont {Z.}~\bibnamefont {Papi\ifmmode~\acute{c}\else \'{c}\fi{}}},\ }\bibfield  {title} {\bibinfo {title} {{Floquet Mechanism for Non-Abelian Fractional Quantum Hall States}},\ }\href {https://doi.org/10.1103/PhysRevLett.121.237401} {\bibfield  {journal} {\bibinfo  {journal} {Phys. Rev. Lett.}\ }\textbf {\bibinfo {volume} {121}},\ \bibinfo {pages} {237401} (\bibinfo {year} {2018})}\BibitemShut {NoStop}%
\bibitem [{\citenamefont {Qin}\ \emph {et~al.}(2022)\citenamefont {Qin}, \citenamefont {Lee},\ and\ \citenamefont {Chen}}]{PhysRevB.106.235405}%
  \BibitemOpen
  \bibfield  {author} {\bibinfo {author} {\bibfnamefont {F.}~\bibnamefont {Qin}}, \bibinfo {author} {\bibfnamefont {C.~H.}\ \bibnamefont {Lee}},\ and\ \bibinfo {author} {\bibfnamefont {R.}~\bibnamefont {Chen}},\ }\bibfield  {title} {\bibinfo {title} {{Light-induced phase crossovers in a quantum spin Hall system}},\ }\href {https://doi.org/10.1103/PhysRevB.106.235405} {\bibfield  {journal} {\bibinfo  {journal} {Phys. Rev. B}\ }\textbf {\bibinfo {volume} {106}},\ \bibinfo {pages} {235405} (\bibinfo {year} {2022})}\BibitemShut {NoStop}%
\bibitem [{\citenamefont {Qin}\ \emph {et~al.}(2024)\citenamefont {Qin}, \citenamefont {Chen},\ and\ \citenamefont {Lee}}]{qin2024light}%
  \BibitemOpen
  \bibfield  {author} {\bibinfo {author} {\bibfnamefont {F.}~\bibnamefont {Qin}}, \bibinfo {author} {\bibfnamefont {R.}~\bibnamefont {Chen}},\ and\ \bibinfo {author} {\bibfnamefont {C.~H.}\ \bibnamefont {Lee}},\ }\bibfield  {title} {\bibinfo {title} {{Light-enhanced nonlinear Hall effect}},\ }\href {https://doi.org/https://doi.org/10.1038/s42005-024-01820-5} {\bibfield  {journal} {\bibinfo  {journal} {Commun Phys}\ }\textbf {\bibinfo {volume} {7}},\ \bibinfo {pages} {368} (\bibinfo {year} {2024})}\BibitemShut {NoStop}%
\bibitem [{\citenamefont {Stegmaier}\ \emph {et~al.}(2024)\citenamefont {Stegmaier}, \citenamefont {Fritzsche}, \citenamefont {Sorbello}, \citenamefont {Greiter}, \citenamefont {Brand}, \citenamefont {Barko}, \citenamefont {Hofer}, \citenamefont {Schwingenschlögl}, \citenamefont {Moessner}, \citenamefont {Lee}, \citenamefont {Szameit}, \citenamefont {Alu}, \citenamefont {Kießling},\ and\ \citenamefont {Thomale}}]{stegmaier2024topologicaledgestatenucleation}%
  \BibitemOpen
  \bibfield  {author} {\bibinfo {author} {\bibfnamefont {A.}~\bibnamefont {Stegmaier}}, \bibinfo {author} {\bibfnamefont {A.}~\bibnamefont {Fritzsche}}, \bibinfo {author} {\bibfnamefont {R.}~\bibnamefont {Sorbello}}, \bibinfo {author} {\bibfnamefont {M.}~\bibnamefont {Greiter}}, \bibinfo {author} {\bibfnamefont {H.}~\bibnamefont {Brand}}, \bibinfo {author} {\bibfnamefont {C.}~\bibnamefont {Barko}}, \bibinfo {author} {\bibfnamefont {M.}~\bibnamefont {Hofer}}, \bibinfo {author} {\bibfnamefont {U.}~\bibnamefont {Schwingenschlögl}}, \bibinfo {author} {\bibfnamefont {R.}~\bibnamefont {Moessner}}, \bibinfo {author} {\bibfnamefont {C.~H.}\ \bibnamefont {Lee}}, \bibinfo {author} {\bibfnamefont {A.}~\bibnamefont {Szameit}}, \bibinfo {author} {\bibfnamefont {A.}~\bibnamefont {Alu}}, \bibinfo {author} {\bibfnamefont {T.}~\bibnamefont {Kießling}},\ and\ \bibinfo {author} {\bibfnamefont {R.}~\bibnamefont {Thomale}},\ }\href {https://doi.org/https://doi.org/10.48550/arXiv.2407.10191} {\bibinfo {title} {{Topological Edge
  State Nucleation in Frequency Space and its Realization with Floquet Electrical Circuits}}} (\bibinfo {year} {2024}),\ \Eprint {https://arxiv.org/abs/2407.10191} {arXiv:2407.10191 [cond-mat.mes-hall]} \BibitemShut {NoStop}%
\bibitem [{\citenamefont {Li}\ \emph {et~al.}(2024)\citenamefont {Li}, \citenamefont {Fu}, \citenamefont {Liu},\ and\ \citenamefont {Wang}}]{PhysRevB.109.045403}%
  \BibitemOpen
  \bibfield  {author} {\bibinfo {author} {\bibfnamefont {R.}~\bibnamefont {Li}}, \bibinfo {author} {\bibfnamefont {P.-H.}\ \bibnamefont {Fu}}, \bibinfo {author} {\bibfnamefont {J.-F.}\ \bibnamefont {Liu}},\ and\ \bibinfo {author} {\bibfnamefont {J.}~\bibnamefont {Wang}},\ }\bibfield  {title} {\bibinfo {title} {{Armchair edge states in shear-strained graphene: Magnetic properties and quantum valley Hall edge states}},\ }\href {https://doi.org/10.1103/PhysRevB.109.045403} {\bibfield  {journal} {\bibinfo  {journal} {Phys. Rev. B}\ }\textbf {\bibinfo {volume} {109}},\ \bibinfo {pages} {045403} (\bibinfo {year} {2024})}\BibitemShut {NoStop}%
\bibitem [{\citenamefont {Guo}\ \emph {et~al.}(2024{\natexlab{b}})\citenamefont {Guo}, \citenamefont {Liu}, \citenamefont {Yu},\ and\ \citenamefont {Liu}}]{PhysRevB.110.L220402}%
  \BibitemOpen
  \bibfield  {author} {\bibinfo {author} {\bibfnamefont {S.-D.}\ \bibnamefont {Guo}}, \bibinfo {author} {\bibfnamefont {Y.}~\bibnamefont {Liu}}, \bibinfo {author} {\bibfnamefont {J.}~\bibnamefont {Yu}},\ and\ \bibinfo {author} {\bibfnamefont {C.-C.}\ \bibnamefont {Liu}},\ }\bibfield  {title} {\bibinfo {title} {{Valley polarization in twisted altermagnetism}},\ }\href {https://doi.org/10.1103/PhysRevB.110.L220402} {\bibfield  {journal} {\bibinfo  {journal} {Phys. Rev. B}\ }\textbf {\bibinfo {volume} {110}},\ \bibinfo {pages} {L220402} (\bibinfo {year} {2024}{\natexlab{b}})}\BibitemShut {NoStop}%
\bibitem [{\citenamefont {Liu}\ \emph {et~al.}(2011)\citenamefont {Liu}, \citenamefont {Jiang},\ and\ \citenamefont {Yao}}]{PhysRevB.84.195430}%
  \BibitemOpen
  \bibfield  {author} {\bibinfo {author} {\bibfnamefont {C.-C.}\ \bibnamefont {Liu}}, \bibinfo {author} {\bibfnamefont {H.}~\bibnamefont {Jiang}},\ and\ \bibinfo {author} {\bibfnamefont {Y.}~\bibnamefont {Yao}},\ }\bibfield  {title} {\bibinfo {title} {{Low-energy effective Hamiltonian involving spin-orbit coupling in silicene and two-dimensional germanium and tin}},\ }\href {https://doi.org/10.1103/PhysRevB.84.195430} {\bibfield  {journal} {\bibinfo  {journal} {Phys. Rev. B}\ }\textbf {\bibinfo {volume} {84}},\ \bibinfo {pages} {195430} (\bibinfo {year} {2011})}\BibitemShut {NoStop}%
\bibitem [{\citenamefont {L\"u}\ \emph {et~al.}(2024)\citenamefont {L\"u}, \citenamefont {Zhang}, \citenamefont {Fu},\ and\ \citenamefont {Liu}}]{PhysRevResearch.6.043108}%
  \BibitemOpen
  \bibfield  {author} {\bibinfo {author} {\bibfnamefont {X.-L.}\ \bibnamefont {L\"u}}, \bibinfo {author} {\bibfnamefont {Y.-C.}\ \bibnamefont {Zhang}}, \bibinfo {author} {\bibfnamefont {P.-H.}\ \bibnamefont {Fu}},\ and\ \bibinfo {author} {\bibfnamefont {J.-F.}\ \bibnamefont {Liu}},\ }\bibfield  {title} {\bibinfo {title} {{Phase diagrams and topological mixed edge states in silicene with intrinsic and extrinsic Rashba effects}},\ }\href {https://doi.org/10.1103/PhysRevResearch.6.043108} {\bibfield  {journal} {\bibinfo  {journal} {Phys. Rev. Res.}\ }\textbf {\bibinfo {volume} {6}},\ \bibinfo {pages} {043108} (\bibinfo {year} {2024})}\BibitemShut {NoStop}%
\bibitem [{\citenamefont {Lin}\ \emph {et~al.}(2023)\citenamefont {Lin}, \citenamefont {Tan}, \citenamefont {Fu},\ and\ \citenamefont {Liu}}]{Lin_Fu}%
  \BibitemOpen
  \bibfield  {author} {\bibinfo {author} {\bibfnamefont {S.-Q.}\ \bibnamefont {Lin}}, \bibinfo {author} {\bibfnamefont {H.}~\bibnamefont {Tan}}, \bibinfo {author} {\bibfnamefont {P.-H.}\ \bibnamefont {Fu}},\ and\ \bibinfo {author} {\bibfnamefont {J.-F.}\ \bibnamefont {Liu}},\ }\bibfield  {title} {\bibinfo {title} {{Interaction-driven Chern insulating phases in the $\alpha$-$T_{3}$ lattice with Rashba spin-orbit coupling}},\ }\href {https://www.sciencedirect.com/science/article/pii/S2589004223016231} {\bibfield  {journal} {\bibinfo  {journal} {iScience}\ }\textbf {\bibinfo {volume} {26}} (\bibinfo {year} {2023})}\BibitemShut {NoStop}%
\bibitem [{\citenamefont {Dai}\ \emph {et~al.}(2024)\citenamefont {Dai}, \citenamefont {Fu}, \citenamefont {Ang},\ and\ \citenamefont {Chen}}]{PhysRevB.110.195409}%
  \BibitemOpen
  \bibfield  {author} {\bibinfo {author} {\bibfnamefont {X.}~\bibnamefont {Dai}}, \bibinfo {author} {\bibfnamefont {P.-H.}\ \bibnamefont {Fu}}, \bibinfo {author} {\bibfnamefont {Y.~S.}\ \bibnamefont {Ang}},\ and\ \bibinfo {author} {\bibfnamefont {Q.}~\bibnamefont {Chen}},\ }\bibfield  {title} {\bibinfo {title} {{Two-dimensional Weyl nodal-line semimetal and antihelical edge states in a modified Kane-Mele model}},\ }\href {https://doi.org/10.1103/PhysRevB.110.195409} {\bibfield  {journal} {\bibinfo  {journal} {Phys. Rev. B}\ }\textbf {\bibinfo {volume} {110}},\ \bibinfo {pages} {195409} (\bibinfo {year} {2024})}\BibitemShut {NoStop}%
\bibitem [{\citenamefont {Dirac}(1928)}]{dirac1928quantum}%
  \BibitemOpen
  \bibfield  {author} {\bibinfo {author} {\bibfnamefont {P.~A.~M.}\ \bibnamefont {Dirac}},\ }\bibfield  {title} {\bibinfo {title} {{The quantum theory of the electron}},\ }\href {https://doi.org/https://doi.org/10.1098/rspa.1928.0023} {\bibfield  {journal} {\bibinfo  {journal} {Proc. R. Soc. Lond. A}\ }\textbf {\bibinfo {volume} {117}},\ \bibinfo {pages} {610} (\bibinfo {year} {1928})}\BibitemShut {NoStop}%
\bibitem [{\citenamefont {Dirac}(1930)}]{Dirac1930-DIRTPO}%
  \BibitemOpen
  \bibfield  {author} {\bibinfo {author} {\bibfnamefont {P.~A.~M.}\ \bibnamefont {Dirac}},\ }\href@noop {} {\emph {\bibinfo {title} {{The Principles of Quantum Mechanics}}}}\ (\bibinfo  {publisher} {Clarendon Press},\ \bibinfo {address} {Oxford},\ \bibinfo {year} {1930})\BibitemShut {NoStop}%
\end{thebibliography}%
\end{document}